\def\timeSpacing{T}
\def\frequencySpacing{F}

\def\timeSpacingVariable{\tau_0}
\def\frequencySpacingVariable{\nu_0}

\def\heisenbergParameter{\xi}

\def\timeDispersion{\sigma_{\rm t}}

\def\timeGravity{\bar{\timeSymbol}}
\def\frequencyGravity{\bar{\frequencySymbol}}
\def\frequencyDispersion{\sigma_{\rm f}}

\def\timeSymbol{t}
\def\frequencySymbol{f}

\def\rollOff{\alpha}
\def\gaussianRollOff{\rho}
\def\truncation{K}
\def\filterLength{T_{\rm filter}}

\def\integrald{{\rm d}}
\def\RMSdelaySpread{\tau_{\rm rms}}
\def\MAXdelay{\tau_{\rm max}}

\def\MAXshift{f_{\rm d_{max}}}

\def\numberOfSubcarrier{N}

\def\ambiguityFunction[#1][#2][#3]{A\left(#1,#2\right)}
\def\interferenceFunction[#1][#2]{I\left(#1,#2\right)}
\def\ambiguityFunctionC[#1][#2][#3]{A^*\left(#1,#2\right)}
\def\ambiguityFunctionReal[#1][#2][#3]{A_{\rm \Re}\left(#1,#2\right)}
\def\timeShift{\tau}
\def\frequencyShift{\nu}
\def\timeShiftOther{\tau_1}
\def\frequencyShiftOther{\nu_1}

\def\prototypeFilterTX[#1]{p_{\rm tx}\left(#1\right)}
\def\prototypeFilterTXc[#1]{p^*_{\rm tx}\left(#1\right)}
\def\prototypeFilterTXv[#1]{p_{\rm tx}}
\def\prototypeFilterRX[#1]{p_{\rm rx}\left(#1\right)}
\def\prototypeFilterRXc[#1]{p^*_{\rm rx}\left(#1\right)}
\def\prototypeFilterRXv[#1]{p_{\rm rx}}

\def\noiseProcess{w(\timeSymbol)}

\def\bitPerSymbol{\beta}
\def\transmittedSignal[#1]{x(#1)}
\def\transmittedSignalD[#1]{\acute{x}(#1)}
\def\receivedSignal{y(\timeSymbol)}
\def\timeIndexTX{m}
\def\frequencyIndexTX{k}
\def\timeIndex{n}
\def\frequencyIndex{l}

\def\symbols[#1][#2]{X_{#1#2}}
\def\symbolsReceived[#1][#2]{\tilde{X}_{#1#2}}
\def\channelProcessTime{h(\timeShift,\timeSymbol)}
\def\channelProcess{H(\timeShift,\frequencyShift)}
\def\channelProcessOther{H^*(\timeShiftOther,\frequencyShiftOther)}

\def\ambiguityTimeShift{\phi}
\def\ambiguityFrequencyShift{\psi}
\def\ambiguityTimeFrequencyShift{\rho}
\def\signalPower{\sigma^2_{\rm S}}
\def\interferencePower{\sigma^2_{\rm I}}
\def\scatteringFunction{S_{H}\left(\timeShift, \frequencyShift\right)}

\def\translatedModulatedFilterTX[#1][#2][#3]{g_{#1#2}(#3)}
\def\translatedModulatedFilter[#1][#2][#3]{\gamma_{#1#2}(#3)}
\def\translatedModulatedFilterC[#1][#2][#3]{\gamma_{#1#2}^*(#3)}

\def\translatedModulatedVectorFilterTX[#1][#2]{p_{#1#2}}
\def\translatedModulatedVectorFilter[#1][#2]{\acute{p}_{#1,#2}}
\def\translatedModulatedVectorFilterC[#1][#2]{\acute{p}_{#1,#2}^*}

\def\vectorBasis{Q}
\def\frameOperator{S}
\def\gramOperator{R}

\def\lattice{\Lambda}
\def\latticeGenerator{L}
\def\latticeDensity{\delta(\lattice)}
\def\bandwidthEfficiency{\epsilon}
\def\latticeVolume{{\rm vol}(\lattice)}
\def\exponentialTerm[#1][#2][#3]{W_{#1}^{#2#3}}

\def\SIR{SIR}

\def\directionParameter{\kappa}

\def\pulseTime[#1]{p_{\rm #1}(\timeSymbol)}
\def\pulseFrequency[#1]{P_{\rm #1}(\frequencySymbol)}

\def\kaiserDesignParameter{\beta}

\def\filterLength{\tau}
\def\filterBand{\sigma}
\def\timeSymbolProlate{t}

\def\gaussian{{\rm gaussian}}

\def\filterCoeffient[#1]{a_{#1}}
\def\filterCoeffientHelp[#1]{k_{#1}}
\def\filterCoefficentIndex{l}
\def\numberOfHermite{N}
\def\hermitePolynomial[#1][#2]{H_{\rm #1}\left({#2}\right)}
\def\hermitePolynomialDegree{n}

\def\numberOfProlate{N}
\def\prolatePolynomial[#1][#2]{H_{\rm #1}\left({#2}\right)}

\def\prolateWave[#1][#2]{\psi_{#1}\left(#2\right)}
\def\prolateEigen[#1]{\lambda_{#1}}

\def\determinantOperator[#1]{
{\rm det}\left({#1}\right)
}

\def\modOperator[#1][#2]{
{\rm mod}\left\{{#1},{#2}\right\}
}

\def\kroneckerDelta[#1][#2]{
{\delta}_{#1#2}
}

\def\deltaDirac[#1]{
{\delta}{(#1)}
}

\def\expectedPowerOperator[#1]{
\mathbb{E}\left[{{\left|{#1}\right|}^2}\right]
}

\def\signOperator[#1]{
{\rm sign}\left\{{#1}\right\}
}

\def\kroneckerOperator[#1]{
{\delta}_{#1}
}

\def\absOperator[#1]{
\left|{#1}\right|
}

\def\expectedOperator[#1]{
\mathbb{E}\left[{#1}\right]
}

\def\ltwonorm[#1]{
\|{#1}\|
}

\def\absSquareOperator[#1]{
{{\left|{#1}\right|}^2}
}

\def\RECformula{
\begin{array}{l}
\pulseTime[]=
\begin{cases}
1~, & |\timeSymbol|\leq\frac{1}{2}\\
0~, & {\rm otherwise}
\end{cases}
\end{array}
}

\def\eRECformula{
\begin{array}{l}
\pulseTime[]=
\begin{cases}
1~, & \absOperator[\timeSymbol] \le \frac{1-\rollOff}{2}\\
\frac{1}{2}+\frac{1}{2}\cos(\frac{\pi }{\alpha}\big(|\timeSymbol|-\frac{1-\alpha}{2}\big)\Big)~,& \frac{1-\rollOff}{2} < 
\absOperator[\timeSymbol] \le \frac{1+\rollOff}{2}\\

0~, & {\rm otherwise}
\end{cases}
\end{array}
}

\def\RCformula{
\begin{array}{l}
\pulseTime[]=\frac{\sin(\pi \timeSymbol)}{\pi \timeSymbol}\frac{\cos(\pi\rollOff t)}{1-4\rollOff^2\timeSymbol^2}
\end{array}
}

\def\Hannformula{
\begin{array}{l}
\pulseTime[]=
\begin{cases}
\frac{1}{2}+\frac{1}{2}\cos(2\pi\timeSymbol)~,& \absOperator[\timeSymbol] \le \frac{1}{2}\\

0~, & {\rm otherwise}
\end{cases}
\end{array}
}

\def\Hammformula{
\begin{array}{l}
\pulseTime[]=
\begin{cases}
\frac{25}{46}+\frac{21}{46}\cos(2\pi\timeSymbol)~,& \absOperator[\timeSymbol] \le \frac{1}{2}\\

0~, & {\rm otherwise}
\end{cases}
\end{array}
}

\def\Blackformula{
\begin{array}{l}
\pulseTime[]=
\begin{cases}
\frac{7938}{18608}+\frac{9240}{18608}\cos(2\pi\timeSymbol)+\frac{1430}{18608}\cos(4\pi\timeSymbol)~,& \absOperator[\timeSymbol] \le \frac{1}{2}\\

0~, & {\rm otherwise}
\end{cases}
\end{array}
}

\def\RRCformula{
\begin{array}{l}
\pulseTime[]=
\begin{cases}
1-\rollOff+4\frac{\rollOff}{\pi}, & \timeSymbol=0\\
\frac{\rollOff}{\sqrt{2}}\Big[\Big(1+\frac{2}{\pi}\Big)
\sin\big(\frac{\pi}{4\rollOff}\Big)+\Big(1-\frac{2}{\pi}\big)
\cos\big(\frac{\pi}{4\rollOff}\big)\Big], & \timeSymbol=\pm \frac{1}{4\rollOff}\\
\frac{\sin\Big((1-\rollOff){\pi \timeSymbol}\Big)+{4\rollOff \timeSymbol}\cos\Big((1+\rollOff){\pi
\timeSymbol}\Big)}{{\pi \timeSymbol}\Big(1-{16\rollOff^2\timeSymbol^2}\Big)}~, &~{\rm otherwise}
\end{cases}
\end{array}
}

\def\PHYformula{
\begin{array}{l}
\pulseTime[]=
\begin{cases}
\filterCoeffient[0]+2\displaystyle\sum_{\filterCoefficentIndex=1}^{\truncation-1}\filterCoeffient[\filterCoefficentIndex]\cos(2\pi{\filterCoefficentIndex\timeSymbol} )\nonumber~,  &\absOperator[\timeSymbol] \le \frac{1}{2}~,\\
0~,  &{\rm otherwise}
\end{cases}
\\~~~~~~~~~~
\filterCoeffientHelp[\filterCoefficentIndex] = (-1)^\filterCoefficentIndex\filterCoeffient[\filterCoefficentIndex]
~,~
\filterCoeffientHelp[0]=-1
~,~
\filterCoeffientHelp[\filterCoefficentIndex]^2+\filterCoeffientHelp[\truncation-\filterCoefficentIndex]^2=1
~,~
\filterCoeffientHelp[0]+2\displaystyle\sum_{\filterCoefficentIndex=1}^{\truncation-1}\filterCoeffientHelp[\filterCoefficentIndex]=0
~,~\\~~~~~~~~~
\displaystyle\sum_{\filterCoefficentIndex=1}^{\truncation-1}\filterCoefficentIndex^q\filterCoeffientHelp[\filterCoefficentIndex]=0 ~,~
q\ge 2~,~ q\in \{2n|n\in\mathbb{Z}\}
\end{array}
}

\def\GAUformula{
\begin{array}{l}
\pulseTime[]=
(2\gaussianRollOff)^{1/4}e^{-\pi\gaussianRollOff\timeSymbol^2}
~,~
\pulseFrequency[]=
p_\gaussian(t,1/\gaussianRollOff)
\end{array}
}

\def\HERformula{
\begin{array}{l}
\pulseTime[]=
\displaystyle\sum_{\filterCoefficentIndex=0}^{\numberOfHermite} \filterCoeffient[4\filterCoefficentIndex]\prolateWave[4\filterCoefficentIndex,\infty,\infty][\timeSymbol]~, \\~~~~~~~~~
\prolateWave[\hermitePolynomialDegree,\infty,\infty][\timeSymbol]=
\hermitePolynomial[\hermitePolynomialDegree][\sqrt{2\pi}\timeSymbol] e^{-\pi \timeSymbol^2}~,~
\hermitePolynomial[\hermitePolynomialDegree][\timeSymbol]=(-1)^\hermitePolynomialDegree e^{\timeSymbol^2}\frac{d^\hermitePolynomialDegree}{\integrald\timeSymbol^\hermitePolynomialDegree}e^{-\timeSymbol^2}~
\end{array}
}

\def\Prolateformula{
\begin{array}{l}
\pulseTime[]=
{\arg\underset{\pulseTime[]}\min}
\left\{ 
\int_{-\infty}^\infty |\pulseFrequency[]|^2\integrald\frequencySymbol-\int_{-\filterBand}^{\filterBand}|\pulseFrequency[]|^2\integrald\frequencySymbol\right\}~.

\end{array}
}

\def\OFDPformula{
\begin{array}{l}
\pulseTime[]=
\displaystyle\sum_{\filterCoefficentIndex=0}^{\numberOfProlate}
\filterCoeffient[2\filterCoefficentIndex]\prolateWave[2\filterCoefficentIndex,\filterLength,\filterBand][\timeSymbol]
\end{array}
}

\def\IOTAformula{
\begin{array}{l}
\pulseTime[]=
\mathcal{F}^{-1}\mathcal{O}_{\tau_0}\mathcal{F}\mathcal{O}_{\upsilon_0}p_{\rm \gaussian}(t)\\~~~~~~~~~
\mathcal{O}_a x(t)=\frac{x(t)}{\sqrt{a\sum_{k=-\infty}^{\infty}{\ltwonorm[x(t-ka)]}^2}}~,~x(t)\in\mathbb{R}
\\~~~~~~~~~
\mathcal{F}^{-1}X(\frequencySymbol) = \displaystyle\int X(f)e^{-j2\pi\frequencySymbol\timeSymbol}\integrald\frequencySymbol~,~\mathcal{F}x(t) = \displaystyle\int x(\timeSymbol)e^{j2\pi\frequencySymbol\timeSymbol}\integrald\timeSymbol
\end{array}
}

\def\EGFformula{
\begin{array}{c}
\pulseTime[]=
\frac{1}{2}\bigg[\displaystyle\sum_{k=0}^\infty d_{k,\gaussianRollOff,\upsilon_0}\Big[p_{\gaussian}(t+k/\upsilon_0,\gaussianRollOff)+p_{\gaussian}(t-k\upsilon_0,\gaussianRollOff)\Big]\bigg]\times\\ \displaystyle\sum_{l=0}^\infty d_{l,1/\gaussianRollOff,\tau_0}\cos(2\pi lt/\tau_0)
\end{array}
}

\def\Kaiserformula{
\begin{array}{l}
\pulseTime[]=
\begin{cases}
\frac{
I_0\left(\beta\sqrt{1-4\timeSymbol^2}\right)}{I_0(\beta)}~, & \absOperator[\timeSymbol]\leq \frac{1}{2} \\
0~, &~{\rm otherwise}
\end{cases}
\\~~~~~~~~~
I_0(x)=1+\sum_{k=1}^\infty\Big[\frac{(x/2)^k}{k!}\Big]^2
\end{array}
}

\def\modifiedKaiserformula{
\begin{array}{l}
\pulseTime[]=
\begin{cases}
\frac{I_0\left(\beta\sqrt{1-4\timeSymbol^2}\right)-1}{I_0(\beta)-1}~, & \absOperator[\timeSymbol]\leq \frac{1}{2} \\
0~, &~{\rm otherwise}
\end{cases}
\end{array}
}

\documentclass[journal]{IEEEtran}

\usepackage{xcolor}
\usepackage{amsfonts}
\usepackage{acronym}
\usepackage{cite}
\usepackage{amsmath,amssymb}
\usepackage[dvips]{graphicx}

\usepackage[caption=false,font=footnotesize]{subfig}
\usepackage[geometry]{ifsym}

\usepackage{multirow}
\usepackage{hhline}

\usepackage{flushend}
\usepackage{rotating}
\begin{document}

\makeatletter
\def\@@acrodef{\@ifstar\@acrodefs\@acrodef}
\newtoks\acro@list
\newcommand{\@acrodef}[2]{%
  \global\acro@list=\expandafter{\the\acro@list\@elt{#1}{#2}}%
  \global\@namedef{acro@#1}{n{#1}{#2}}}
\newtoks\acro@resetlist
\newcommand{\@acrodefs}[2]{%
  \global\acro@resetlist=\expandafter{\the\acro@resetlist\@elt{#1}}%
  \@acrodef{#1}{#2}}
\def\acro@doresetlist{\begingroup
  \def\@elt##1{\expandafter\expandafter\expandafter
    \acro@reset\csname acro@##1\endcsname}\the\acro@resetlist\endgroup}
\def\acro@reset#1#2#3{\global\@namedef{acro@#2}{n{#2}{#3}}}
\newcommand{\acro}[1]{\expandafter\expandafter\expandafter
  \use@acro\csname acro@#1\endcsname}
\def\use@acro#1#2#3{\ifx n#1
  #3 (#2)\global\@namedef{acro@#2}{o{#2}{#3}}%
  \else
  #2%
\fi}
\newcommand{\listofacronyms}[1][tabular]{%
  \begingroup\def\@elt##1##2{##1&##2\\}%
  \@ifundefined{chapter}{\section*}{\chapter*}{\listacronymname}
  \noindent\begin{#1}{@{}p{6em}p{\dimexpr\columnwidth-2\tabcolsep-6em\relax}@{}}
    \the\acro@list
  \end{#1}\endgroup}
\providecommand\listacronymname{List of acronyms}
\newenvironment{acronyms}{\let\acrodef\@@acrodef}{}
\newenvironment{acronyms*}{\let\acrodef\@@acrodef}{\listofacronyms}
\def\g@preto@macro#1#2{\toks0=\expandafter{#1}%
  \toks2={#2}\xdef#1{\the\toks2 \the\toks0 }}
\@ifundefined{chapter}
  {\g@preto@macro\section\acro@doresetlist}
  {\g@preto@macro\chapter\acro@doresetlist}
\makeatother

\title{\huge A Survey on Multicarrier Communications: Prototype Filters, Lattice Structures, and Implementation Aspects}

\author{\authorblockN{Alphan \c{S}ahin$^1$, {\.I}smail G\"{u}ven\c{c}$^2$, and H\"{u}seyin Arslan$^1$\\}
\authorblockA{$^1$Department of Electrical Engineering, University of South Florida, Tampa, FL, 33620}\\
\authorblockA{$^2$Department of Electrical and Computer Engineering, Florida International University, Miami, FL, 33174}\\
Email: {\tt alphan@mail.usf.edu}, {\tt iguvenc@fiu.edu}, {\tt arslan@usf.edu}
}

\maketitle

\begin{abstract}

Due to their numerous advantages, communications over multicarrier schemes constitute an appealing approach for broadband wireless systems. Especially, the strong penetration of orthogonal frequency division multiplexing (OFDM) into the communications standards has triggered heavy investigation on multicarrier systems, leading to re-consideration of different  approaches as an alternative to OFDM. The goal of the present survey is not only to provide a unified review of waveform design options for multicarrier schemes, but also to pave the way for the evolution of the multicarrier schemes from the current state of the art to future technologies. In particular, a generalized framework on multicarrier schemes is presented, based on what to transmit, i.e., symbols, how to transmit, i.e., filters, and where/when to transmit, i.e., lattice. Capitalizing on this framework, different variations of orthogonal, bi-orthogonal, and non-orthogonal multicarrier schemes are discussed. In addition, filter design for various multicarrier systems is reviewed considering four different design perspectives: energy concentration, rapid decay, spectrum nulling, and channel/hardware characteristics. Subsequently, evaluation tools which may be used to compare different filters in multicarrier schemes are studied.  Finally, multicarrier schemes are evaluated from the view of the practical implementation issues, such as lattice adaptation, equalization, synchronization, multiple antennas, and hardware impairments.


\end{abstract}

\acrodef{HCF}{half-cosine function}
\acrodef{VSB}{vestigial side-band modulation}
\acrodef{GFDM}{Generalized frequency division multiplexing}
\acrodef{PSWF}{Prolate spheroidal wave function}
\acrodef{OFDM}{orthogonal frequency division multiplexing}
\acrodef{OFDMA}{orthogonal frequency division multiple accessing}
\acrodef{BFDM}{Bi-orthogonal frequency division multiplexing}
\acrodef{DFT}{discrete Fourier transformation}
\acrodef{IDFT}{inverse discrete Fourier transformation}
\acrodef{IFFT}{inverse fast Fourier transformation}
\acrodef{FBMC}{filter bank multicarrier}
\acrodef{CP}{cyclic prefix}
\acrodef{PAPR}{peak-to-average-power ratio}
\acrodef{QAM}{quadrature amplitude modulation}
\acrodef{OQAM}{offset quadrature amplitude modulation}
\acrodef{FFT}{fast Fourier transformation}
\acrodef{RRC}{root-raised cosine}
\acrodef{CFO}{carrier frequency offset}
\acrodef{SIR}{signal-to-interference ratio}
\acrodef{SNR}{signal-to-noise ratio}
\acrodef{SINR}{signal-to-interference-plus-noise ratio}
\acrodef{ICI}{inter-carrier interference}
\acrodef{ISI}{inter-symbol interference}
\acrodef{PPN}{polyphase network}
\acrodef{WSSUS}{wide-sense stationary uncorrelated scattering}
\acrodef{SEM}{spectral emission mask}
\acrodef{BWA}{broadband wireless access}
\acrodef{PSD}{power spectral densitie}
\acrodef{MIMO}{multiple-input multiple-output}
\acrodef{DSL}{digital subscriber lines}
\acrodef{OFDP}{optimal finite duration pulse}
\acrodef{FMT}{Filtered multitone}
\acrodef{SMT}{Staggered multitone}
\acrodef{CMT}{Cosine-modulated multitone}
\acrodef{IOTA}{Isotropic orthogonal transform algorithm}
\acrodef{RMS}{root mean square}
\acrodef{MMSE}{minimum mean square error}
\acrodef{MLD}{maximum likelihood detection}
\acrodef{STBC}{space-time block coding}
\acrodef{TO}{timing offset}
\acrodef{CFO}{carrier frequency offset}
\acrodef{IAM}{interference approximation method}
\acrodef{PN}{phase noise}
\acrodef{RF}{radio-frequency}
\acrodef{CPM}{continuous phase modulation}
\acrodef{ADC}{analog-to-digital converter}
\acrodef{PA}{power amplifier}
\acrodef{CCDF}{complementary cumulative distribution function}
\acrodef{SC-FDMA}{single carrier frequency division multiple accessing}
\acrodef{CP-OFDM}{Cyclic Prefix-OFDM}
\acrodef{ZP-OFDM}{Zero Padded-OFDM}
\acrodef{FPGA}{field-programmable gate array}
\acrodef{STTC}{space-time trellis coding}
\acrodef{BER}{bit error rate}
\acrodef{ZP}{zero-padded}
\acrodef{FDE}{frequency domain equalization}
\acrodef{TDE}{time domain equalization}
\acrodef{LS}{least square}
\acrodef{SC-FDE}{single carrier frequency domain equalization}
\acrodef{FB-S-FBMC}{filter-bank-spread-filter-bank multicarrier}
\acrodef{LTE}{Long Term Evolution}
\acrodef{AWGN}{additive white Gaussian noise}
\acrodef{RD}{random demodulator}
\acrodef{FTN}{faster-than Nyquist}
\acrodef{PRS}{partial-response signaling}
\acrodef{MAP}{maximum a posteriori}
\acrodef{DFE}{decision feedback equalizer}
\acrodef{FIR}{finite impulse response}
\acrodef{LMS}{least-mean-square}
\acrodef{RLS}{recursive-least-squares}
\acrodef{PHYDYAS}{Physical layer for dynamic access}
\acrodef{MLSE}{maximum likelihood sequence estimator}
\acrodef{OOB}{out-of-band radiation}
\acrodef{EGF}{Extended Gaussian function}
\acrodef{AIC}{Akaike information criterion}
\acrodef{SIC}{successive interference cancellation}
\acrodef{CDMA}{code division multiple accessing}

\section{Introduction}

The explosion of mobile applications and data usage in the recent years necessitate the development of adaptive, flexible, and efficient radio access technologies. To this end, multicarrier techniques have been extensively used over the last decade for broadband wireless communications. This wide interest is primarily due to their appealing characteristics, such as the support for multiuser diversity, simpler equalization, and adaptive modulation and coding techniques.

Among many other multicarrier techniques, \ac{OFDM} dominates the current broadband wireless communication systems. On the other hand, \ac{OFDM} also  suffers from several shortcomings such as high spectral leakage, stringent synchronization requirements, and susceptibility to frequency dispersion. Transition from the existing \ac{OFDM}-based multicarrier systems to the next generation radio access technologies may follow two paths.  In the first approach, existing \ac{OFDM} structure is preserved, and its shortcomings are addressed through appropriate solutions \cite{Hwang_2009}. Considering backward compatibility advantages with existing technologies, this approach has its own merits.  The second approach follows a different rationale based on a generalized framework for multicarrier schemes \cite{24_farhang2011ofdm, SPMAG2013}, which may lead to different techniques than \ac{OFDM}. In this survey, we choose to go after the second approach since it provides a wider perspective for multicarrier schemes, with \ac{OFDM} being a special case. 
Based on this strategy, the goals of the paper are listed as follow:
\begin{itemize}
	\item To provide a unified framework for multicarrier schemes along
with Gabor systems by emphasizing their basic elements: what to transmit, i.e.,
symbols, how to transmit, i.e., filters, and where/when to transmit, i.e., lattices;
	\item To extend the understanding of existing multicarrier
schemes by identifying the relations to each other;
	\item To review the existing prototype  filters in the literature considering their utilizations in multicarrier schemes;
	\item To understand the trade-offs between different multicarrier schemes in practical scenarios;
	\item To pave the way for the further developments by providing a wider perspective  on multicarrier schemes.
\end{itemize}

The survey is organized as follow: First, preliminary concepts and the terminology  are presented in Section \ref{sec:preliminary}. Various multicarrier schemes are provided in Section \ref{sec:multicarrier}, referring to the concepts introduced in Section \ref{sec:preliminary}. 
Then, known prototype filters are identified and their trade-offs are discussed in Section \ref{sec:prototypFilterDesign}. 
Useful tools and metrics to evaluate the filter performances are investigated in Section \ref{sec:toolsForPrototypeFilterAnalyses},
 transceiver design issues for multicarrier schemes are investigated in Section \ref{sec:txdesign}, and, finally, the paper is concluded in Section \ref{sec:conc}.

\section{Preliminary Concepts: Symbols, Lattices, and Filters}
\label{sec:preliminary}

The purpose of this section is to provide preliminary concepts related with multicarrier schemes along with the notations used throughout the survey. 
Starting from the basics, symbols, lattices, and filters are discussed in detail within the framework of Gabor systems.
For a comprehensive treatment on the same subject, we refer the reader to the books by I. Daubechies
\cite{Daubechies_1992}, H.G. Feichtinger and T. Strohmer \cite{feichtinger1998gabor}, and O. Christensen \cite{christensen2003}. 
Also, the reader who wants to reach the development of Gabor systems from the mathematical point of view may refer to the surveys in \cite{Benedetto_1994,Janssen_2001, Christensen_2001frames, heil_2007, SPMAG2013}. Besides, it is worth noting  \cite{Gabor_1946, Daubechies_1990,Daubechies_1991,janssen_1994_duality, 2_le1995coded,77_kozek1998nonorthogonal,95_Strohmer_ACHA_2001, 39_strohmer2003optimal,Werther_2005, Kutyniok_2005,88_jung2007wssus,80_matz2007analysis,117_Han_TSP_2007, 86_han2009wireless} constitute the key research papers which construct a bridge between the Gabor theory and its applications on communications. These studies also reveal how the Gabor theory changes the understanding of multicarrier schemes, especially, within the two last decades. Additionally,\cite{26_Du_KTH_2007, 97_Sondergaard_Thesis_2007, Jung_Thesis_2007,Du2008} are the recent complete reports and theses based on Gabor systems.

\subsection{Fundamentals}
In the classical paper by C. Shannon \cite{shannon_1998}, a geometrical representation of communication systems is presented. According to this representation, messages and corresponding signals are points in two function spaces: message space and signal space. While a transmitter maps every point in the message space into the signal space, a receiver does the reverse operation. As long as the mapping is one-to-one from the message space to the signal space, a message is always recoverable at the receiver. 
Based on this framework, a waveform corresponds to a specific structure in the signal space and identifies the formation of the signals. 
Throughout this survey, the signal space is considered as a time-frequency plane where time and frequency constitute its coordinates, which is a well-known notation for representing one dimensional signals in two dimensions \cite{Gabor_1946, Daubechies_1990}. 
When the structure in signal space relies on multiple simultaneously-transmitted subcarriers, it corresponds to a multicarrier scheme. It is represented by
\begin{align}
\transmittedSignal[\timeSymbol] = \sum_{\timeIndexTX = -\infty}^{\infty} \sum_{\frequencyIndexTX=0}^{\numberOfSubcarrier-1} \symbols[\timeIndexTX][\frequencyIndexTX]\translatedModulatedFilterTX[\timeIndexTX][\frequencyIndexTX][\timeSymbol]~,
\label{Eq:transmittedSignal}
\end{align}
where  $\timeIndexTX$ is the time index, $\frequencyIndexTX$ is the subcarrier index, $\symbols[\timeIndexTX][\frequencyIndexTX]$ is the symbol (message) being transmitted, 
 $\numberOfSubcarrier$ is the number of subcarriers, and $\translatedModulatedFilterTX[\timeIndexTX][\frequencyIndexTX][\timeSymbol]$ is the {synthesis function} which maps  $\symbols[\timeIndexTX][\frequencyIndexTX]$ into the signal space. 
The family of $\translatedModulatedFilterTX[\timeIndexTX][\frequencyIndexTX][\timeSymbol]$ is  referred to as a {\em Gabor system}, when it is given by
\begin{align}
\translatedModulatedFilterTX[\timeIndexTX][\frequencyIndexTX][\timeSymbol]&=\prototypeFilterTX[\timeSymbol - \timeIndexTX \timeSpacingVariable]e^{j2\pi\frequencyIndexTX \frequencySpacingVariable\timeSymbol}~,
\label{Eq:lattice}
\end{align}
where $\prototypeFilterTX[\timeSymbol]$ is the prototype filter (also known as pulse shape, Gabor atom), $\timeSpacingVariable$ is the symbol spacing in time, and $\frequencySpacingVariable$ is the subcarrier spacing. 
A Gabor system implies that a single pulse shape is considered as a prototype and others are derived from the prototype filter via some translations in time and modulations in frequency, as given in \eqref{Eq:lattice}. The coordinates of the filters form a two dimensional structure in the time-frequency plane, known as lattice. 
Assuming a linear time-varying multipath channel $\channelProcessTime$, the received signal is obtained as
\begin{align}
\receivedSignal =& \int_{\timeShift}\channelProcessTime\transmittedSignal[\timeSymbol-\timeShift]{\integrald\timeShift}+ \noiseProcess
,
\end{align}
where  $\noiseProcess$ is the \ac{AWGN}.
Then, the symbol $\symbolsReceived[\timeIndex][\frequencyIndex]$ located on time index $\timeIndex$ and subcarrier index $\frequencyIndex$ is obtained by the projection of the received signal onto the analysis function $\translatedModulatedFilter[\timeIndex][\frequencyIndex][\timeSymbol]$ as
\begin{align}
\symbolsReceived[\timeIndex][\frequencyIndex]&=
\langle \receivedSignal, \translatedModulatedFilter[\timeIndex][\frequencyIndex][\timeSymbol]\rangle   \triangleq\int_{\timeSymbol} \receivedSignal \translatedModulatedFilterC[\timeIndex][\frequencyIndex][\timeSymbol] \integrald\timeSymbol~,
\label{Eq:receivedSymbol}
\end{align}
where
\begin{align}
\translatedModulatedFilter[\timeIndex][\frequencyIndex][\timeSymbol]&=\prototypeFilterRX[\timeSymbol - \timeIndex \timeSpacingVariable]e^{j2\pi\frequencyIndex\frequencySpacingVariable\timeSymbol}~.
\label{Eq:latticeRX}
\end{align}
Similar to \eqref{Eq:transmittedSignal},  $\translatedModulatedFilter[\timeIndex][\frequencyIndex][\timeSymbol]$ given in \eqref{Eq:latticeRX} is
obtained by a prototype filter $\prototypeFilterRX[\timeSymbol]$ translated in both time and frequency, constructing another Gabor system at the receiver.

\begin{figure}[!t]
\centering
\subfloat[A block diagram for  communications via multicarrier schemes. Both the transmitter and  the receiver construct Gabor systems.]{\includegraphics[width=3.5in]{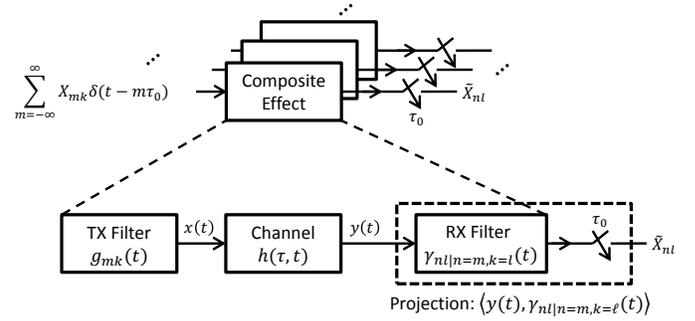} 
\label{Fig:block}}\\
\subfloat[Sampling the time-frequency plane. Modulated pulses are placed into the time-frequency plane, based on the locations of samples. In the illustration, the product of $1/\timeSpacingVariable\frequencySpacingVariable$ corresponds to the lattice density.]{\includegraphics[width=3.4in]{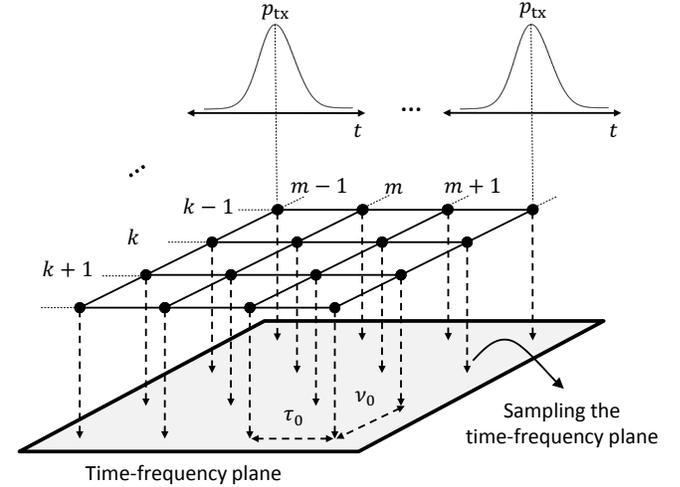}
\label{Fig:gabor}}\\
\caption{Utilization of the prototype filters at the transmitter and the
receiver.} \label{Fig:multicarrier}
\end{figure}
 
 The equations \eqref{Eq:transmittedSignal}-\eqref{Eq:latticeRX} correspond to  basic model for  multicarrier schemes illustrated in \figurename~\ref{Fig:multicarrier}\subref{Fig:block}, without stressing the variables in the equations.
In a crude form, a multicarrier scheme can be represented by  a specific set of equations constructed in the time-frequency plane, i.e., these equations are synthesized at the transmitter and analyzed at the receiver. In the following subsections, symbols, filters, and lattices in a multicarrier system are discussed in detail.

\subsection{Symbols}
Without loss of generality,  the transmitted symbols are denoted by $\symbols[\timeIndexTX][\frequencyIndexTX]\in\mathbb{C}$, where $\mathbb{C}~$ is the set of all complex numbers. As a special case, it is possible to limit the set of $\symbols[\timeIndexTX][\frequencyIndexTX]$ to real numbers, i.e. $\symbols[\timeIndexTX][\frequencyIndexTX]\in\mathbb{R}$, where $\mathbb{R}~$ is the set of all real numbers. One may choose 
$\symbols[\timeIndexTX][\frequencyIndexTX]$ as a modulation symbol or a part of the modulation symbol, e.g. its real or imaginary part or a partition after a spreading operation. In addition, it is reasonable to consider finite number of elements in the set, based on the limited number of modulation symbols in digital communications. 
Note that the set of the symbols may be  important for the perfect reconstruction of the symbols since its properties may lead one-to-one mapping from message space to the signal space \cite{shannon_1998}, as in signaling over Weyl-Heisenberg frames, faster-than-Nyquist signaling, or partial-response signaling \cite{86_han2009wireless,117_Han_TSP_2007,Kabal_1975,Mazo_1975}.

\subsection{Filters }
In digital communication, symbols are always associated with {pulse shape}s (also known as {filter}s). A pulse shape essentially corresponds to an energy distribution which indicates the density of the symbol energy (in time, frequency, or any other domain). Hence, it is one of the determining factors for the dispersion characteristics of the signal. At the receiver side, the dispersed energy due to the transmit pulse shape is coherently combined via receive filters. Thus, the transmit and receive filters jointly determine the amount of the energy transfered from the transmitter to  the receiver. Also, they determine the correlation between the points in the lattice, which identify the structure of the multicarrier scheme, i.e., orthogonal, bi-orthogonal, or non-orthogonal.

\def\orthobiortho{\rho}
\label{sec:bi-orthogonality}

\subsubsection{Matched Filtering}
If the prototype filter employed at the receiver is the same as the one that the transmitter utilizes, i.e., $\prototypeFilterTX[\timeSymbol]=\prototypeFilterRX[\timeSymbol]$, this approach corresponds to matched filtering, which maximizes \ac{SNR}. As opposed to matched filtering, one may also use different prototype filters at the transmitter and receiver, i.e., $\prototypeFilterTX[\timeSymbol]\neq\prototypeFilterRX[\timeSymbol]$ \cite{77_kozek1998nonorthogonal}.

\subsubsection{Orthogonality of Scheme}
If the synthesis functions and the analysis functions do not produce any correlation between the different points in the lattice, i.e., $
\langle \translatedModulatedFilterTX[\timeIndexTX][\frequencyIndexTX][\timeSymbol], \translatedModulatedFilter[\timeIndex][\frequencyIndex][\timeSymbol]\rangle = \kroneckerOperator[\timeIndexTX\timeIndex]\kroneckerOperator[\frequencyIndexTX\frequencyIndex]$, where $\kroneckerOperator[]$ is the Kronecker delta function, the scheme is either orthogonal or bi-orthogonal. Otherwise, the scheme is said to be non-orthogonal,  i.e., $
\langle \translatedModulatedFilterTX[\timeIndexTX][\frequencyIndexTX][\timeSymbol], \translatedModulatedFilter[\timeIndex][\frequencyIndex][\timeSymbol]\rangle \neq \kroneckerOperator[\timeIndexTX\timeIndex]\kroneckerOperator[\frequencyIndexTX\frequencyIndex]$ . 
While orthogonal schemes dictate to the use of the same prototype filters at the transmitter and receiver, bi-orthogonal schemes allow to use different prototype filters at the transmitter and the receiver.

A nice interpretation on orthogonality and bi-orthogonality is provided in \cite{39_strohmer2003optimal}. Let $\gramOperator$ be a Gram matrix given by $\gramOperator\triangleq\vectorBasis\vectorBasis^{\rm H}$ where $\vectorBasis^{\rm H}$ is a block-circulant matrix in which the columns consist of the modulated-translated vectors generated by an initial filter $p(t)$. Then, the relation between the filters at the transmitter and the receiver for orthogonal and bi-orthogonal schemes can be investigated by
\begin{align}
\symbols[\timeIndexTX][\frequencyIndexTX]&= \gramOperator\gramOperator^{-1}\symbols[\timeIndexTX][\frequencyIndexTX]
=\gramOperator^{-\orthobiortho}\gramOperator\gramOperator^{-(1-\orthobiortho)}\symbols[\timeIndexTX][\frequencyIndexTX]\nonumber
\\&=\gramOperator^{-\orthobiortho}\vectorBasis\vectorBasis^{\rm H}
\gramOperator^{-(1-\orthobiortho)}\symbols[\timeIndexTX][\frequencyIndexTX]\nonumber
\\&=\underbrace{\gramOperator^{-\orthobiortho}\vectorBasis\times \overbrace{(
\gramOperator^{-(1-\orthobiortho)}\vectorBasis)^{\rm H}\symbols[\timeIndexTX][\frequencyIndexTX]}^{\rm Transmitted~signal}}_{\rm Received~symbol}
\label{Eq:ortohogonalAnotherRepresentationVector}
\end{align}
where $[\cdot]^{H}$ is the Hermitian operator and $\orthobiortho$ is the weighting parameter to characterize orthogonality and bi-orthogonality. Using \eqref{Eq:ortohogonalAnotherRepresentationVector}, the prototype filters at the transmitter and the receiver can be obtained from the first rows of $\gramOperator^{-\orthobiortho}\vectorBasis$ and $\gramOperator^{-(1-\orthobiortho)}\vectorBasis$, respectively, which yields
\begin{align}
\prototypeFilterRXv[\timeSymbol]=\sum_{m,k} \gramOperator^{-\orthobiortho} p_{mk}
\label{Eq:gramRX}
\end{align}
and
\begin{align}
\prototypeFilterTXv[\timeSymbol]=\sum_{m,k}\gramOperator^{-(1-\orthobiortho)} p_{mk}~,
\label{Eq:gramTX}
\end{align}
where $p_{mk}$ is the column vector generated by modulating and translating $p$. As a simpler approach, it is also possible to calculate \eqref{Eq:gramRX} and \eqref{Eq:gramTX} as
\begin{align}
\prototypeFilterRXv[\timeSymbol]=\frameOperator^{-\orthobiortho} p
\label{Eq:gramRXS}
\end{align}
and
\begin{align}
\prototypeFilterTXv[\timeSymbol]=\frameOperator^{-(1-\orthobiortho)} p
\label{Eq:gramTXS}
\end{align}
respectively, where $\frameOperator=\vectorBasis^{\rm H}\vectorBasis$.
While the choice $\orthobiortho=1/2$ leads to an orthogonal scheme, $\orthobiortho\rightarrow0$ or $\orthobiortho\rightarrow1$ result in bi-orthogonal schemes \cite{39_strohmer2003optimal}. 
When $\orthobiortho=1$,  minimum-norm dual pulse shape is obtained.

Note that orthogonal schemes maximize \ac{SNR} for \ac{AWGN} channel \cite{39_strohmer2003optimal} since they assure matched filtering. On the contrary, bi-orthogonal schemes may offer better performance for dispersive channels, as stated in \cite{77_kozek1998nonorthogonal}. 
In addition, when the scheme has receive filters which are not orthogonal to each other, i.e.,  $
\langle \translatedModulatedFilter[\timeIndex'][\frequencyIndex'][\timeSymbol], \translatedModulatedFilter[\timeIndex][\frequencyIndex][\timeSymbol]\rangle \neq \kroneckerOperator[\timeIndex'\timeIndex]\kroneckerOperator[\frequencyIndex'\frequencyIndex]$,  the noise samples becomes correlated, as in non-orthogonal and bi-orthogonal schemes\cite{77_kozek1998nonorthogonal}.

\subsubsection{Localization}
The localization of a prototype filter characterizes the variances of the energy in time and frequency. While the localization in time is measured
by $\ltwonorm[\timeSymbol{\pulseTime[]}]^2$, the localization in frequency is obtained as $\ltwonorm[\frequencySymbol{\pulseFrequency[]}]^2$, where $\ltwonorm[\cdot]$ is the $L^2$-norm and $\pulseFrequency[]$ is the Fourier transformation of $\pulseTime[]$. The functions where $\ltwonorm[\frequencySymbol{\pulseFrequency[]}]\ltwonorm[\timeSymbol{\pulseTime[]}]\rightarrow\infty$ are referred as non-localized filters; otherwise, they are referred as localized filters.

\subsection{Lattices}
A lattice corresponds to an algebraic set which contains the coordinates of the filters in the time-frequency plane \cite{26_Du_KTH_2007,39_strohmer2003optimal, Kutyniok_2005, 88_jung2007wssus, 117_Han_TSP_2007}. In other words, it is a set  generated by sampling the time-frequency plane as illustrated in \figurename~\ref{Fig:multicarrier}\subref{Fig:gabor}. It determines the bandwidth efficiency and the reconstruction properties of a  multicarrier scheme. Without loss of generality, a lattice $\lattice$ can be described by a non-unique generator matrix $\latticeGenerator$ as, 
\begin{align}
\latticeGenerator
=
{
\begin{bmatrix}
x  & y \\
0 & z
\end{bmatrix}}~,
\end{align}
where $x,z\neq0$. The generator matrix contains the coordinates of the first two identifying points of the lattice in its column vectors, i.e., $(0,x)$ and $(y,z)$ \cite{39_strohmer2003optimal}. The locations of other points are calculated by applying $\latticeGenerator$ to $[\timeIndexTX~\frequencyIndexTX]^{T}$, where $[\cdot]^{T}$ is the transpose operation.

\subsubsection{Lattice Geometry}
Generator matrix $\latticeGenerator$ determines the lattice geometry. For example, the choice   
\begin{align}
\latticeGenerator
=
{
\begin{bmatrix}
\timeSpacing  & 0 \\
0 & \frequencySpacing
\end{bmatrix}}~,
\label{Eq:rectangular}
\end{align} 
yields a rectangular structure as in \eqref{Eq:lattice} and \eqref{Eq:latticeRX}, with a symbol duration of $\timeSpacing$ and subcarrier width $\frequencySpacing$. Similarly, a hexagonal (or quincunx) pattern \cite{26_Du_KTH_2007, 39_strohmer2003optimal, 117_Han_TSP_2007} is obtained when 
\begin{align}
\latticeGenerator
=
{
\begin{bmatrix}
\timeSpacing  & 0.5\timeSpacing \\
0 & \frequencySpacing
\end{bmatrix}}~.
\label{Eq:hexagonal}
\end{align}
Lattice geometry identifies the distances between the points indexed by the integers $\timeIndexTX$ and $\frequencyIndexTX$. For example, assuming that $\frequencySpacing = 1/\timeSpacing$, while the minimum distance between the points  is 1 for the rectangular lattice in \eqref{Eq:rectangular}, it is $\sqrt{1.25}$ for the quincunx lattice in \eqref{Eq:hexagonal}\cite{26_Du_KTH_2007}. 

\subsubsection{Lattice Density/Volume}
\label{sssec:latticeDensity}
Lattice density can be obtained as
\begin{align}
\latticeDensity=\frac{1}{\latticeVolume}=\frac{1}{\absOperator[{\determinantOperator[\latticeGenerator]}]},
\label{eq:latticeDensity}
\end{align}
where $\absOperator[\cdot]$ is the absolute value of its argument and $\latticeVolume$ is the volume of the lattice $\lattice$ calculated via determinant operation $\determinantOperator[\cdot]$. It identifies not only the bandwidth efficiency of the scheme as
\begin{align}
\bandwidthEfficiency=\bitPerSymbol\latticeDensity,
\label{eq:efficiency}
\end{align}
where $\bitPerSymbol$ is the bit per volume, but also the perfect reconstruction of the symbols at the receiver. In order to clarify the impact of the lattices on the perfect reconstruction of the symbols, the concept of basis is needed to be investigated along with Gabor systems.

A set of linearly independent vectors is called a basis if these vectors are able to represent all other vectors for a given space. While including an extra vector to the basis spoils the linear independency, discarding one from the set destroys the completeness. From communications point of view, having linearly independent basis functions is a conservative condition since it allows one-to-one mapping from symbols to constructed signal without introducing any constraints on the symbols. Representability of the space with the set of $\{\translatedModulatedFilterTX[\timeIndexTX][\frequencyIndexTX][\timeSymbol]\}$  is equivalent to the completeness property, which is important in the sense of reaching Shannon's capacity \cite{2_le1995coded, 77_kozek1998nonorthogonal}. 
Gabor systems provide an elegant relation between the linear independence and the completeness properties based on the lattice density. This relation for Gabor systems is given as follows\cite{97_Sondergaard_Thesis_2007,77_kozek1998nonorthogonal,56_bolcskei1999design,80_matz2007analysis}:
\begin{itemize}
    \item Undersampled case $(\latticeDensity<1)$: Gabor system cannot be a complete basis since the time-frequency plane is not sampled sufficiently. However, this case gives linearly independent basis functions. Well-localized prototype filters can be utilized, but the bandwidth efficiency of the Gabor system degrades with decreasing $\latticeDensity$.

\item 
Critically-sampled case $(\latticeDensity=1)$: It results in a
complete Gabor System. Bases exist, but they cannot
utilize well-localized prototype filters according to the
Balian-Low theorem \cite{Daubechies_1990}. This theorem states that there
is no well-localized function in both time and frequency
for a Gabor basis where $\latticeDensity=1$. It dictates the use of non-localized functions,
e.g., rectangular and sinc functions. A consequence of
Balian-Low theorem can also be observed when the dual
filters are calculated as in \eqref{Eq:gramRX} and \eqref{Eq:gramTX}. If one attempts
to utilize a well-localized function, e.g., Gaussian, when
$\latticeDensity=1$, the Gram matrix $\gramOperator$ in \eqref{Eq:ortohogonalAnotherRepresentationVector} becomes ill-conditioned.
Hence, the calculation of the dual pulse shape becomes
difficult. The degree of ill-conditioning can be measured via the condition
number of $\gramOperator$. As stated in \cite{39_strohmer2003optimal}, the condition number of $\gramOperator$ approaches to infinity for Gaussian pulses when $\latticeDensity\rightarrow 1$.

\item Oversampled case $(\latticeDensity>1)$: It yields an overcomplete set of functions. Gabor system cannot be a basis, but it may be a {\em frame}\footnote{{\em Frames} are introduced in 1952 by Duffin and Schaeffer \cite{Duffin_1952}, as an extension of the concept of a basis. They can include more than the required elements to span a space. This issue corresponds to an overcomplete system, which causes non-unique representations \cite{christensen2003,strang1996wavelets}. }  with well-localized pulse shapes. However, since the Gabor system is overcomplete, representation of a signal might not be unique. Note that non-unique representations do not always imply loss of one-to-one mapping from modulation symbols to signal constructed. For example, a finite number of modulation symbols may be useful to preserve the one-to-one relation between the signal space and the message space \cite{86_han2009wireless}.

\end{itemize}

\subsection{A Combined Approach: Lattice Staggering}
\label{subsec:staggering}
\begin{figure*}[!t]
\centering
\graphicspath{{D:/Calismalar/Project_FBMC/paper_fbmc_survey/figures/fbmc_v6/}}
\subfloat[Real part of the multicarrier signal (in-phase component).]{\includegraphics[width=3.5in]{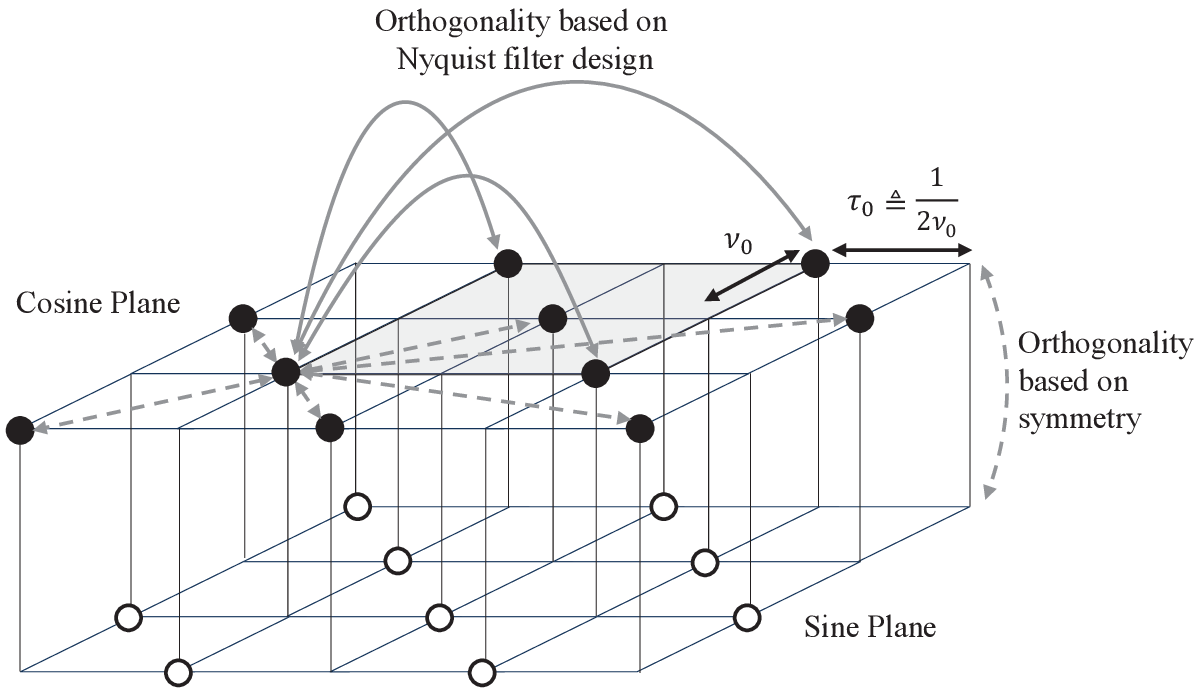}
\label{fig:inPhase}}
\subfloat[Imaginary part of the multicarrier signal (quadrature component)]{\includegraphics[width=3.56in]{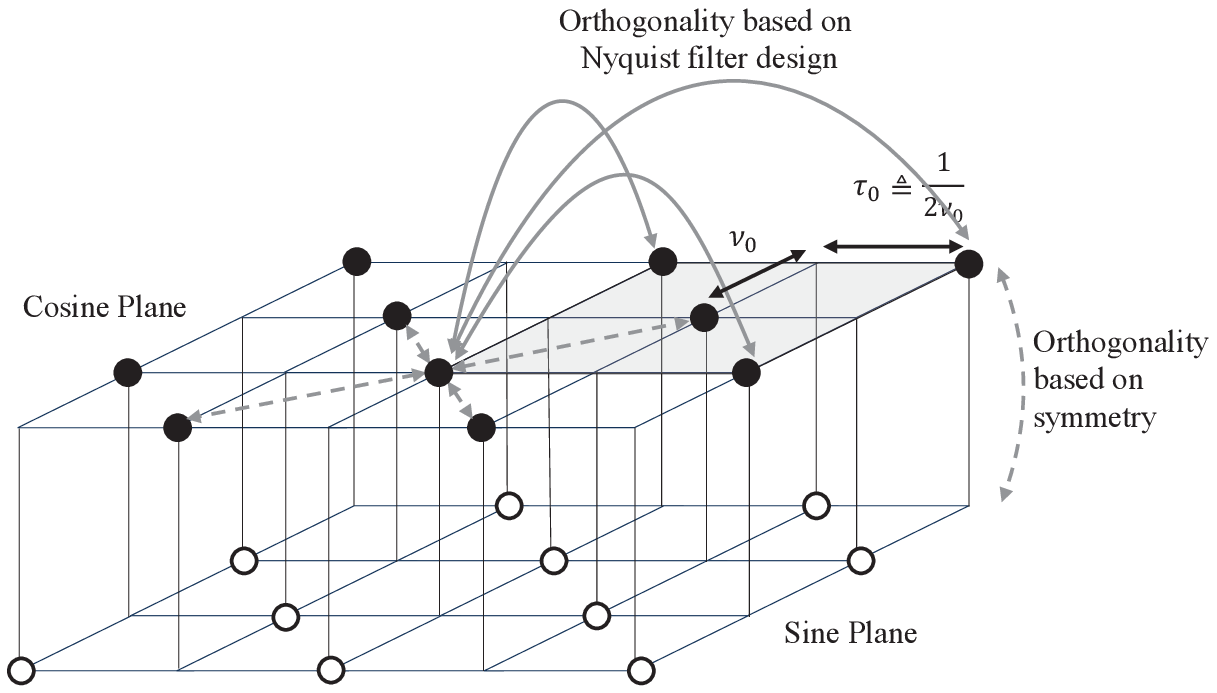} 
\label{fig:quadrature}}
\\
\subfloat[Illustration for the orthogonality based on  even-symmetric filters. Inner product of three functions is zero when $\timeSpacingVariable$ is set to $1/2\frequencySpacingVariable$. ]{\includegraphics[width=3.8in]{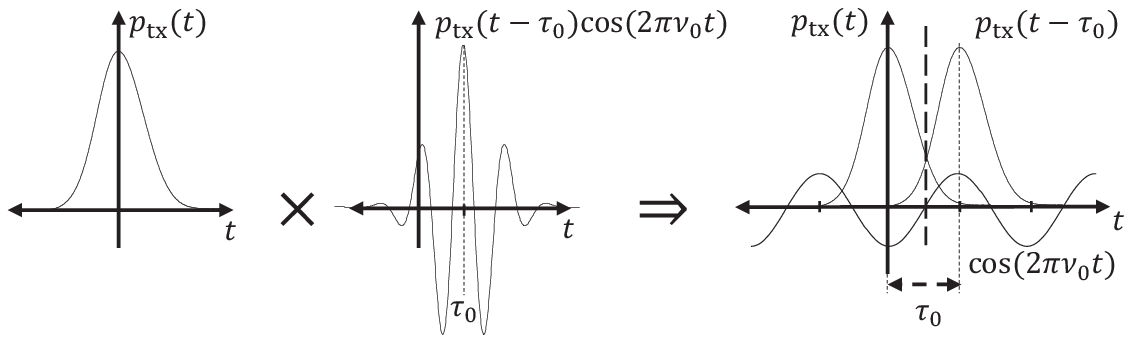} 
\label{fig:inherent}}

\caption{Lattice staggering.} 
\label{Fig:TFL_Lattice}
\end{figure*}

It is possible to circumvent the restriction of Balian-Low theorem on the filter design with lattice staggering\footnote{In the literature, this approach appears with different names, e.g. \ac{OQAM}, staggered modulation. Rather than indicating a specific modulation, it is referred as {\em lattice staggering} throughout the study.} \cite{Daubechies_1991,2_le1995coded,56_bolcskei1999design, gabor_book_chap,27_farhang2010cosine}. 
 It is a methodology that generates  inherent orthogonality between the points in the lattice for  real domain  through mandating symmetry conditions on the prototype filter. Since the inherent orthogonality does not rely on the cross-correlation between the filters, it relaxes the conditions for the filter design. It is worth noting that the real domain may be either the imaginary portion or the real portion of the complex domain. Thus, in lattice staggering, the real  and imaginary parts of the scheme are treated separately. However, processing in real domain does not imply that the real and imaginary parts do not contaminate each other. Indeed, they interfere, but the contamination is orthogonal to the desired part.

The concept of lattice staggering is illustrated in detail in \figurename~\ref{Fig:TFL_Lattice}. The lattices on real and imaginary parts of the scheme are given in \figurename~\ref{Fig:TFL_Lattice}\subref{fig:inPhase} and \figurename~\ref{Fig:TFL_Lattice}\subref{fig:quadrature}, respectively. They also correspond to the lattices on in-phase and quadrature branches in baseband. While the filled circles represent the locations of the filters on the cosine plane, the empty circles show the locations of the filters on the sine plane. Cosine and sine planes indicate that the filters on those planes are modulated with either cosine or sine functions. First, consider the lattice given in the cosine plane of \figurename~\ref{Fig:TFL_Lattice}\subref{fig:inPhase}. According to the Euler's formula, a pulse modulated with a complex exponential function includes components on both cosine and sine planes. Hence, when the filters on this lattice are modulated with complex exponential functions, there will be same lattice on the cosine and sine planes, as illustrated in \figurename~\ref{Fig:TFL_Lattice}\subref{fig:inPhase} and \figurename~\ref{Fig:TFL_Lattice}\subref{fig:quadrature}. It is important to observe that the cross-correlation among the points indicated by arrows in the cosine plane of \figurename~\ref{Fig:TFL_Lattice}\subref{fig:inPhase} is always zero, when the filters are even-symmetric and the symbol spacing is selected as $\timeSpacingVariable=1/2\frequencySpacingVariable$. This is because of the fact that the integration of a function which contains a cosine function multiplied with a symmetric function about the cosine's zero-crossings yields zero, as illustrated in \figurename~\ref{Fig:TFL_Lattice}\subref{fig:inherent}. From the communications point of view, this structure allows to carry only one real symbol without interference. Considering the same structure on the imaginary part by staggering the same lattice, illustrated in \figurename~\ref{Fig:TFL_Lattice}\subref{fig:quadrature}, another real symbol could be transmitted. Although transmitting on the imaginary and real parts leads to contaminations on the sine planes, these contaminations are always orthogonal to the corresponding cosine planes when the filter is an even-symmetric function.

Lattice staggering induces an important result for the filter design: In order to obtain an orthogonal or biorthogonal scheme using lattice staggering, the correlation of the transmit filter and the receive filter should provide nulls in time and frequency at the multiples of $2\timeSpacingVariable$ and $2\frequencySpacingVariable$. In other words, the unit area between the locations of the nulls in time-frequency plane is $2\timeSpacingVariable\times2\frequencySpacingVariable$, which is equal to $2$ since $\timeSpacingVariable$ is set to $1/2\frequencySpacingVariable$. This is because of the fact that even-symmetrical filters provide inherent orthogonality between some of the diagonal points in lattice when $\timeSpacingVariable=1/2\frequencySpacingVariable$ even though the filters do not satisfy Nyquist criterion, as illustrated in \figurename~\ref{Fig:TFL_Lattice}. Also, this approach circumvents the Balian-Low theorem since $\latticeDensity=1/\timeSpacingVariable\frequencySpacingVariable=2$.  Hence, lattice staggering allows orthogonal and bi-orthogonal schemes with well-localized filters while maintaining bandwidth efficiency as $\bandwidthEfficiency=\bitPerSymbol$. 
From the mathematical point of view, lattice staggering corresponds to utilizing Wilson bases on each real and imaginary parts of the scheme, which is equivalent to the linear combinations of two Gabor systems where $\latticeDensity= 2$ \cite{Benedetto_1994,Christensen_2001frames, 95_Strohmer_ACHA_2001, 39_strohmer2003optimal}.

%

\subsection{Summary}
\begin{figure*}[t]
\begin{center}
\includegraphics[width = 7.2in]{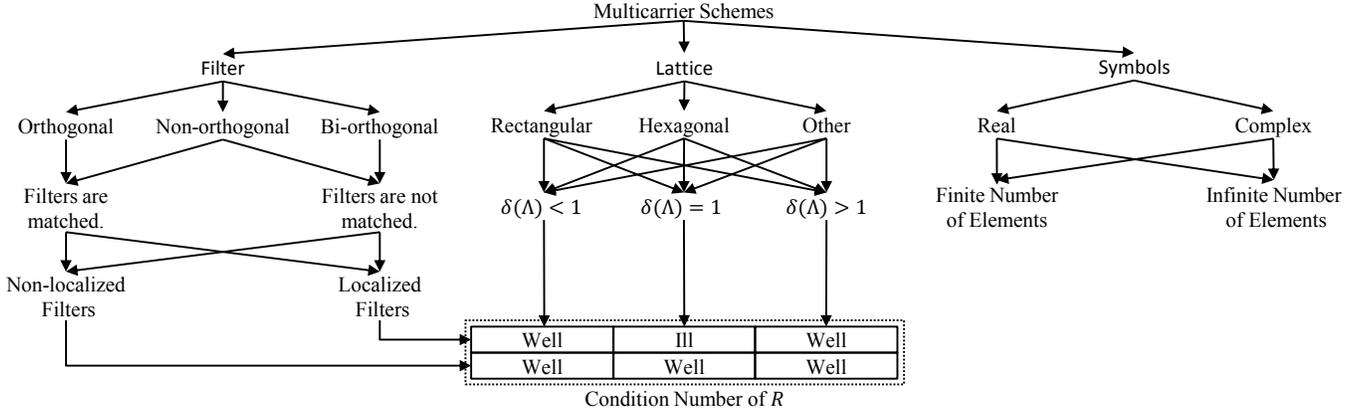}
\caption{Multicarrier schemes based on lattices, filters, and symbols.}
\label{table:classification}
\end{center}
\end{figure*}
As a summary,
 the relations between the fundamental elements of a multicarrier scheme, i.e. symbols, lattice, and filter are given in \figurename~\ref{table:classification}. One can determine these elements based on Gabor theory, considering the needs of the communication system. For example, let the signaling be an orthogonal scheme which is based on a rectangular lattice geometry without lattice staggering. Then, the localization of the filter is determined according to the statements of Gabor theory. For instance, equipping this system with well-localized filters yields an ill-conditioned $\gramOperator$ when $\latticeDensity=1$. Other inferences can also be obtained by following \figurename~\ref{table:classification}.


\section{Multicarrier Schemes }
\label{sec:multicarrier}
In this section, the concepts introduced in Section \ref{sec:preliminary} are harnessed and they are associated with known multicarrier schemes. Rather than discussing the superiorities of schemes to each other, the relations between the orthogonal, bi-orthogonal, and non-orthogonal schemes within the framework of Gabor theory are emphasized. 
An interpretation of the spreading operation in multicarrier systems (e.g., as in \ac{SC-FDMA}) is also provided in the context of Gabor systems. Finally,  milestones for multicarrier schemes reviewed for completeness.

\subsection{Orthogonal Schemes}
The schemes that fall into this category have orthogonal basis functions at both the  transmitter and the  receiver and follow matched filtering approach. The phrase of {\em orthogonal frequency division multiplexing} is often used for a specific scheme that is based on rectangular filters. However, there are other multicarrier schemes which provide orthogonality. We begin by describing the orthogonal schemes which do not consider lattice staggering:
\begin{itemize}
    \item {\em Plain \& \ac{ZP-OFDM}:}
Plain \ac{OFDM} is an orthogonal scheme which is equipped with rectangular filters at the transmitter and the receiver when $\latticeDensity =1$. In order to combat with multipath channel, one can provide guard interval between \ac{OFDM} symbols, known as \ac{ZP-OFDM} \cite{Muquet_2002}. It corresponds to stretching the lattice in time domain, which yields $\latticeDensity <1$.
	\item {\em \ac{FMT}:} \ac{FMT}
 is an orthogonal scheme where the filters do not overlap in frequency domain. There is no specific filter associated with \ac{FMT}. Instead of the guard intervals in \ac{ZP-OFDM}, guard bands between the subcarriers can be utilized in order to obtain more room for the filter localization in frequency domain. Hence, it is based on a lattice where $\latticeDensity \le 1$. For more details we refer the reader to the studies in \cite{15_cherubini1999filtered, 16_cherubini2002filtered, 22_benvenuto2002equalization,49_tonello2005performance, 28_farhang2010signal,24_farhang2011ofdm}.
	\item{\em Lattice-OFDM:} Lattice-OFDM is the optimum orthogonal scheme for time- and frequency- dispersive channels in the sense of minimizing interference between the symbols  in the lattice \cite{39_strohmer2003optimal}. It relies on different lattice geometries and orthogonalized Gaussian pulses, depending on the channel dispersion characteristics. For more details, we refer the reader to Section \ref{SubSec:LatticeAndFilterAdapatations}.
\end{itemize}

The orthogonal schemes which consider lattice staggering are given as follow:

\begin{itemize}
    \item {\em \ac{SMT} and \ac{CMT}:}
Both schemes exploit the lattice staggering approach to obtain flexibility on the filter design when $\bandwidthEfficiency=\bitPerSymbol$ \cite{24_farhang2011ofdm, 27_farhang2010cosine}, where $\bandwidthEfficiency$ is the bandwidth efficiency and $\bitPerSymbol$ is the bit per volume, introduced in \eqref{eq:efficiency}. In these schemes, the symbols are real numbers due to the lattice staggering, however, as a special case, they are either real or imaginary part of the modulation symbols. The main difference between the \ac{SMT} and the \ac{CMT} is the modulation type. While \ac{SMT} uses \ac{QAM} type signals, \ac{CMT} is dedicated to \ac{VSB}. Yet, \ac{SMT} and \ac{CMT} are
structurally identical; it  is possible to
synthesize one from another by applying a frequency
shift operation and proper symbol placement \cite{27_farhang2010cosine}.

\end{itemize}

\begin{figure}[!t]
\centering
\subfloat[Plain OFDM.]{\includegraphics[width=1.9in]{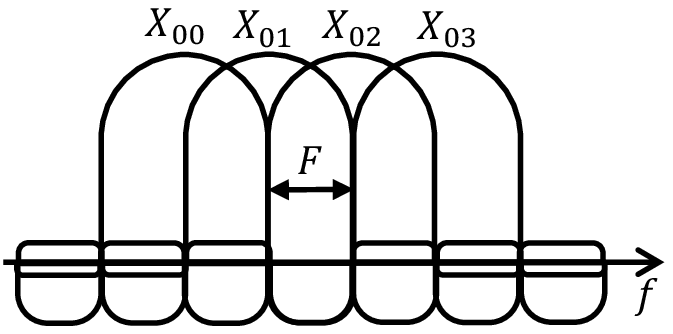}
\label{fig:ofdm}}
\subfloat[FMT.]{\includegraphics[width=1.57in]{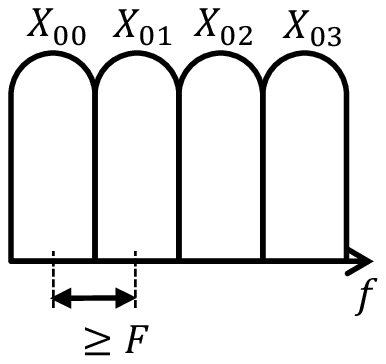}
\label{fig:fmt}}\\
\subfloat[SMT.]{\includegraphics[width=1.9in]{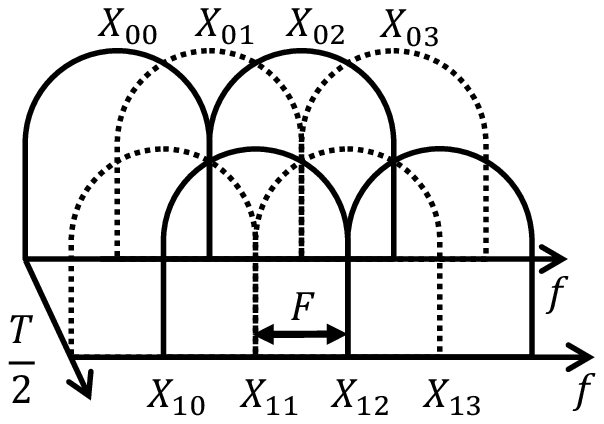}
\label{fig:smt}}
\subfloat[CMT.]{\includegraphics[width=1.57in]{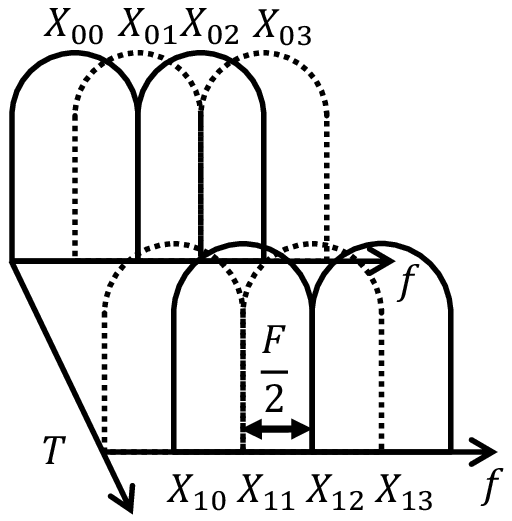}
\label{fig:cmt}}
\caption{Illustrations of various orthogonal multicarrier schemes. }
\label{Fig:OFDM_FMT_SMT_CMT}
\end{figure}

Illustrations for Plain/\ac{ZP-OFDM}, \ac{FMT}, \ac{SMT}, and \ac{CMT} in time and frequency are provided in \figurename~\ref{Fig:OFDM_FMT_SMT_CMT}.

\subsection{Bi-orthogonal Schemes}
These schemes do not follow matched filtering approach and do not have to contain orthogonal basis functions at the transmitter and the receiver. However, transmit and receive filters are mutually orthogonal to each other.

\begin{itemize}
    \item {\em \ac{CP-OFDM}:}
Plain \ac{OFDM} is often utilized with \ac{CP} to combat with the multipath channels. \ac{CP} induces a lattice where $\latticeDensity >1$. At the same time, it results in a longer rectangular filter at the transmitter, compared to one at the receiver.  Therefore, \ac{CP-OFDM} does not follow matched filtering and constructs a bi-orthogonal scheme \cite{hlawatsch2011wireless,88_jung2007wssus,SPMAG2013}. Yet, it provides many benefits, e.g. single-tap equalization and simple synchronization. 
\item {\em Windowed-OFDM:} OFDM has high \ac{OOB} due to the rectangular filter. In order mitigate the out-of-band radiation, one may consider to smooth the transition between \ac{OFDM} symbols. This operation smooths the edges of rectangular filter, and commonly referred as windowing. If the windowing is performed with an {\em additional} guard period, a bi-orthogonal scheme where $\latticeDensity >1$ is obtained.

 \item {\em \ac{BFDM}:}  In \cite{77_kozek1998nonorthogonal}, it is stated that extending the rectangular filter as in \ac{CP}-\ac{OFDM} is likely to be a suboptimal solution under doubly dispersive channels, since this approach does not treat the time and frequency dispersions equally. As an alternative to \ac{CP-OFDM}, by allowing different filters at the transmitter and the receiver, \ac{BFDM} with properly designed filters can reduce the interference contribution from other symbols in doubly dispersive channels. In \cite{77_kozek1998nonorthogonal}, the design is given based on a prototype filter constructed with Hermite-Gaussian function family and a rectangular lattice geometry when $\latticeDensity=1/2$. In order to maintain the bandwidth efficiency, one can utilize \ac{BFDM} with lattice staggering, as investigated in \cite{gabor_book_chap,Siclet_2000}.

\item {\em Signaling over Weyl-Heisenberg Frames:} Main motivation  is to exploit  overcomplete Gabor frames with well-localized pulses and finite number of symbols for digital signal transmission. It is a unique approach that allows a scheme where $\latticeDensity >1$ with the perfect reconstruction property, which exploits subspace classifications \cite{86_han2009wireless}.
\end{itemize}


It is interesting to examine bi-orthogonal schemes from the point of equalizers. We refer the reader to the related discussion provided in Section \ref{subsec:equalization}.

\subsection{Non-orthogonal Schemes}
The schemes that fall into this category do not contain orthogonal basis functions at the transmitter or the receiver. Also, there is no bi-orthogonal relation between the filters at the transmitter and the filters at the receiver.

\begin{itemize}
    \item {\em Generalized Frequency Division Multiplexing:}
\ac{GFDM} is a non-orthogonal scheme which allows correlation between the points in the lattice in order to be able to utilize well-localized filters when $\latticeDensity=1$ \cite{Fettweis_2009}. Also, it utilizes complex symbols and \ac{CP} along with tail biting in the pulse shape.
At the receiver side, successive interference cancellation is applied to remove the interference between the symbols \cite{82_datta2011fbmc}.
	\item {\em Concentric Toroidal Pulses:} By exploiting the orthogonality between Hermite-Gaussian functions, concentric toroidal pulses are introduced to increase the bandwidth efficiency of the transmission \cite{93_aldirmaz2010spectrally}. 
Four Hermite pulses are combined on each point in the lattice and each Hermite pulse carries one symbol. Although Gaussian-Hermite functions are orthogonal among each other, the pulses between neighboring points are not orthogonal.
	\item {\em Faster-than-Nyquist  \& Partial Response Signaling:}
When $\latticeDensity >1$, certain conditions may yield the reconstruction of the transmitted symbols. This issue firstly is investigated by Mazo in 1975 as  faster-than-Nyquist by addressing the following question: to what extent can the symbols be packed more
than the Nyquist rate without loss in \ac{BER} performance?
It is shown that the symbol spacing can be reduced to 0.802$T$ without suffering any loss in minimum Euclidean distance between the synthesized signals for binary modulation symbols and sinc pulse  \cite{Mazo_1975}. In other words, \ac{BER} performance is still achievable with {\em optimal receivers} even when the symbols are transmitted at a rate greater than the Nyquist rate.
The minimum symbol spacing that keeps the minimum Euclidean distance is later on referred as the {\em Mazo limit}  in the literature. By generalizing faster-than-Nyquist approach to other pulses, various Mazo limits are obtained for \ac{RRC} pulses with different roll-off factors in \cite{Liveris_2003}. For example,  when roll-off is set to 0, 0.1, 0.2, and, 0.3, Mazo limits are derived as 0.802, 0.779, 0.738, and, 0.703 respectively. The faster-than-Nyquist approach is extended to multicarrier schemes by allowing interference in time and frequency in \cite{Rusek_2005,Rusek_2006,Rusek_TWC_2009}, which show that two dimensional signaling is more bandwidth efficient than one dimensional signaling. 
Another way of developing a scheme where $\latticeDensity >1$ is to transmit correlated symbols. This approach corresponds to partial-response signaling and introduced in \cite{Kabal_1975}. Similar to the faster-than-Nyquist signaling and partial-response signaling, in \cite{117_Han_TSP_2007}, Weyl-Heisenberg frames  ($\latticeDensity > 1$) is examined considering a hexagonal lattice geometry and sequence detector is employed at the receiver for symbol detection.
\end{itemize}

\subsection{Multicarrier Schemes with Spreading Approaches}
Spreading operation is commonly used to reduce the \ac{PAPR} in multicarrier schemes, in which modulation symbols are mapped to the multiple points in the lattice.
One way to interpret and generalize the spreading operation in multicarrier systems (e.g., as in \ac{SC-FDMA} \cite{A11_Sari_ComMag_1995} and \ac{FB-S-FBMC}  \cite{40_ihalainen2009filter}) is to consider another Gabor system that spreads the energy of the modulation symbols into multiple subcarriers. In other words, as opposed to using single Gabor system at the transmitter and receiver, two Gabor systems combined with serial-to-parallel conversions  are employed at the transmitter and the receiver. For example, it can be said that \ac{SC-FDMA}, which allows better \ac{PAPR} characteristics and \ac{FDE} along with \ac{CP} utilization \cite{A10_Sari_GLOBECOM_1994, A12_Falconer_ComMag_2002,A8_Hyung_VTMag_2006,A9_Berardinelli_WirelessComm_2008}, employs an extra Gabor system equipped with a rectangular filter and $\latticeDensity=1$ to spread the modulation symbols at the transmitter (i.e., \ac{DFT}) and de-spread them at the receiver (i.e., inverse \ac{DFT}). On the contrary, in \ac{OFDM}, since there is no spreading of the modulation symbols in frequency domain, employed prototype filter for spreading is a Dirac function. 

\subsection{Milestones  for Orthogonal Schemes}
Having discussed the different variations of multicarrier systems in the earlier subsections, this subsection provides a brief history on the development of aforementioned multicarrier systems.  Earlier works related to orthogonal multicarrier schemes actually date back to 1960s~\cite{A1_Chang_BST_1966,14_saltzberg1967performance}, which utilize a bank of filters for parallel data transmission. In \cite{A1_Chang_BST_1966}, Chang presented the orthogonality condition for the multicarrier scheme schemes  considering band-limited filters. This condition basically indicates that the subcarriers can be spaced half of the symbol rate apart without any interference. This  scheme  has then been re-visited by Saltzberg in 1967~\cite{14_saltzberg1967performance} by showing the fact that Chang's condition is also true when the time and frequency axes are interchanged, based on \ac{OQAM}. Indeed, Chang and Saltzberg exploits the lattice staggering for their multicarrier schemes which includes the basics of \ac{CMT} and \ac{SMT}.
However, the idea of parallel transmission suggested in \cite{A1_Chang_BST_1966} and \cite{14_saltzberg1967performance} were {\em unreasonably expensive and complex} for large number of data channels at that time. 
In~\cite{A2_Weinstein_TCT_1971}, through the use of \acp{DFT}, Weinstein and Ebert eliminated the banks of subcarrier oscillators to allow simpler implementation of the multicarrier schemes.
This approach has been later named as \ac{OFDM}, and it has become more and more popular after 1980s due to its efficient implementation through \ac{FFT} techniques and \ac{FDE} along with \ac{CP} utilization \cite{Peled_1980} compared to other multicarrier schemes. 
On the other hand, Weinstein's \ac{DFT} method in~\cite{A2_Weinstein_TCT_1971} limits the flexibility on different baseband filter utilization while modulating or demodulating the subcarriers, but instead used a time windowing technique to cope with the spectral leakage. In~\cite{29_hirosaki1981orthogonally}, by extending Weinstein's method, Hirosaki showed that different baseband filters may also be digitally implemented through \ac{DFT} processing by using a \ac{PPN} \cite{78_bellanger1976digital}, \cite{79_vaidyanathan1990multirate}. 
Several other developments over the last two decades have demonstrated low complexity and efficient implementations of lattice staggering, paving the way for its consideration in the next generation wireless standards (see e.g., \cite{2_le1995coded,4_Siohan_TSP_2002,24_farhang2011ofdm}, and the references listed therein).

\section{Filter Design}

\label{sec:prototypFilterDesign}

In a multicarrier scheme, a prototype filter  determines the correlation between the symbols and the robustness of the scheme against dispersive channels. This issue induces to design prototype filters which are suitable for communications  in time-selective and frequency-selective channels. The goal of this section is to review the filters available in the literature. In order to reveal the connections between the filters, we categorized the filters based on their design criteria: 1) energy concentration \cite{39_strohmer2003optimal, Landau_1961,Slepian_1961,Landau_1962,Slepian_1964,Slepian_1978, Slepian_1983, Kaiser_1980,Halpern_1979,105_Vahlin_TC_1996,Walter2005432,Moore2004208}, 2) rapid-decay \cite{ 106_Martin_TCS_1998,3_bellanger2001specification,Mirabbasi_2002, 84_haas1997time }, 3) spectrum-nulling, and 4)  channel characteristics and hardware. Analytical expressions of the investigated filters are given in \tablename~\ref{table:waveform_comp}. For more detailed discussions on the  discussed filters, we refer the reader to the review papers \cite{Harris_1978,Geckinli_1978,24_farhang2011ofdm} and the books \cite{28_farhang2010signal, strang1996wavelets}.

\def\derivationOrder{q}
\def\discussionWidth{5cm}
\def\titleWidth{2.2cm}
\begin{table*}
\caption{Analytical expressions of known prototype filters in the literature.}
\centering
 \begin{tabular}{l||l|p{1.9in}}

 \textbf{Filter} &\textbf{Analytical Model} & \textbf{Comments}  \\ \hline \hline
\parbox{\titleWidth}{ Rectangular}  & $\RECformula$ & \parbox{\discussionWidth}{It  distributes the symbol energy uniformly in time domain.  It is the prototype filter for \ac{CP}-\ac{OFDM} scheme.} \\ \cline{1-3}
\parbox{\titleWidth}{Hanning\\(Raised-cosine)} &  $\Hannformula$  & \parbox{\discussionWidth}{The function itself and its first derivative are continuous. Hence, the power of the sidelobes fall at $1/\absOperator[\omega]^3$ per octave.} \\ \cline{1-3} 
\parbox{\titleWidth}{Exact Hamming} &  $\Hammformula$  & \parbox{\discussionWidth}{Exact Hamming filter places zero at the position of the first sidelobe.} \\ \cline{1-3}
\parbox{\titleWidth}{Exact Blackman} &  $\Blackformula$  & \parbox{\discussionWidth}{Exact Blackman filter places zeros at the positions of the the third and fourth sidelobes.} \\ \cline{1-3}

\parbox{\titleWidth}{ Tapered-cosine-in-time (Tukey)} &  $\eRECformula$  & \parbox{\discussionWidth}{It is the rectangular filter where the edges are tapered by convolving a rectangular function with a cosine lobe. When $\rollOff=1$, it corresponds to Hanning filter.} \\ \cline{1-3}
\parbox{\titleWidth}{ Tapered-cosine-in-frequency (Tukey)} &  $\RCformula$  & \parbox{\discussionWidth}{It distributes the symbol energy uniformly in frequency domain when $\rollOff=0$.}  \\ \cline{1-3}
\parbox{\titleWidth}{ Root-raised-cosine}  &  $\RRCformula$  & \parbox{\discussionWidth}{It corresponds to tapered-cosine-in-frequency after matched filtering.} \\ \cline{1-3}
\parbox{\titleWidth}{Mirabbasi-Martin}  & $\PHYformula$ & \parbox{\discussionWidth}{It provides rapid-decaying. Power of the sidelobes fall at $1/\absOperator[\omega]^{(\derivationOrder+3)}$ per octave, where $\derivationOrder$  is the derivation order. Last equation is utilized to construct a complete set of equations.} \\ \cline{1-3}
\parbox{\titleWidth}{Prolate } &   $\Prolateformula$   & \parbox{\discussionWidth}{It is the optimally-concentrated pulse in frequency for a given filter length and bandwidth.} \\ \cline{1-3}	
\parbox{\titleWidth}{Optimal finite\\ duration pulses} &   $\OFDPformula$   & \parbox{\discussionWidth}{It is the optimally-concentrated pulse  for a given duration and bandwidth, which also satisfies Nyquist criterion in both time and frequency.} \\ \cline{1-3}	
\parbox{\titleWidth}{ Kaiser} &   $\Kaiserformula$   & \parbox{\discussionWidth}{Kaiser filter has very similar time-frequency characteristics of prolate filter. Although it is suboptimum solution for concentration problem, its formulation is given in closed-from.} \\ \cline{1-3}
\parbox{\titleWidth}{Modified Kaiser} &   $\modifiedKaiserformula$   & \parbox{\discussionWidth}{In order to provide faster decaying, Kaiser window is modified to obtain zeros at $\absOperator[\timeSymbol] = 1/2$.} \\ \cline{1-3}
\parbox{\titleWidth}{ Gaussian}  &  $\GAUformula$  & \parbox{\discussionWidth}{It is the optimally-concentrated filter when there are no restrictions on filter length and bandwidth  and $\gaussianRollOff=1$.}   \\ \cline{1-3}

\parbox{\titleWidth}{ IOTA} &   $\IOTAformula$   & \parbox{\discussionWidth}{IOTA yields optimally-concentrated function when there are no restrictions on filter length and bandwidth. It also fulfills Nyquist criterion after matched filtering.} \\ \cline{1-3}

\parbox{\titleWidth}{ Hermite} &   $\HERformula$   & \parbox{\discussionWidth}{By deforming the Gaussian filter with the high-order Hermite functions, it obtains zero-crossings to satisfy Nyquist criterion. It has similar characteristics with IOTA. }\\ \cline{1-3}

\parbox{\titleWidth}{ Extended \\ Gaussian }&   $\EGFformula$   & \parbox{\discussionWidth}{It is a generalized family based on Gaussian function which gives the closed-form expression of the filter derived via IOTA.} \\ \cline{1-3}

\end{tabular}
\label{table:waveform_comp}
\end{table*}

\subsection{Design Criterion: Energy Concentration}
In practice, limiting a pulse shape in time decreases the computational complexity and reduces the communications latency, which are inversely proportional to the filter length. However, using shorter or truncated filter may cause high sidelobes in the frequency domain. 
\acp{PSWF} address this energy-concentration trade-off problem through obtaining a time-limited pulse with minimum out-of-band leakage or a band-limited pulse with maximal concentration within given interval. There are severals ways to characterize \acp{PSWF} \cite{Walter2005432}. A convenient definition  for the prototype filter design is that
\acp{PSWF},  $\{\prolateWave[n,\filterLength,\filterBand][\timeSymbolProlate]\}$, is a family that includes the orthogonal functions which are optimal in terms of  the energy concentration of a $\filterBand$-bandlimited function on the interval  $[-\filterLength,\filterLength]$, where $n$ is the function order. In the family, $\prolateWave[0,\filterLength,\filterBand][\timeSymbolProlate]$ is the most concentrated pulse and the concentration of the functions decreases with the function order. In other words, $\prolateWave[n,\filterLength,\filterBand][\timeSymbolProlate]$ is the most concentrated function after $\prolateWave[n-1,\filterLength,\filterBand][\timeSymbolProlate]$ and it is also orthogonal to $\prolateWave[n-1,\filterLength,\filterBand][\timeSymbolProlate]$. Hence, if one provides the filter length and the bandwidth (where the pulse should be concentrated) as the design constraints, the optimum pulse becomes $\prolateWave[0,\filterLength,\filterBand][\timeSymbolProlate]$ constructed based on these constraints.

\acp{PSWF} have many appealing properties \cite{Slepian_1961, Slepian_1983}. For example, they are the eigenfunctions of the operation of {\em first-truncate-then-limit-the-bandwidth}. Therefore, these functions can pass through this operation without any distortion or filtering effect excluding the scaling with a real coefficient, i.e, eigenvalue, which also corresponds to the energy after this operation. Assuming that the energy of the pulse is 1, eigenvalues will always be less than 1. Also, \acp{PSWF} correspond to an important family when $\filterLength=\filterBand\rightarrow\infty$, known as Hermite-Gaussian functions which are the eigenfunctions of Fourier transformation. Hermite-Gaussian functions provide optimum concentration in time and frequency at the same time. Hence, they are able to give isotropic (same) responses in time and frequency. We also refer the reader to the detailed discussions on the properties of \acp{PSWF} in \cite{ Landau_1961,Landau_1962,Slepian_1964,Slepian_1978,Walter2005432,Moore2004208}

In the following subsections, the prototype filters that target time-frequency concentration are discussed. Their characteristics are inherently related with the \acp{PSWF}.

\subsubsection{Prolate Window}
Prolate window addresses the energy concentration in frequency for a given filter length and bandwidth. In time domain, its expression corresponds to $\prolateWave[0,\filterLength,\filterBand][\frequencySymbol]$ or $0$th order Slepian sequence in time for the discrete case \cite{Slepian_1978}.
This issue is explained as a sidelobe minimization problem in \cite{24_farhang2011ofdm}, as shown in \tablename~\ref{table:waveform_comp}. The time and frequency characteristics of prolate window are given in \figurename~\ref{fig:ConcenratedPulses}.

\subsubsection{Kaiser Function}
An efficient solution for a filter with finite length is proposed by Jim Kaiser by employing Bessel functions to achieve an approximation to the prolate window \cite{Kaiser_1980,Harris_1978}. It offers a suboptimal solution for the out-of-band leakage. A favorable property of Kaiser filter is its flexibility to control the sidelobes and stop-band attenuation, through a single design parameter $\kaiserDesignParameter$ with a closed-form expression. The expression is given in \tablename~\ref{table:waveform_comp} where $I_0(x)$ denotes the zeroth order modified Bessel function of the first kind. 

\subsubsection{Optimal Finite Duration Pulses}
Although prolate window is an optimally-concentrated filter in terms of minimum sidelobe energy for a given filter length and bandwidth, it does not satisfy the Nyquist criterion that ensures zero-interference between the points in the lattice. Considering this fact, Vahlin exploits \acp{PSWF} to realize a new family which is referred as \ac{OFDP} \cite{105_Vahlin_TC_1996} by generalizing the optimization procedure given for  single carrier, presented in \cite{Halpern_1979}. The aim is to achieve a Nyquist filter in both time and frequency with the maximum energy in the main lobe for a given bandwidth and filter length. In order to develop these pulses, as summarized in \tablename~\ref{table:waveform_comp}, Vahlin chooses the signal representation of  \acp{OFDP} as the linear combinations of the \acp{PSWF} and formulates the constraints as an optimization problem to find the weights $\filterCoeffient[\filterCoefficentIndex]$ for $\filterCoefficentIndex$th \ac{PSWF} for a given interval, using Lagrange multipliers and calculus of variations. 
Since only  the even-indexed prolate functions are even-symmetrical, only $\filterCoeffient[2\filterCoefficentIndex]$ are considered through the optimization procedure. By applying similar optimization procedure, \ac{OFDP} has been utilized in \cite{4_Siohan_TSP_2002}. Also, another optimization procedure which is based on deriving the composite matched filtering response of \ac{OFDP} filter instead \ac{OFDP} itself is suggested in \cite{Nigam_2010} to reduce the optimization complexity.

\subsubsection{Gaussian Function}
It is a prolate window when $\filterLength=\filterBand\rightarrow\infty$.
It is utilized with a parameter $\gaussianRollOff$ in \cite{39_strohmer2003optimal, 26_Du_KTH_2007,49_tonello2005performance, 64_du2008pulse}  to control the filter localization.
Using the properties of Fourier transform, one may show that the frequency response of a Gaussian function is also another Gaussian function~\cite{26_Du_KTH_2007}.
When $\gaussianRollOff=1$, it yields identical responses in time and frequency and it corresponds to optimally-concentrated pulse among all functions. On the other hand, since it has no zero-crossings, Gaussian filter does not satisfy the Nyquist criterion. In other words, it introduces interference from one point to other points in the lattice.

\subsubsection{Isotropic Orthogonal Transform Algorithm}
\ac{IOTA} targets to obtain a filter which preserves the optimum concentration property of Gaussian filter, and orthogonalizes it to prevent interference to
neighboring points in the lattice \cite{2_le1995coded}. Starting with the Gaussian function, orthogonalized pulse is obtained as shown in \tablename~\ref{table:waveform_comp} where $\mathcal{F}$ and $\mathcal{F}^{-1}$ are the operators for Fourier transform and its inverse, respectively, and $\mathcal{O}_a$ is the orthogonalization operator. This operation also corresponds to the orthogonalization process in \eqref{Eq:gramRXS} and \eqref{Eq:gramTXS} \cite{ 39_strohmer2003optimal}. 
The constructed prototype filter fulfills the Nyquist criterion and may yield isotropic response in time and frequency. For example, when $\gaussianRollOff=1$, $\timeSpacingVariable = \sqrt{2}\timeSpacing/2$ and $\frequencySpacingVariable = \sqrt{2}\frequencySpacing/2$, the pulse shape becomes identical to its Fourier transform as shown in \figurename~\ref{fig:ConcenratedPulses}.

\subsubsection{Extended Gaussian Function}
\ac{EGF} corresponds to the analytical expression of the function obtained via \ac{IOTA}\cite{Roche_1997} as given in  \tablename~\ref{table:waveform_comp}, where $0.528\frequencySpacingVariable^2\leq\gaussianRollOff\leq 7.568\frequencySpacingVariable^2$ and $d_{k,\gaussianRollOff,\frequencySpacingVariable}$ are real valued coefficients. It is shown that these coefficients can be computed (for finite number of coefficients $b_{k,j}$) as
\begin{align}
d_{k,\gaussianRollOff,\upsilon_0}=\sum_{j=0}^{j_k}b_{k,j}e^{-(\pi\gaussianRollOff/2 \frequencySpacingVariable^2)(2j+k)}~,~0\leq k\leq K~,
\end{align}
where \cite{107_Siohan_TSP_2000} lists the coefficients $b_{k,j}$ for $0\leq k\leq 14$ and $0\leq j\leq 7$.  For $\gaussianRollOff=1$, $\timeSpacingVariable = \sqrt{2}\timeSpacing/2$ and $\frequencySpacingVariable = \sqrt{2}\frequencySpacing/2$, EGF gives the identical responses in time and frequency. In addition,
one may truncate \acp{EGF} into  one symbol duration to handle the latency drawback due to the filter length  while maintaining filter localization through some optimization procedures~\cite{Du2008,4_Siohan_TSP_2002}.

\subsubsection{Hermite Filter}
\begin{figure}[!t]
\centering
\subfloat[Energy distribution in time.]{\includegraphics[width=3.2in]{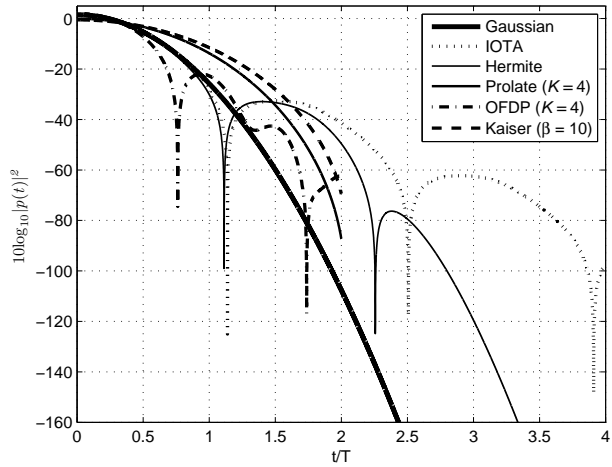}
\label{fig:ConcenratedPulses_timeResponse}}\\
\subfloat[Energy distribution in frequency.]{\includegraphics[width=3.2in]{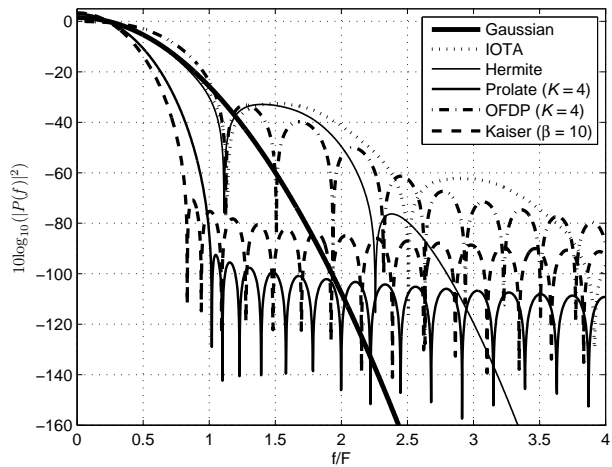}
\label{fig:ConcenratedPulses_freqResponse}}\\
\caption{Time and frequency characteristics of the filters, designed based on energy concentration.}
\label{fig:ConcenratedPulses}
\end{figure}

Hermite filter is obtained from the linear combinations of Hermite-Gaussian functions. By deforming the Gaussian filter with the high-order Hermite functions, zero-crossings are provided to satisfy Nyquist criterion \cite{84_haas1997time}. It has similar characteristics with \ac{IOTA} and yields isotropic response, which can be observed in \figurename~\ref{fig:ConcenratedPulses}. Advanced forms of the Hermite-Gaussian combinations also consider the dispersion characteristics of communication medium, which is discussed in Section \ref{sssec:OptimedPulses}.

\subsection{Design Criterion: Rapid-Decay}
Even though \acp{PSWF} provide the optimum solution for the energy concentration problem, they do not address the rapid-decaying of the sidelobes, as can be seen in  \figurename~\ref{fig:ConcenratedPulses}. Decaying of the sidelobes is related with the smoothness of a filter. It is known that smoothness is measured   by the number of continuous derivatives. If the $m$th derivative of a function is impulsive, the sidelobes of the function in frequency falls at $|\omega|^{-m}$ or $6m$~dB/octave, where $\omega$ is the angular frequency \cite{Harris_1978,Mirabbasi_2002,Singla_2010}.

\subsubsection{Raised-Cosine Function (Hanning Filter)}
The Hanning filter whose shape is captured through  a period of cosine function, is a smooth function in time. The zeroth and first order derivatives of the Hanning filter are continuous. Hence, sidelobes fall at $1/\absOperator[\omega]^3$ per octave, which corresponds to $18$~dB/octave.

\subsubsection{Tapered-Cosine Function (Tukey Filter)}
Tapered-cosine function is a filter where the time-frequency localization is controlled by the roll-off factor $(\rollOff)$.
While tapered-cosine function where $\rollOff=0$ results in rectangular shape, the shape becomes a raised-cosine function, i.e., Hanning filter, when $\rollOff=1$. Hence, it provides a function family where the decaying range is between $6$~dB/octave to $18$~dB/octave.

\subsubsection{Root-Raised-Cosine Function}
\ac{RRC} is typically utilized to satisfy Nyquist criterion after matched filtering. It is derived from the raised-cosine filter. \ac{RRC} where $\rollOff=1$ is known as \ac{HCF}, which is employed in ~\cite{2_le1995coded,4_Siohan_TSP_2002,24_farhang2011ofdm,27_farhang2010cosine,64_du2008pulse}.
It provides a good compromise for time/frequency behavior; its relaxed transition bands allow approximation through a relatively short time-domain filter, while achieving high attenuation in the stop-band~\cite{24_farhang2011ofdm}. On the other hand, \ac{RRC} filter with $\rollOff=0$ becomes a sinc function. While sinc function results in minimal bandwidth, it is very susceptible to truncation in time.

\subsubsection{Mirabbasi-Martin Filter}

\begin{figure}[!t]
\centering
\subfloat[Energy distribution in time.]{\includegraphics[width=3.2in]{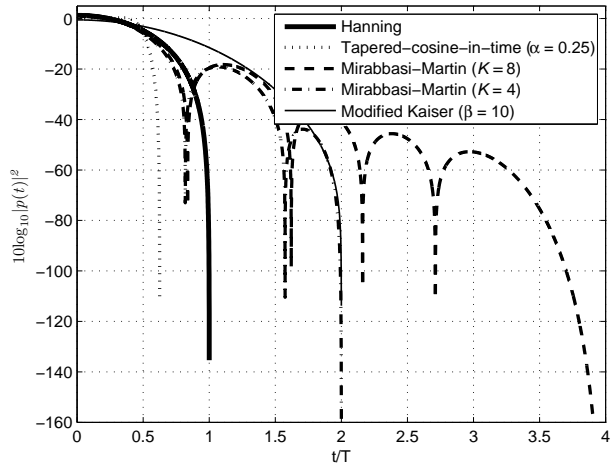}
\label{fig:rapidDecaying_timeResponse}}\\
\subfloat[Energy distribution in frequency.]{\includegraphics[width=3.2in]{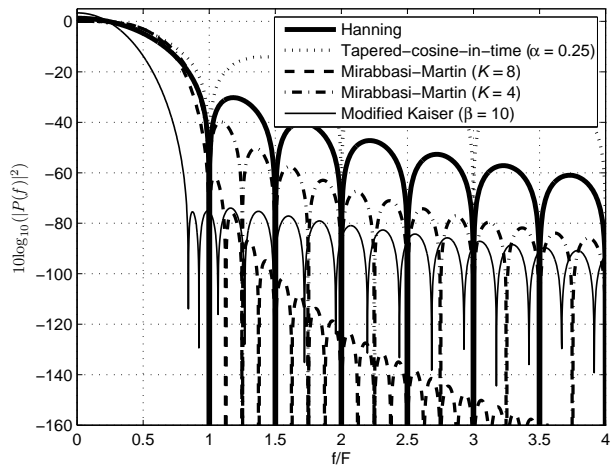}
\label{fig:rapidDecaying_freqResponse}}\\
\caption{Time and frequency characteristics of the filters, designed based on rapid-decaying property.}
\label{fig:rapidDecaying}
\end{figure}

Mirabbasi-Martin function is a filter where its coefficients are calculated by ensuring that the derivatives of the function is continuous \cite{106_Martin_TCS_1998, Mirabbasi_2002}. Obtained filter coefficients  are also specified for different $\truncation$ in \tablename~\ref{Table:PHYDYAS_COEFF}. While the decaying rate is  $1/\absOperator[\omega]^{5}$  per octave when $\truncation=6$, it is $1/\absOperator[\omega]^{7}$ per octave when $\truncation=8$ \cite{Mirabbasi_2002}. Its efficacy on decaying can be observed in  \figurename~\ref{fig:rapidDecaying}. It has been investigated further in~\cite{3_bellanger2001specification, Mirabbasi_2003} for multicarrier schemes, and was then subsequently adopted to be used in the European \ac{PHYDYAS} project on \ac{FBMC}~\cite{36_schaich2010filterbank,44_bellanger2010fbmc}. A similar approach based on polynomial functions is also proposed in \cite{Singla_2010}.

\begin{table}[t]
\caption{Filter coefficients for Mirabbasi-Martin \cite{Mirabbasi_2002}.} \label{Table:PHYDYAS_COEFF}
\begin{center}
\begin{tabular}{c||p{0.6in}|p{0.6in}|p{0.6in}|p{0.6in}}
   & $K=3$ & $K=4$ & $K=6$ & $K=8$ \\
\hline  \hline  $\filterCoeffient[0]$  & $1$ & $1$ & $1$ & $1$  \\
\hline  $\filterCoeffient[1]$  & $0.91143783$ & $0.97195983$ & $0.99818572$ & $0.99932588$  \\
\hline  $\filterCoeffient[2]$  & $0.41143783$ & $0.70710678$ & $0.94838678$ & $0.98203168$  \\
\hline  $\filterCoeffient[3]$  & & $0.23514695$ & $0.70710678$ & $0.89425129$  \\
\hline  $\filterCoeffient[4]$  & & & $0.31711593$ & $0.70710678$  \\
\hline  $\filterCoeffient[5]$  & & & $0.06021021$ & $0.44756522$ \\
\hline  $\filterCoeffient[6]$  & & & & $0.18871614$  \\
\hline  $\filterCoeffient[7]$  & & & & $0.03671221$  \\
\hline
\end{tabular}
\end{center}
\end{table}

\subsubsection{Modified Kaiser Function} Since Kaiser window is not a continuous function, it decays at the rate of $1/|\omega|$. In order to provide faster decaying, Kaiser window is modified to obtain zeros at $\absOperator[\timeSymbol] = 1/2$ \cite{Geckinli_1978}.

\begin{figure}[!t]
\centering
\subfloat[Energy distribution in time.]{\includegraphics[width=3.2in]{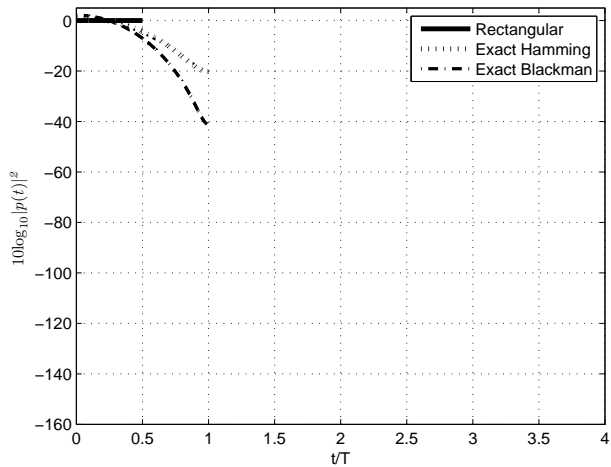}
\label{fig:spectrumNulling_timeResponse}}\\
\subfloat[Energy distribution in frequency.]{\includegraphics[width=3.2in]{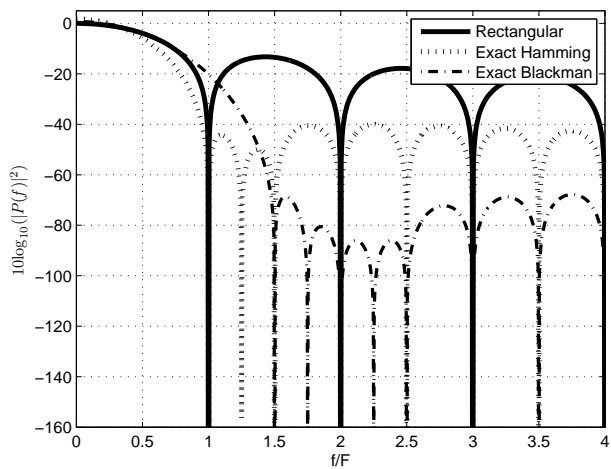}
\label{fig:spectrumNulling_freqResponse}}\\
\caption{Time and frequency characteristics of the filters, designed based on spectrum-nulling approach and rectangular function. }
\label{fig:spectrumNulling}
\end{figure}

\subsection{Design Criterion: Spectrum-nulling}

 As it can be realized, some of the functions in \tablename~\ref{table:waveform_comp} follow a general form of the low-pass filters as
\begin{align}
\pulseTime[]&=
\begin{cases}
\filterCoeffient[0]+2\displaystyle\sum_{\filterCoefficentIndex=1}^{\truncation-1}\filterCoeffient[\filterCoefficentIndex] \cos\left(2\pi{\filterCoefficentIndex\timeSymbol} \right)~, & \absOperator[\timeSymbol] \le \frac{1}{2}~,\\
0~, & {\rm otherwise}
\end{cases}
\label{eq:lowPass}
\end{align}
where $\truncation$ is the number of bins in frequency and $\filterCoeffient[\filterCoefficentIndex]$ is the $\filterCoefficentIndex$th filter coefficient. The form given in \eqref{eq:lowPass} is a generalized characterization to fulfill the spectrum-nulling criteria, since it corresponds to summation of sinc functions in the frequency domain. By exploiting the coherent cancellation of the tails of the sinc functions, one may systematically  place zeros in the spectrum. This approach may not directly yield good prototype filters for a multicarrier scheme because of the equal-ripple characteristics of the sidelobes. However, it is worth noting that the design criterion of these filters has some similarities with the existing sidelobe suppression techniques for the \ac{OFDM}-based schemes, e.g. cancellation carriers \cite{Brandes_2006}.
The following filters are designed based on this approach:

\subsubsection{Hamming Filter} Exact Hamming filter places a null at the position of the first sidelobe as shown in \figurename~\ref{fig:spectrumNulling}.

\subsubsection{Blackman Filter}
Exact Blackman filter places nulls at the positions of the third and fourth sidelobes, as shown in \figurename~\ref{fig:spectrumNulling}.

\begin{figure}[!t]
\centering
\subfloat[Energy distribution in time.]{\includegraphics[width=3.2in]{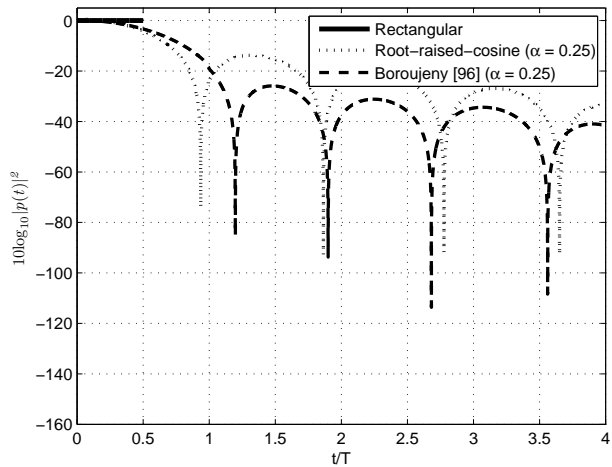}
\label{fig:channelRobustPulses_timeResponse}}\\
\subfloat[Energy distribution in frequency.]{\includegraphics[width=3.2in]{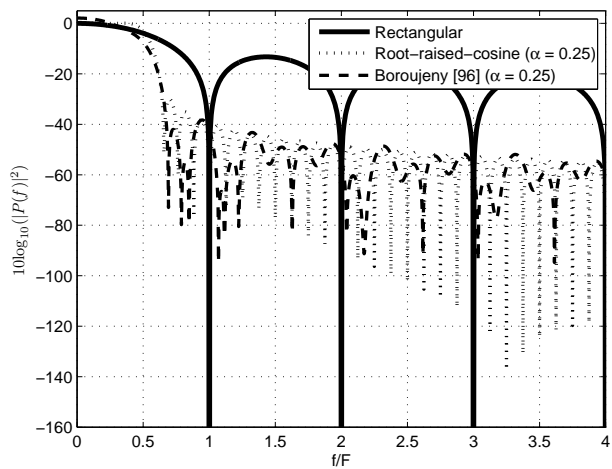}
\label{fig:channelRobustPulses_freqResponse}}\\
\caption{Time and frequency characteristics of the filters, designed based on channel constraints. }
\label{fig:channelRobustPulses}
\end{figure}

\subsection{Design Criterion: Channel  Characteristics and Hardware}
\label{subsec:channelBasedPulses}
Radio channel may hurt the orthogonality of the multicarrier schemes due to its dispersive  characteristics in frequency and time,  and causes \ac{ISI} and \ac{ICI}. In order to combat with \ac{ISI} and \ac{ICI}, one may design  pulse optimized to minimize  interference  among the points in the lattice. 

\subsubsection{Rectangular Function}
Rectangular function distributes the symbol energy uniformly in time domain. It is the prototype filter for conventional \ac{CP-OFDM} scheme. It provides an effective solution to combat with \ac{ICI} and \ac{ISI} in time-invariant multipath channels via an extension on its duration at the transmitter. This approach exploits the uniformity of rectangular function.

\subsubsection{Channel-based Pulses}
\label{sssec:OptimedPulses}
Prototype filters can be designed to perform best for given channel characteristics.
One of the methods is to use of optimally weighted combinations of the Hermite-Gaussian functions to construct new pulse shapes that are suitable for doubly dispersive channels.
For example, in 
 \cite{89_amini2010isotropic}, the nulls in the ambiguity surface, which correspond to zero-interference regions, are widened using the Hermite-Gaussian functions to increase the robustness of the scheme in doubly dispersive channels.
In \cite{77_kozek1998nonorthogonal}, Hermite-Gaussian family is utilized as a basis to minimize \ac{ICI} and \ac{ISI}, while in \cite{81_trigui2007optimum}, they are utilized to maximize the \ac{SIR} considering biorthogonal schemes.

Based on different considerations, e.g. maximum \ac{SINR} \cite{88_jung2007wssus} and  minimum \ac{ISI} and \ac{ICI} \cite{75_schafhuber2002pulse,80_matz2007analysis}, optimum pulses can be constructed using the Gaussian function itself an initial filter. 
As opposed to data communications, in \cite{linhao_2009}, two prototype filter design procedures are proposed by relaxing the orthogonality constraint of the prototype filter for preamble transmission. 

In addition  to above considerations, one may design a prototype filter  considering the hardware constraints. For example, a prototype filter which addresses the \ac{PAPR} and timing jitter problems by minimizing the tails of the prototype filter is proposed in \cite{69_farhang2008square}, which is shown in \figurename~\ref{fig:channelRobustPulses}.

\section{Evaluation Metrics and Tools for Multicarrier Schemes}
\label{sec:toolsForPrototypeFilterAnalyses}

In this section, various tools and metrics are introduced in order to have a better understanding of the prototype filters and characterize their performance for multicarrier communications. These tools are instrumental for assessing the performances of different prototype filters along with multicarrier schemes. First,  Heisenberg uncertainty parameter and direction parameter are given. Then, ambiguity surface is introduced and its usefulness on the evaluation of interference characteristics of multicarrier schemes is discussed. Further details about these tools and metrics may be found in \cite{Daubechies_1992,Harris_1978, Geckinli_1978, 24_farhang2011ofdm, Du2008}.

\subsection{Heisenberg Uncertainty Parameter}
Time-frequency localization of a filter is measured by the Heisenberg uncertainty parameter $\heisenbergParameter$ which is given by
\begin{align}
\heisenbergParameter \equiv \frac{\ltwonorm[{\pulseTime[]}]^2}{4\pi \timeDispersion\frequencyDispersion}\leq 1~,
\end{align}
where
\begin{align}
\timeDispersion = \sqrt{\displaystyle\int_{\mathbb{R}}(t-\timeGravity)^2|\pulseTime[]|^2\integrald\timeSymbol}~,&\\
\frequencyDispersion = \sqrt{\displaystyle\int_{\mathbb{R}}(f-\frequencyGravity)^2|\pulseFrequency[]|^2\integrald\frequencySymbol}~,
\end{align}
$\timeDispersion$ is the time dispersion (or the standard deviation of the energy in time), and $\frequencyDispersion$ is the frequency dispersion (or the standard deviation of the energy in frequency), and $\timeGravity$ and $\frequencyGravity$ are the mean values of the supports of the pulse in time and frequency, respectively~\cite{24_farhang2011ofdm,26_Du_KTH_2007,39_strohmer2003optimal}. 
Filters with good localization characteristics have a Heisenberg parameter closer to 1. Heisenberg parameter is exactly equal to 1 with the Gaussian filter where $\gaussianRollOff =1$.

\subsection{Direction Parameter}
The direction parameter $\directionParameter \in [0,\infty)$, which shows how a pulse shape lies in time-frequency plane, is given in \cite{64_du2008pulse} as
\begin{align}
\directionParameter = \frac{\timeDispersion}{\frequencyDispersion}~.
\end{align}
For example, while $\directionParameter$ is equal to 0 for the rectangular
filter, Gaussian filter where $\gaussianRollOff=1$ yields $\directionParameter = 1$ because of its isotropic dispersion. A
larger $\directionParameter$ gives a pulse stretched more along the time axis compared to the frequency axis.

\subsection{Ambiguity Function}
In order to obtain the correlation between the points in the lattice, the projection of transmitter and receiver prototype filters should be calculated at every integer multiple of symbol spacing in both time and frequency as
$
\langle \translatedModulatedFilterTX[\timeIndexTX][\frequencyIndexTX][\timeSymbol],\translatedModulatedFilter[\timeIndex][\frequencyIndex][\timeSymbol]\rangle
$.
By using fractional values instead of using $\timeIndex$ and $\frequencyIndex$, ambiguity function is obtained as \cite{2_le1995coded,81_trigui2007optimum,26_Du_KTH_2007,24_farhang2011ofdm} 
\begin{align}
\ambiguityFunction[\ambiguityTimeShift][\ambiguityFrequencyShift][\ambiguityTimeFrequencyShift]\triangleq& \int_{-\infty}^\infty
\prototypeFilterTX[\timeSymbol+{\frac{\ambiguityTimeShift}{2}}] \prototypeFilterRXc[\timeSymbol-{\frac{\ambiguityTimeShift}{2}} ]
e^{-j2\pi\ambiguityFrequencyShift \timeSymbol}{\rm d}\timeSymbol
~.
\label{Eq:realAmbiguity}
\end{align}
The properties of ambiguity function are given below:

\begin{itemize}
    \item Ambiguity function is a two dimensional correlation function in the time-frequency plane. It gives an intuitive demonstration of the robustness against ICI/ISI due to different impairments, such as time and frequency selectivity of the propagation channel
\cite{2_le1995coded,81_trigui2007optimum,26_Du_KTH_2007,24_farhang2011ofdm}.
\item Ambiguity function yields real values in case of even-symmetric prototype filters. 

\item One may express Nyquist criterion in terms of the ambiguity function \cite{24_farhang2011ofdm} as
\begin{align}
\ambiguityFunction[\timeIndex\timeSpacingVariable][\frequencyIndex\frequencySpacingVariable][]
=\begin{cases}
1~, & n=l=0~\\
0~, & {\rm otherwise}
\end{cases}~.
\end{align}


\item Similar to time-frequency localization of a filter, ambiguity function also cannot be concentrated arbitrarily. A similar expression to Heisenberg parameter is defined for ambiguity functions in \cite{39_strohmer2003optimal}.
\end{itemize}

\def\ambiguityFigureSize{2.25in}
\begin{figure*}[!t]
\centering
\subfloat[TX: Rectangular, RX: Rectangular ($\truncation=1$).]{\includegraphics[width=\ambiguityFigureSize]{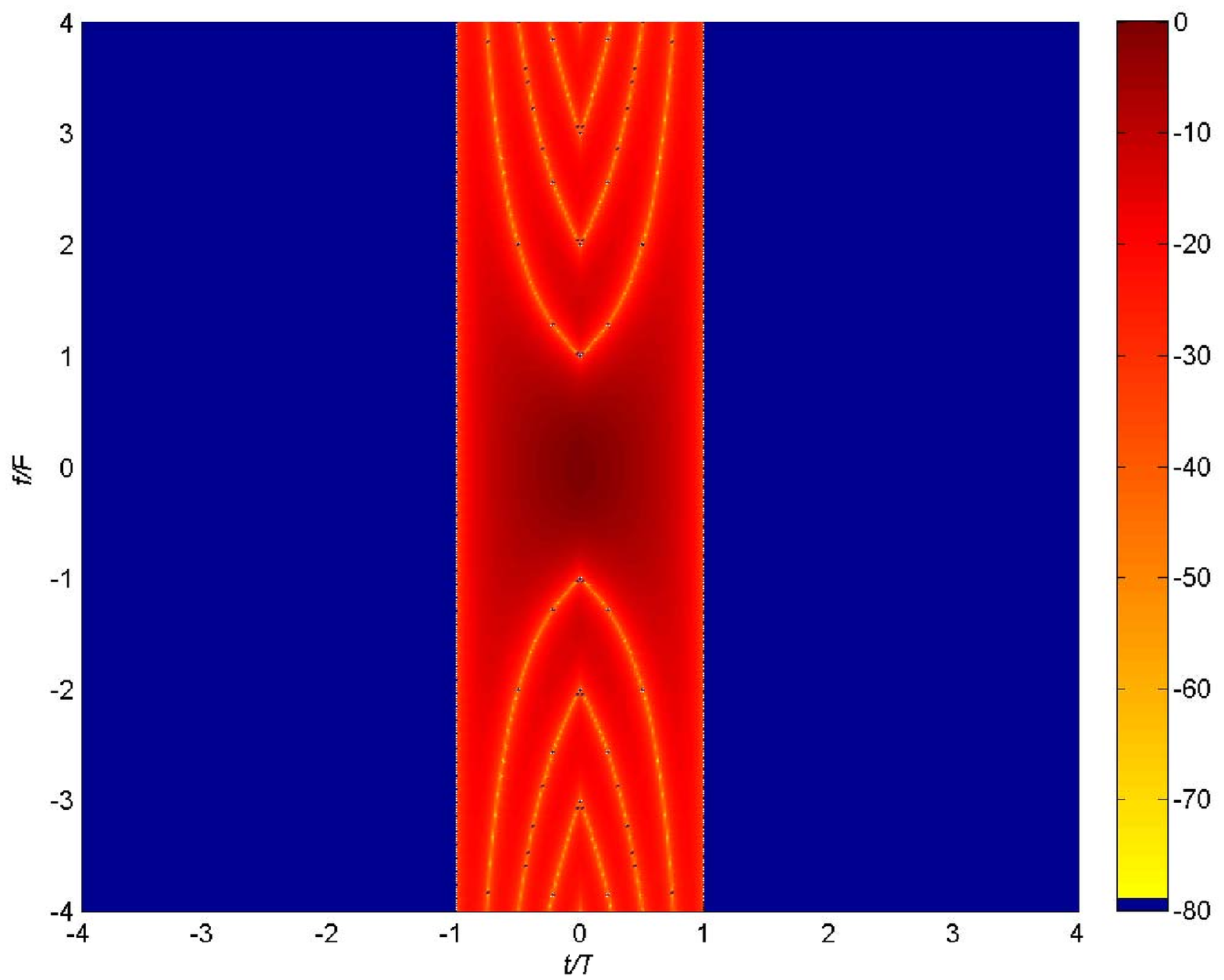}
\label{fig:rectangular}}~~
\subfloat[TX: Rectangular (extended-in-time), RX: Rectangular ($\truncation=1$).]{\includegraphics[width=\ambiguityFigureSize]{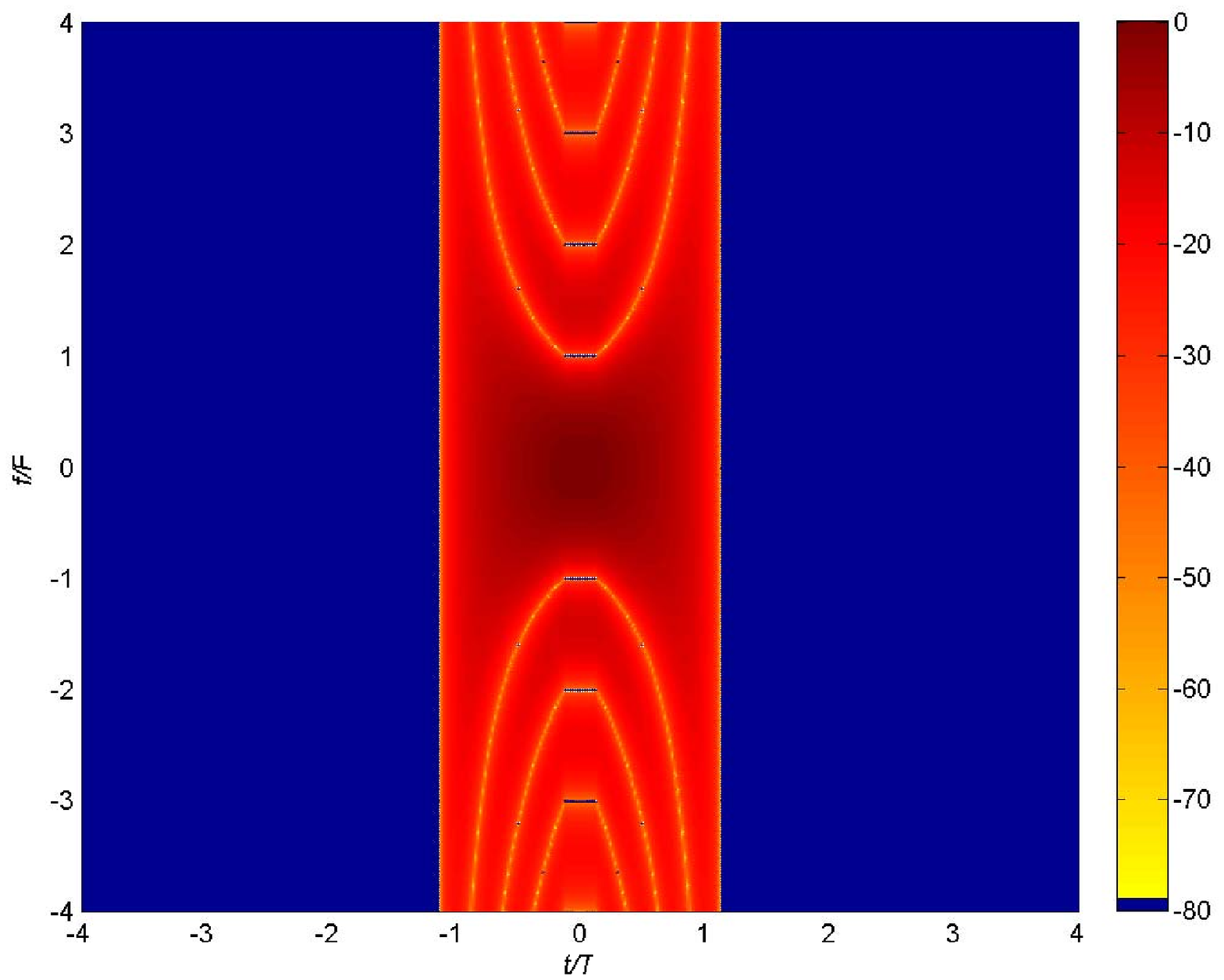}
\label{fig:erectangular}}~~
\subfloat[TX: Half-cosine, RX: Half-cosine ($\truncation=16$).]{\includegraphics[width=\ambiguityFigureSize]{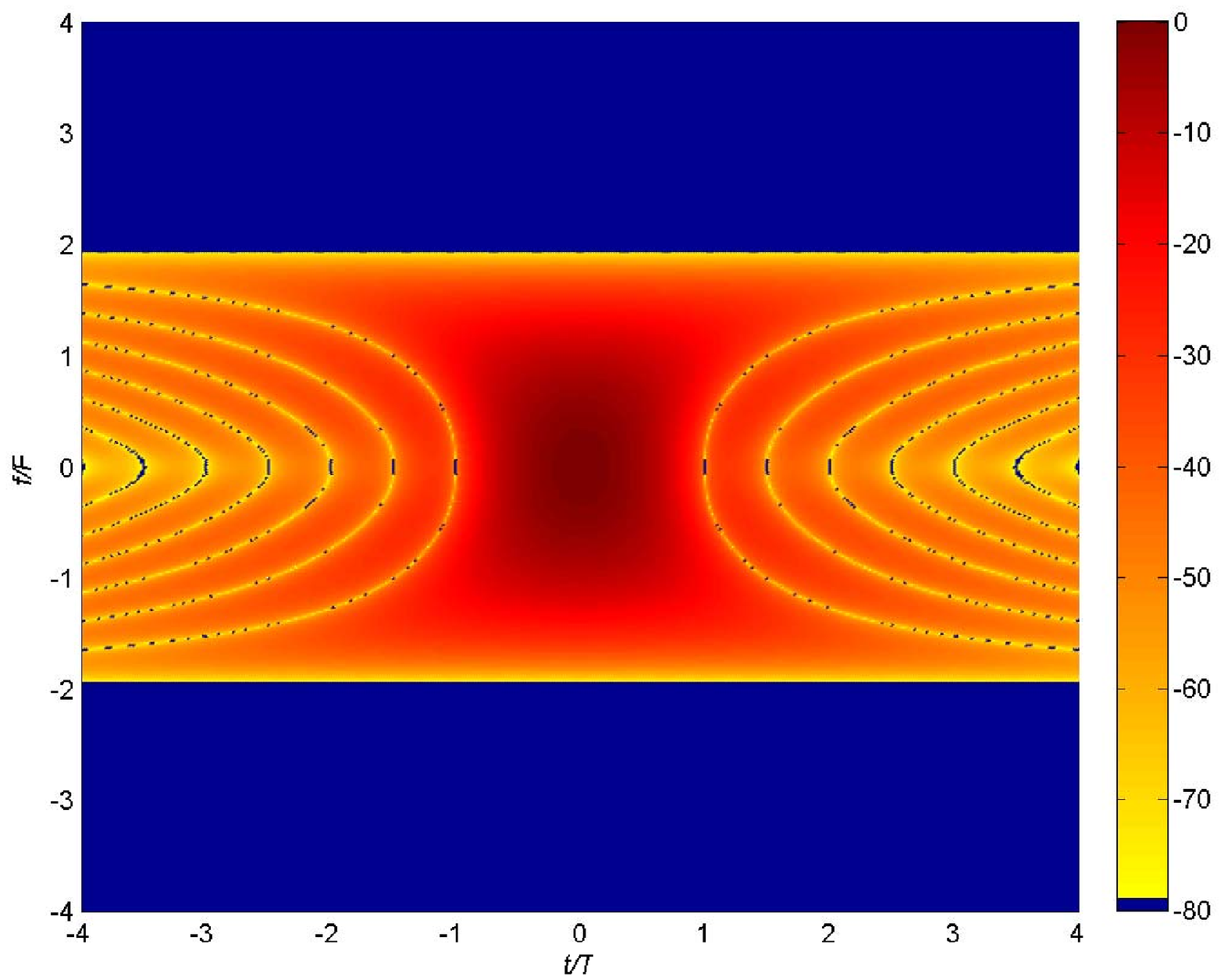}
\label{fig:hc}}
\vspace{-3mm}
\\
\subfloat[TX: Sinc, RX: Sinc ($\truncation=128$).]{\includegraphics[width=\ambiguityFigureSize]{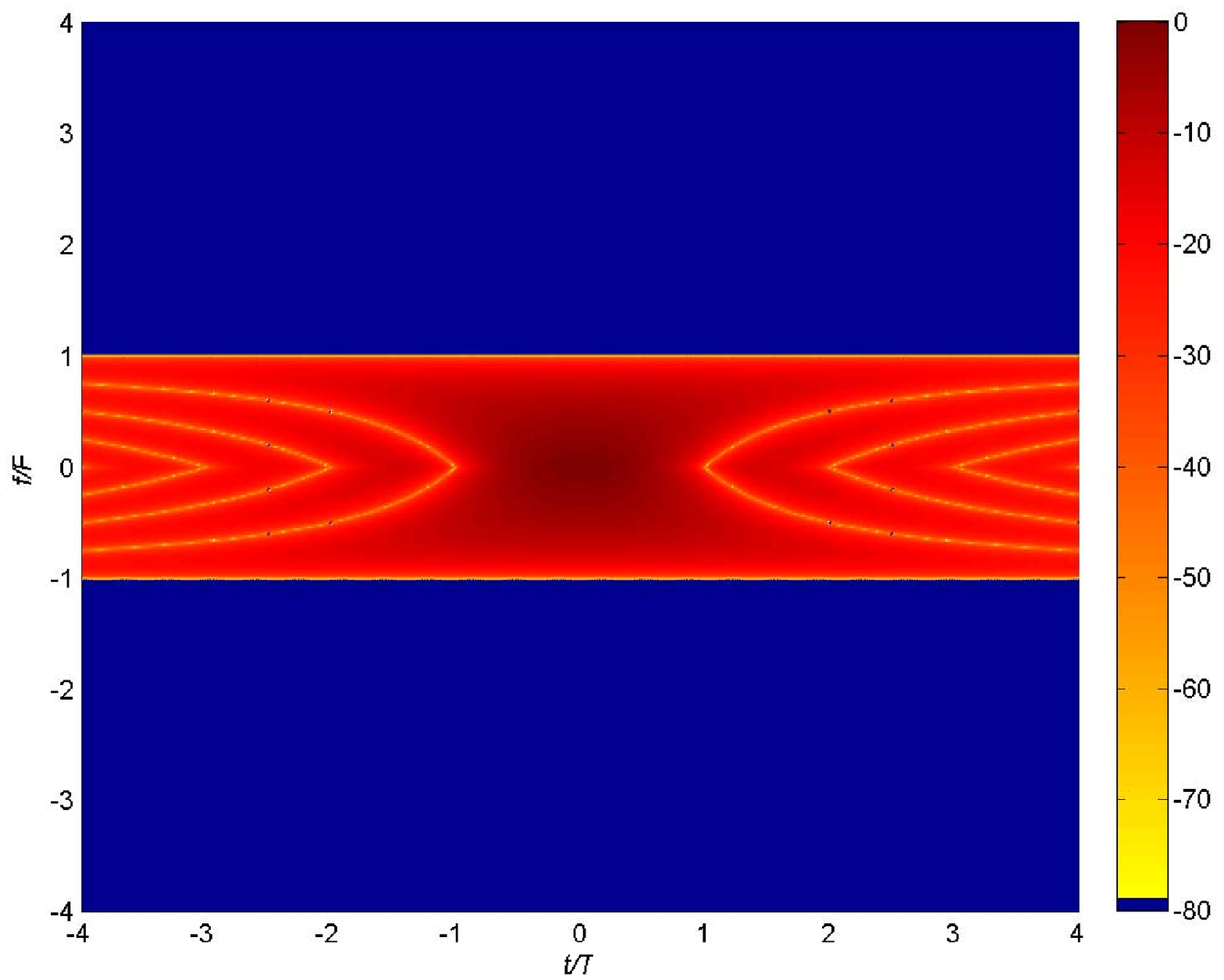}
\label{fig:sinc}}~~
\subfloat[TX: Sinc (extended-in-frequency), RX: Sinc ($\truncation=128$).]{\includegraphics[width=\ambiguityFigureSize]{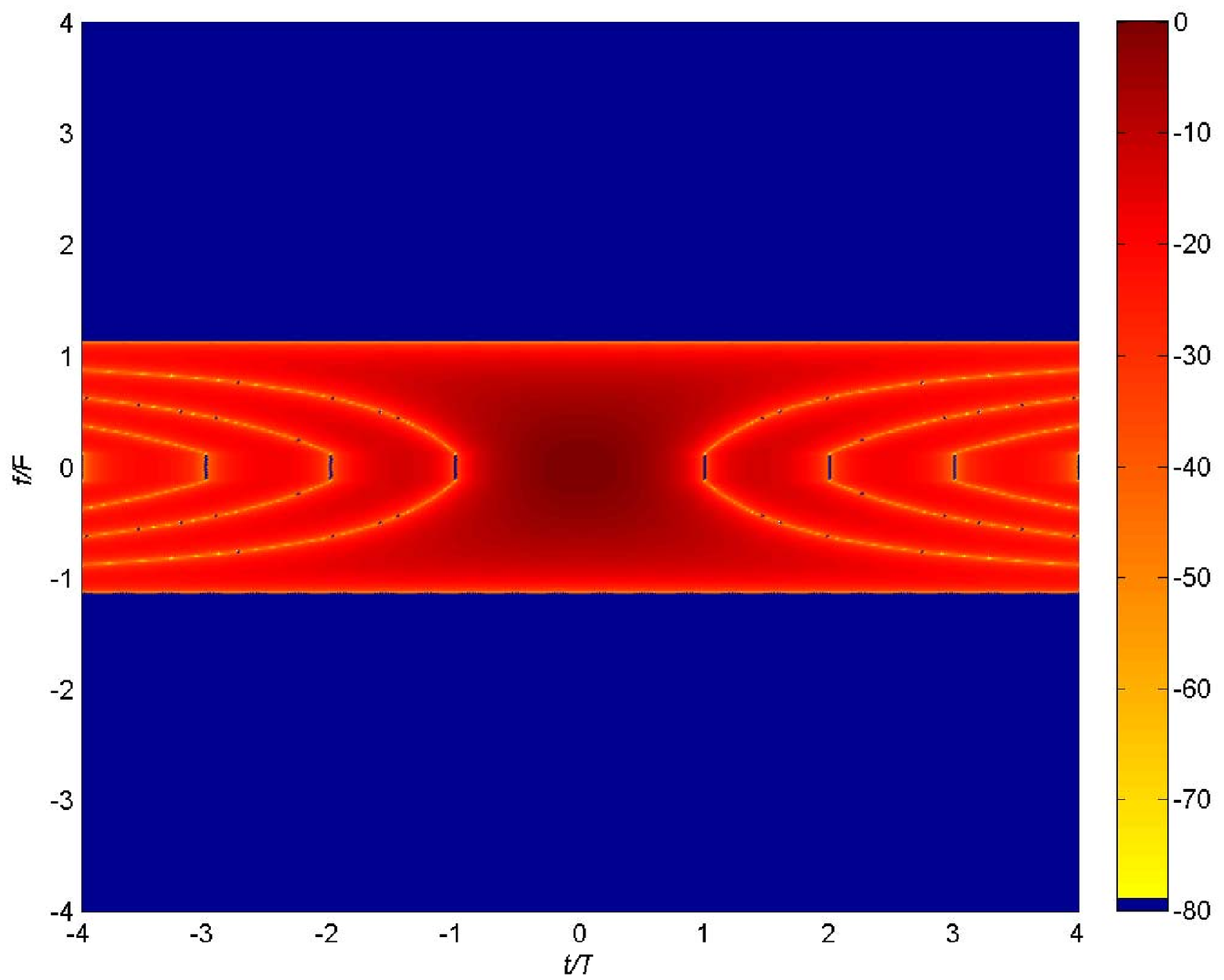}
\label{fig:esinc}}~~
\subfloat[TX: Mirabbasi-Martin, RX: Mirabbasi-Martin ($\truncation=8$).]{\includegraphics[width=\ambiguityFigureSize]{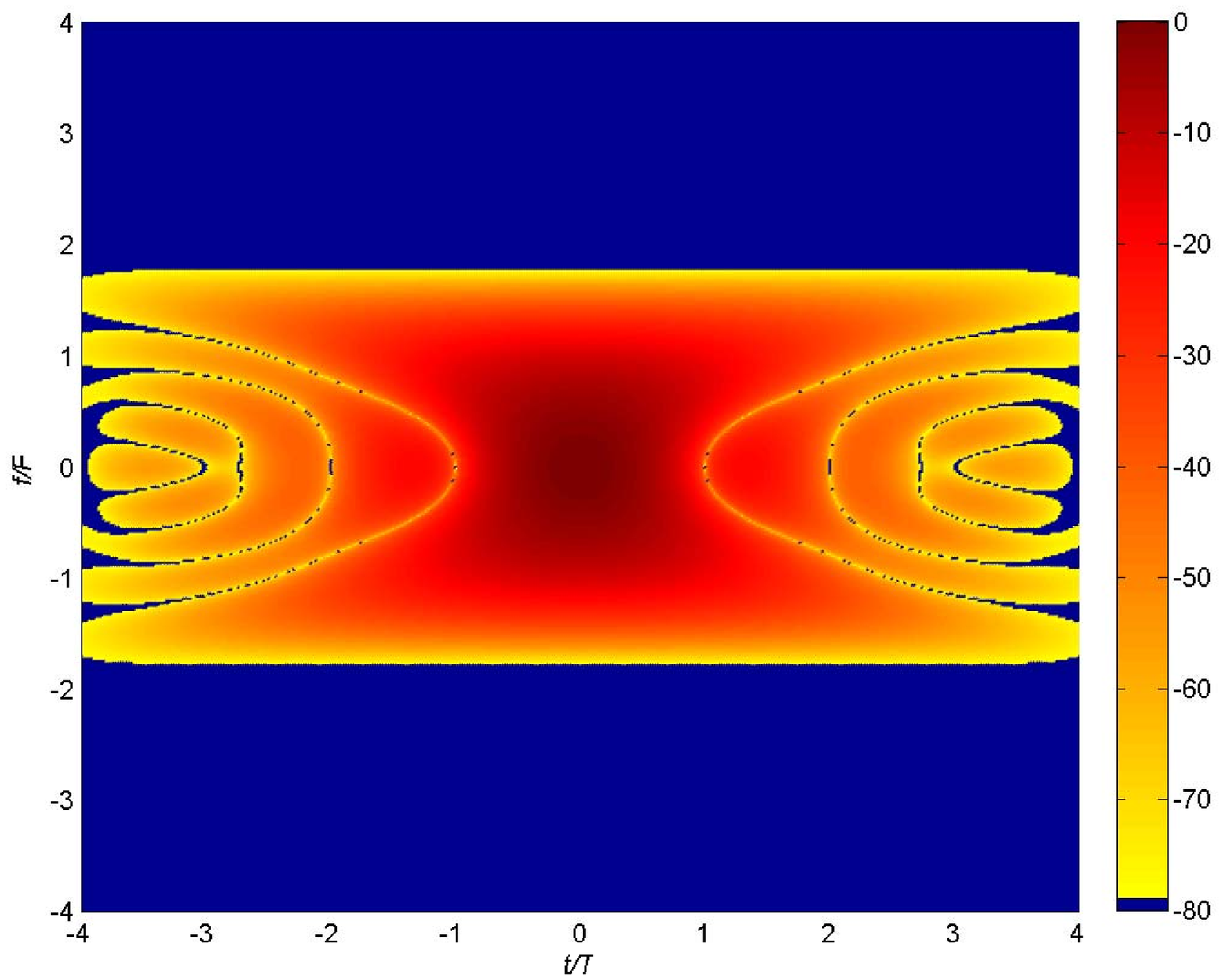}
\label{fig:mm8}}
\vspace{-3mm}
\\
\subfloat[TX: Prolate, RX: Prolate ($\truncation=4$).]{\includegraphics[width=\ambiguityFigureSize]{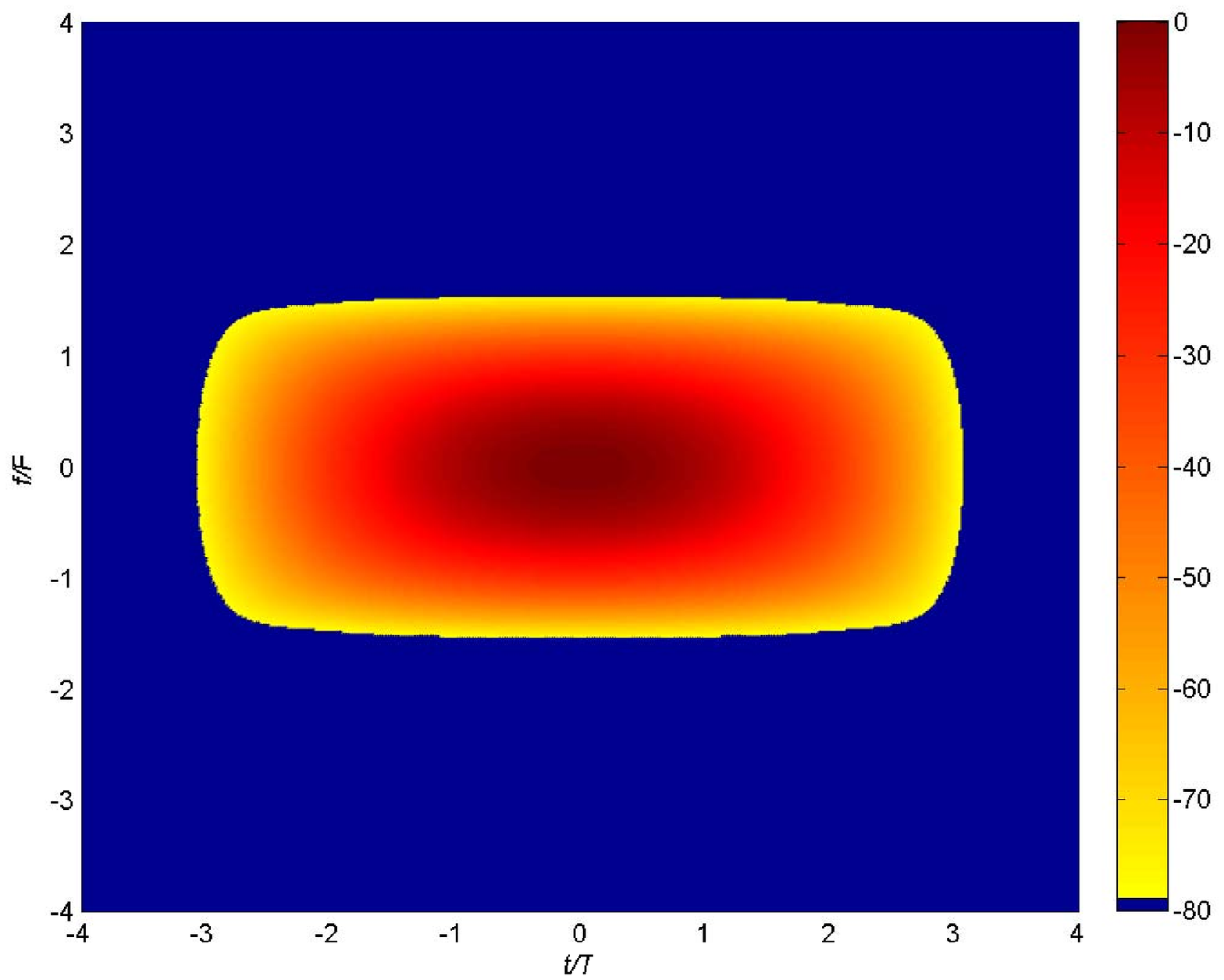}
\label{fig:prolate}}~~
\subfloat[TX: OFDP, RX: OFDP ($\truncation=4$).]{\includegraphics[width=\ambiguityFigureSize]{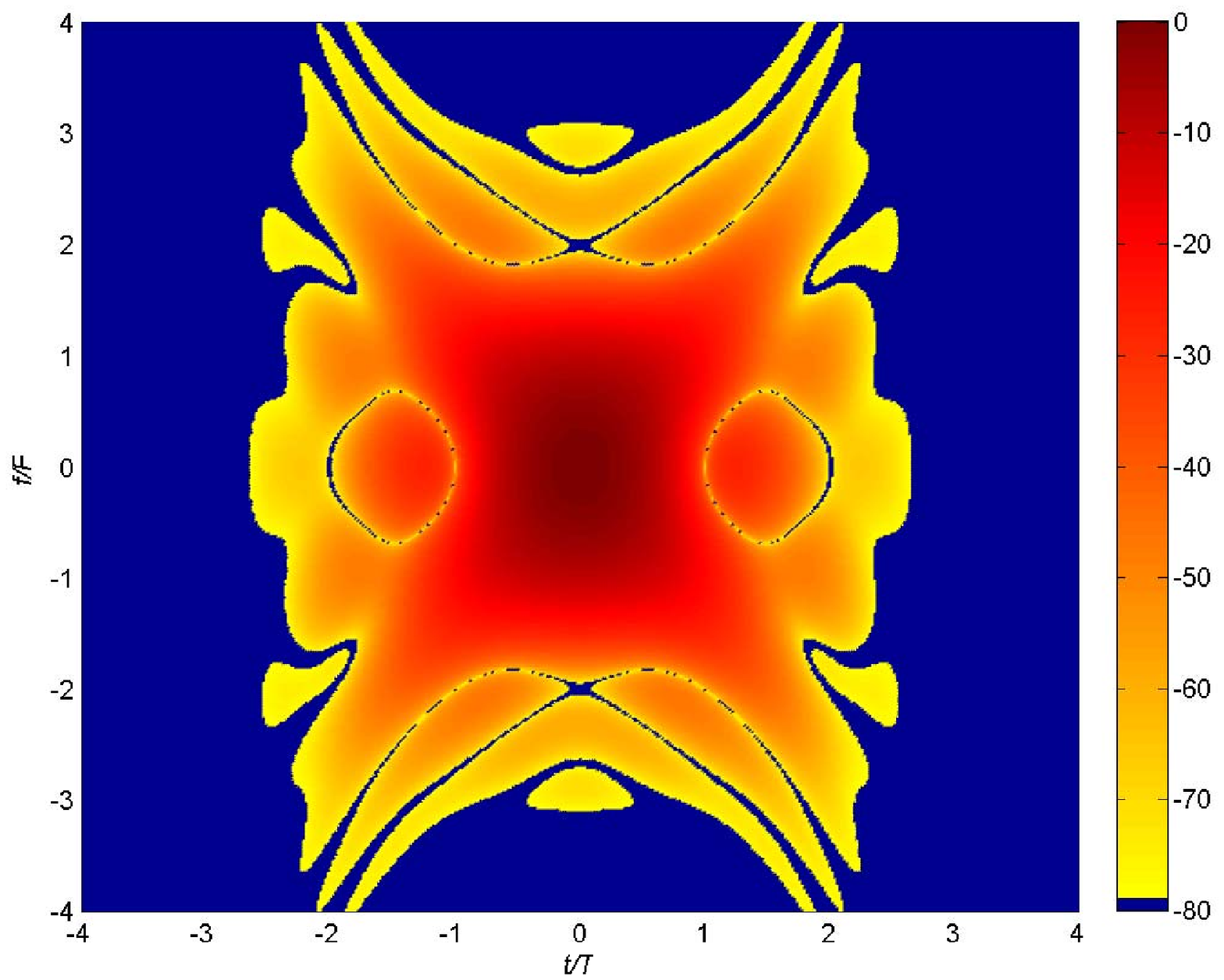}
\label{fig:ofdpk415}}~~
\subfloat[TX: Mirabbasi-Martin, RX: Mirabbasi-Martin ($\truncation=4$).]{\includegraphics[width=\ambiguityFigureSize]{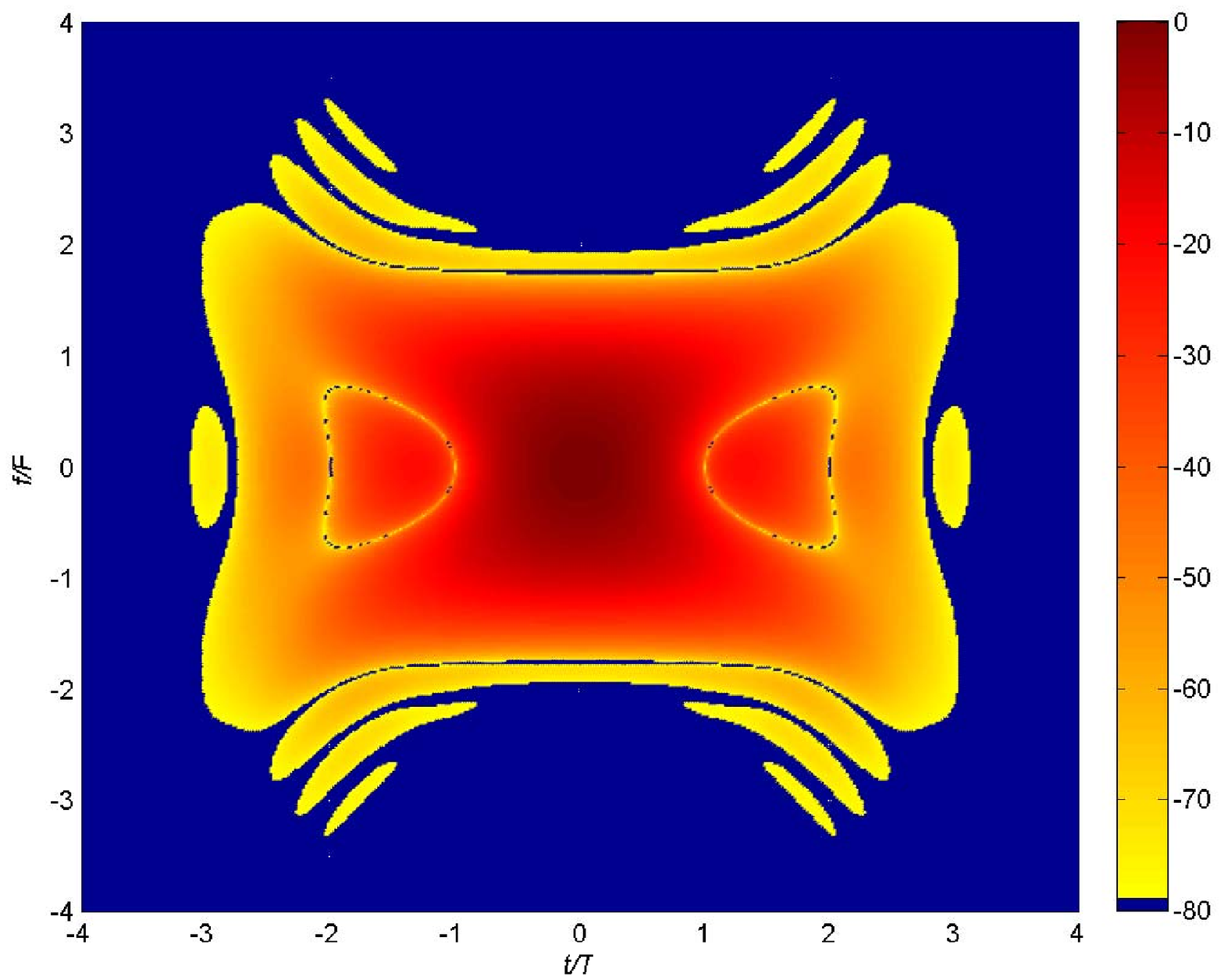}
\label{fig:mm4}}
\vspace{-3mm}
\\
\subfloat[TX: Gaussian, RX: Gaussian ($\truncation=4$).]{\includegraphics[width=\ambiguityFigureSize]{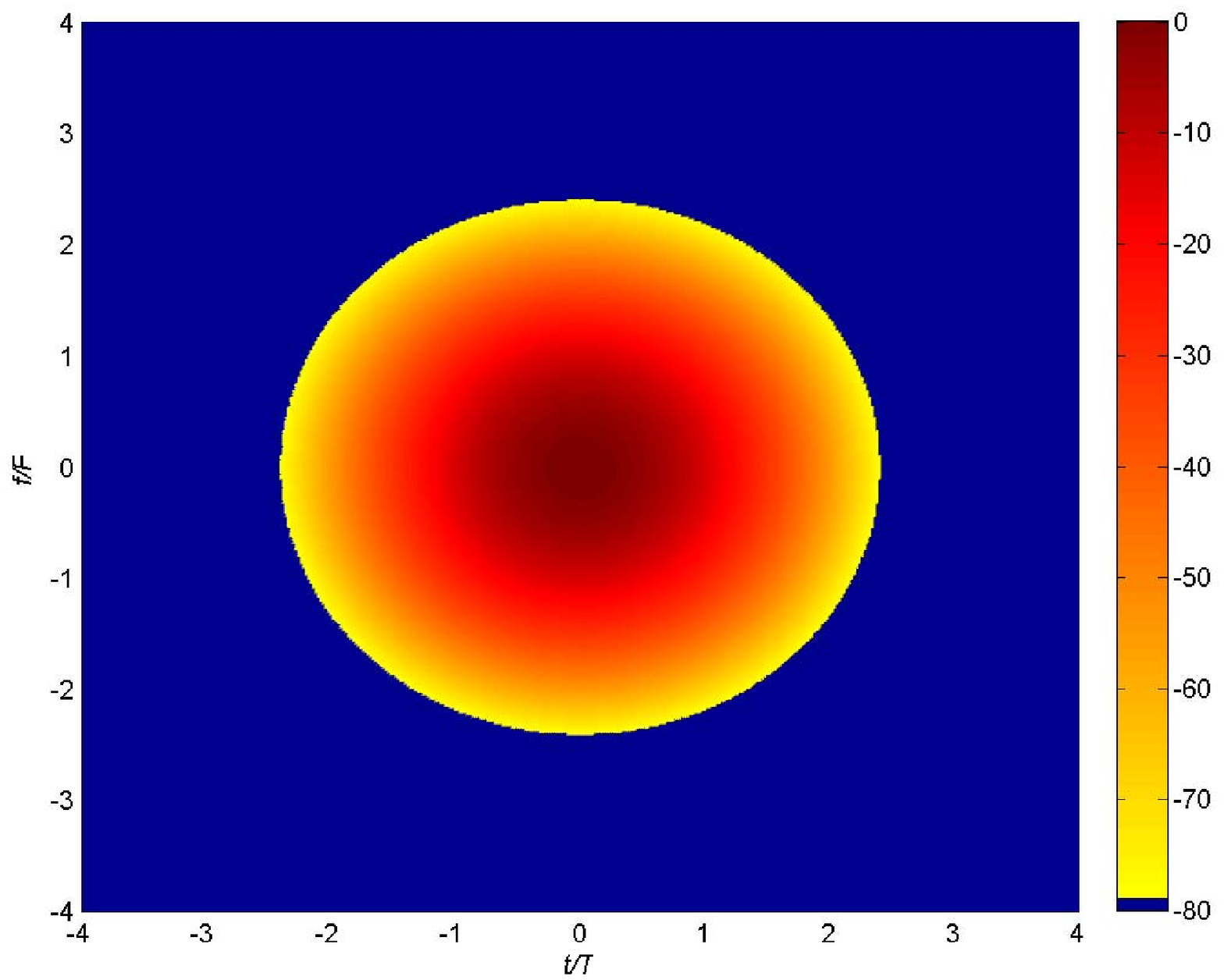}
\label{fig:gsn}}~~
\subfloat[TX: IOTA, RX: IOTA ($\truncation=8$).]{\includegraphics[width=\ambiguityFigureSize]{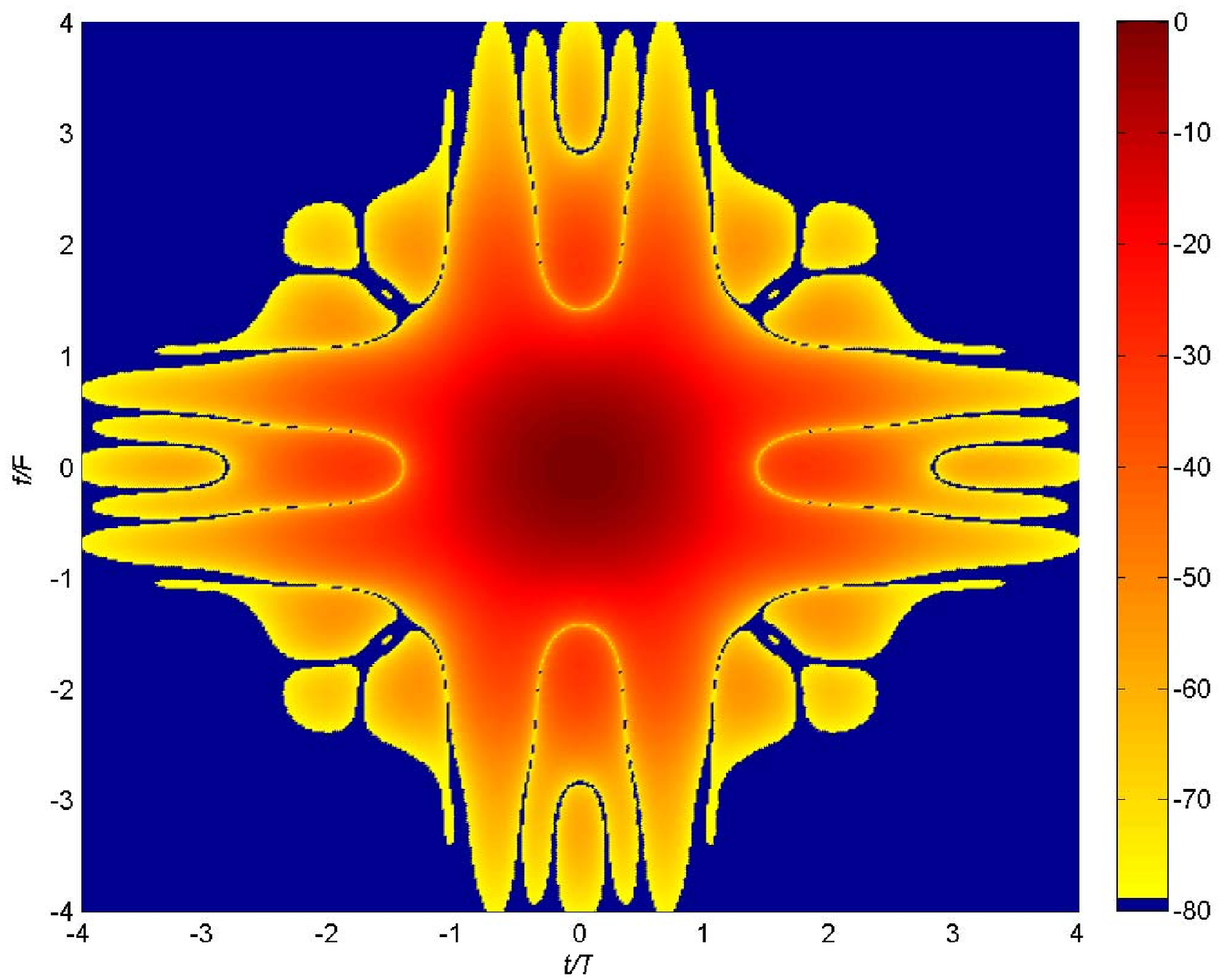}
\label{fig:iota}}~~
\subfloat[TX: Hermite, RX: Hermite ($\truncation=8$).]{\includegraphics[width=\ambiguityFigureSize]{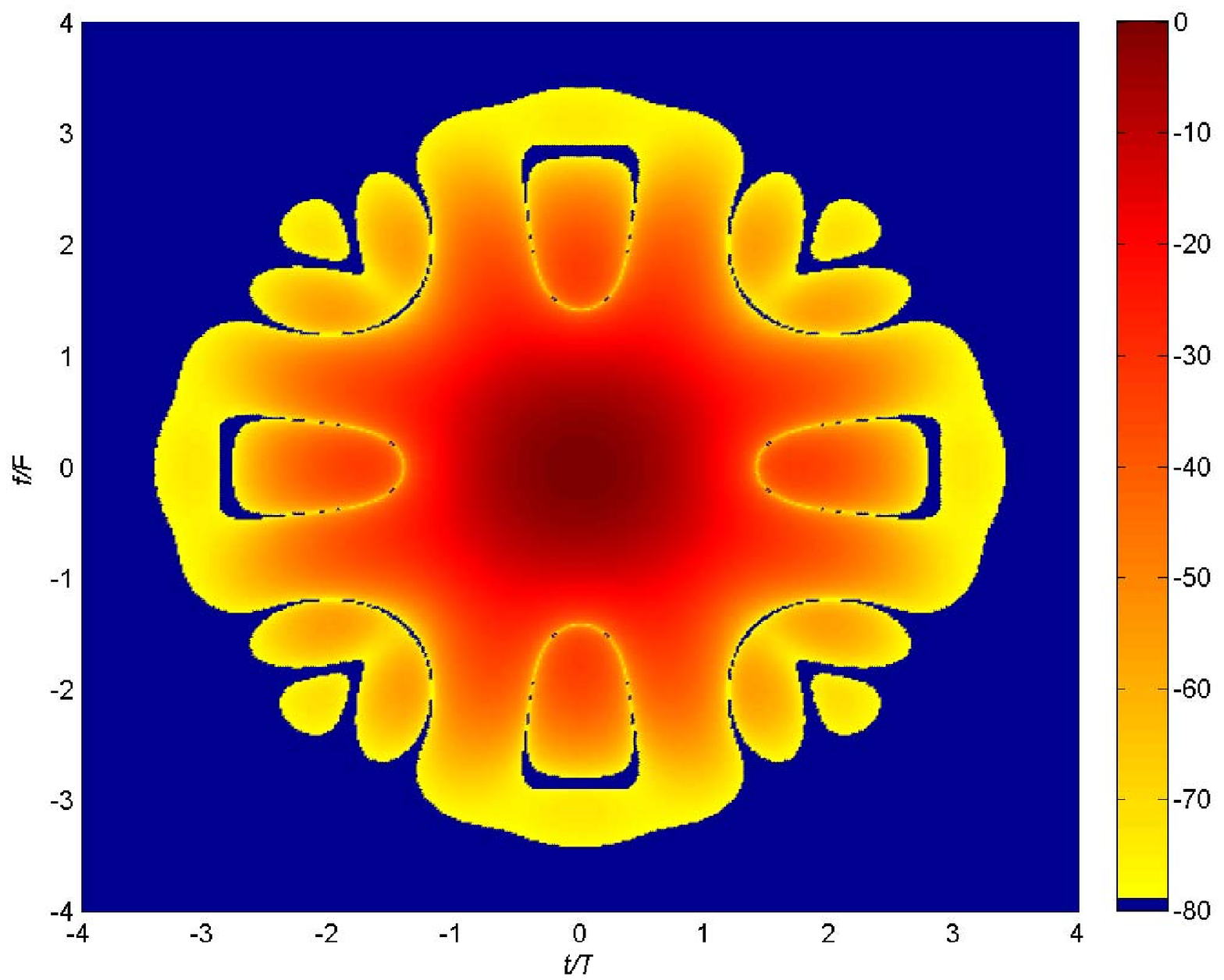}
\label{fig:her}}~~
\caption{Ambiguity surfaces with known prototype filters in the literature ($10\log_{10}(\absOperator[{\ambiguityFunction[\ambiguityTimeShift][\ambiguityFrequencyShift][\ambiguityTimeFrequencyShift]}]^2)$).}
\label{fig:amb}
\end{figure*}

Ambiguity surfaces for several filter pairs are provided in \figurename~\ref{fig:amb}. Without loss of generality, the filter length is set to $\truncation\times\timeSpacing$, where $\timeSpacing=1/\frequencySpacing$ and $\truncation$ is an integer number. In \figurename~\ref{fig:amb}\subref{fig:rectangular}, both transmitter and receiver utilize a rectangular filter, which corresponds to plain \ac{OFDM}. As it can be seen the nulls are located at the integer multiples of $\timeSpacing$ and $\frequencySpacing$.  In \figurename~\ref{fig:amb}\subref{fig:erectangular}, rectangular filter at the transmitter is intentionally extended in time compared to the one at the receiver, which corresponds to  the use of a  \ac{CP}. Hence, the nulls in the ambiguity surface are extended in time domain, around $\timeSymbol=0$. The dual response of extension-in-time is given with sinc filters in \figurename~\ref{fig:amb}\subref{fig:sinc} and \figurename~\ref{fig:amb}\subref{fig:esinc} by extending the filter in frequency domain. Hence, same impact of cyclic prefix is obtained in frequency, instead of time. In  \figurename~\ref{fig:amb}\subref{fig:hc}, ambiguity surface is obtained for the half-cosine function, i.e. \ac{RRC} where $\rollOff=1$. Since it has band-limited characteristic, it lies on the time domain. Also, it fulfills the Nyquist criterion as there are nulls located at the integer multiples of $\timeSpacing$ and $\frequencySpacing$. In \figurename~\ref{fig:amb}\subref{fig:mm8} and \figurename~\ref{fig:amb}\subref{fig:mm4}, Mirabbasi-Martin functions are investigated for $\truncation=8$ and $\truncation=4$, respectively. It can be seen that Nyquist criterion does not always hold for Mirabbasi-Martin filter, especially in frequency.
The ambiguity surface of prolate filter, given in \figurename~\ref{fig:amb}\subref{fig:prolate}, is similar to the one obtained for the Gaussian pulse in \figurename~\ref{fig:amb}\subref{fig:gsn}. Although the prolate window provides optimum concentration for a given filter length and bandwidth, it does not satisfy the Nyquist criterion.
Since \acp{OFDP} in \figurename~\ref{fig:amb}\subref{fig:ofdpk415} are derived from the proper combinations of prolate sequences, they are also concentrated while satisfying the Nyquist criterion. One may observe the nulls at the multiples of $\timeSpacing$ and $\frequencySpacing$ in \figurename~\ref{fig:amb}\subref{fig:ofdpk415}.
The ambiguity surfaces of Gaussian, \ac{IOTA}, and Hermite-Gaussian combinations are given in \figurename~\ref{fig:amb}\subref{fig:gsn},   \figurename~\ref{fig:amb}\subref{fig:iota}, and \figurename~\ref{fig:amb}\subref{fig:her}, respectively. Gaussian filter provides a circular ambiguity surface without any nulls in the surface. However, the localization of Gaussian filter is the best compared to the other filters and decays fast in both time and frequency. Hermite-Gaussian combinations and IOTA filters provide localized ambiguity functions, while satisfying Nyquist criterion at the integer multiples of $\sqrt{2}\timeSpacing$ and $\sqrt{2}\frequencySpacing$. The main reason of the selections of $\sqrt{2}\timeSpacing$ and $\sqrt{2}\frequencySpacing$ is to obtain a prototype filter with identical responses in time and frequency, which is also suitable for the schemes with lattice staggering investigated in Section \ref{subsec:staggering}.

\subsection{Signal-to-Interference Ratio  in Dispersive Channels}
\label{subsec:orthogonalityParameter}
It is possible to write the average \ac{SIR} performance as
\begin{align}
\SIR = \frac{\signalPower}{\interferencePower}~,
\label{Eq:SIR}
\end{align}
where $\signalPower$ and $\interferencePower$ are the power of the desired part and the interference leaking from other symbols, respectively. \ac{SIR}, as defined here, should ideally be infinity for orthogonal and biorthogonal schemes, since no other interference sources are considered. However, orthogonality can be still spoiled by not only  due to the lack of exact representation of the filter in digital domain, e.g. truncation, but also due to the dispersive channel.  In the literature, statistical characteristics of the channel are generally described with \ac{WSSUS} assumption 
\cite{85_bello1963characterization,91_molnar1996wssus} given by
\begin{align}
&\expectedOperator[\channelProcess] = 0~,\label{eq:channelAverage}\\
&\expectedOperator[\channelProcess\channelProcessOther] = \scatteringFunction\delta(\timeShift\textrm{-}\timeShiftOther)\delta(\frequencyShift\textrm{-}\frequencyShiftOther)~,
\label{eq:channelAutocorrelation}
\end{align}
where $\scatteringFunction$ is the channel scattering function and $\expectedOperator[\cdot]$ is the expected value operator. The term {\em wide sense stationary} means that the  statical characteristics of the first two moments of the channel scattering function do not change with time, and are only related with the time difference as in \eqref{eq:channelAutocorrelation}. The term of {\em uncorrelated scattering} implies that one of the delay components of the received signal is uncorrelated with all the other delay components. However, note that the \ac{WSSUS} assumption ignores the non-stationary characteristics due to the distance dependent path loss, shadowing, delay drift, and the correlation between the reflected rays from the same physical objects \cite{90_matz2005statistical}. In addition, \ac{WSSUS} assumption is not valid for short-term channel characteristics, especially, when the specular reflections dominate over diffuse scattering \cite{99_Kozek_SPAWC_1997}. Yet, \ac{WSSUS} assumption is widely used in the system models to characterize the wireless channels because of its simplicity. For example, exponential decaying multipath with Jakes Doppler spectrum \cite{Rappaport2001} or ITU models \cite{itu_channels} for different environments are commonly used for the channel scattering function for terrestrial communications. 

\begin{figure}[!t]
\centering
{\includegraphics[width=2.5in]{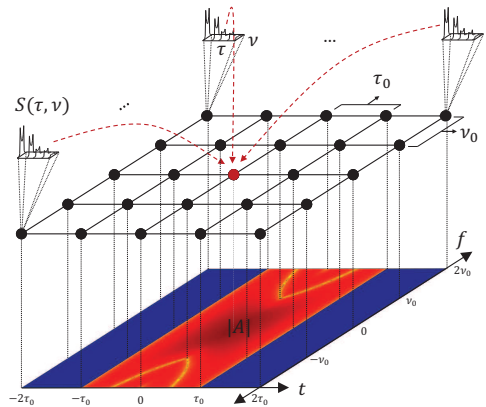}
\label{fig:qam}}
\caption{Interference analysis using the ambiguity function of rectangular filter in Fig. 9(a).
}
\label{fig:qam_oqam}
\end{figure}

Considering \ac{WSSUS} assumption, analytical expressions of $\signalPower$ and $\interferencePower$ can be obtained as follows. Assume that the impact of \ac{ICI} and \ac{ISI} on each subcarrier is statistically equal to each other, all subcarriers are utilized, $\symbols[\timeIndexTX][\frequencyIndexTX]$ are independent identically distributed
with zero mean, and $\langle\prototypeFilterTX[\timeSymbol], \prototypeFilterRX[\timeSymbol]\rangle=1$. Then, the power of desired part and the power of interference from other symbols in the lattice are obtained as
\begin{align}
\signalPower &= 
\int_{\timeShift}\int_{\frequencyShift}\scatteringFunction\absSquareOperator[{\ambiguityFunction[\timeShift][\frequencyShift][0]}]{\integrald\frequencyShift}{\integrald\timeShift}~,
\label{Eq:QAM_desired}
 \\
\interferencePower &= \sum_{\substack{( \timeIndexTX,\frequencyIndexTX)\neq(0,0)}}\int_{\timeShift}\int_{\frequencyShift} \scatteringFunction\absSquareOperator[
{\ambiguityFunction[\timeIndexTX\timeSpacingVariable+\timeShift][\frequencyIndexTX\frequencySpacingVariable+\frequencyShift][\frequencyIndexTX\frequencySpacingVariable\timeShift]}]{\integrald\frequencyShift}{\integrald\timeShift}~,
\label{Eq:QAM_interference}
\end{align}
respectively. 
The illustration of \eqref{Eq:QAM_interference} is given in \figurename~\ref{fig:qam_oqam}, where the absolute value of the ambiguity function at the grid points (black circles) are equal to zero in time-invariant single-path channel, i.e., $\interferencePower=0$. Due to time-variant multipath channel, the desired symbol (red circle at the middle) observes \ac{ISI}/\ac{ICI} from the neighboring points in the lattice, which is captured though \eqref{Eq:QAM_interference}.

\section{Practical  Implementation Aspects}
\label{sec:txdesign}

This section addresses several issues that have not been directly addressed in the earlier sections and that are related to implementation of multicarrier schemes. Representative references are provided to other related work in the literature and some of the key insights are highlighted. For a more comprehensive treatment and a larger body of literature related to these problems, the reader is referred to the corresponding references given in \tablename~\ref{Table:FBMC_Taxonomy}. This table provides a classification of various references in the literature related to prototype filter design. 
While each row of the table includes a specific topic, commonly used prototype filters in the literature are listed in each column of \tablename~\ref{Table:FBMC_Taxonomy}. References for different prior work on schemes are placed in the table based on the prototype filters used in each reference, as well as the scope of the corresponding reference. 
The goal is to capture a {\em research activity map} of this area, which is useful for understanding the key contributions and open research directions.

\subsection{Lattice and Filter Adaptations}
\label{SubSec:LatticeAndFilterAdapatations}
Lattice and filter parameters that maximize the robustness of the multicarrier schemes against doubly-dispersive channels are given by
\begin{align}
\underbrace{\frac{\timeDispersion}{\frequencyDispersion}\propto \frac{\RMSdelaySpread}{\MAXshift}}_{\rm pulse~ adaptation}~,~\underbrace{\frac{\timeSpacingVariable}{\frequencySpacingVariable}\propto \frac{\RMSdelaySpread}{\MAXshift}}_{\rm grid~adaptation}~,
\label{eq:matching}
\end{align}
where $\RMSdelaySpread$ is the \ac{RMS} delay spread of the channel and $\MAXshift$ is the maximum Doppler frequency. While {\em grid adaptation} corresponds to aligning $\timeSpacingVariable$ and $\frequencySpacingVariable$ with $\RMSdelaySpread$ and $\MAXshift$ when the lattice geometry is rectangular, {\em pulse adaptation} is equivalent to the dilation of the pulse depending on the channel dispersion  in time and frequency. The rationale behind these adaptations is to match the proportion of the channel dispersion to the pulse dispersion in the time and frequency.  
In an early study, \cite{2_le1995coded}, the pulse adaptation is given without theoretical explanations. In a later work,  the equality of  $\timeDispersion / \frequencyDispersion = {\MAXdelay}/{\MAXshift}$ is obtained theoretically in \cite{feichtinger1998gabor} when $\MAXdelay\MAXshift\rightarrow0$, where $\MAXdelay$ is the maximum excess delay of the channel. This identity is utilized for orthogonal schemes in \cite{39_strohmer2003optimal,Du2008,64_du2008pulse}, for biorthogonal schemes in \cite{77_kozek1998nonorthogonal,80_matz2007analysis} ($\latticeDensity \le 1$) and \cite{86_han2009wireless} ($\latticeDensity > 1$), for non-orthogonal scheme \cite{117_Han_TSP_2007} ($\latticeDensity > 1$) and improved by including adaptation on the lattice geometry in \cite{86_han2009wireless, 117_Han_TSP_2007, 39_strohmer2003optimal,Du2008}. 
For example, in \cite{86_han2009wireless}, both pulse adaptation and lattice adaption are jointly optimized to adapt to the doubly-dispersive channel characteristics based on \ac{WSSUS} assumption. Considering a rectangular geometry for the lattice, the optimum conditions based on Gaussian filter for the pulse and grid adaptations are derived as
\begin{align}
&\scatteringFunction = \deltaDirac[\timeShift]\deltaDirac[\frequencyShift] \Rightarrow\frac{\timeDispersion}{\frequencyDispersion}= \frac{\MAXdelay}{\MAXshift} 
\label{eq:idealScattering}
\\ 
&\scatteringFunction = 
\frac{1}{2\MAXdelay\MAXshift} \Rightarrow
\frac{\timeDispersion}{\frequencyDispersion}= \frac{\timeSpacingVariable}{\frequencySpacingVariable}= \frac{\MAXdelay}{\MAXshift} 
\label{eq:rectangularScattering}
\\
& \scatteringFunction = \frac{e^{-\frac{\timeShift}{\RMSdelaySpread}}\times\frac{1}{\pi\MAXshift}}{\RMSdelaySpread\sqrt{1-\frac{\frequencyShift^2}{\MAXshift^2}}} \Rightarrow
\frac{\timeDispersion}{\frequencyDispersion}= \frac{\timeSpacingVariable}{\frequencySpacingVariable} = \frac{1.5\RMSdelaySpread}{\MAXshift}
\label{eq:expUScattering}
\end{align}
where \eqref{eq:rectangularScattering} corresponds to a doubly dispersive channel with a uniform delay power profile and uniform Doppler power spectrum, while \eqref{eq:expUScattering} corresponds to a doubly dispersive channel with an exponential delay power profile and U-shape Doppler power spectrum.

In \cite{39_strohmer2003optimal}, a theoretical framework to adapt the pulse shape and lattice geometry to different channel conditions is introduced under the name of Lattice-OFDM.
It is emphasized that \ac{OFDM} with a rectangular lattice is a suboptimal solution for doubly dispersive channels. Instead of rectangular symbol placement, if the lattice is constructed with a hexagonal shape, i.e. neighboring symbols located at each corner of hexagon, better protection against \ac{ISI} and \ac{ICI} compared to rectangular lattices is achieved since the minimum distance between symbols increases. 
Therefore, the immunity of the scheme against time-frequency dispersion is increased without losing from the bandwidth efficiency. For exponential decaying and Jake's Doppler channel model, the improvement in \ac{SINR} is in the range of $1$-$2$ dB in case of forcing the Gaussian filter to be a Nyquist filter.
Additionally, considering dilation and chirp operations, the pulse adaptation based on orthogonalized Gaussian functions is combined with the hexagonal lattice structures. In~\cite{117_Han_TSP_2007}, Lattice-OFDM is re-investigated for the lattices where $\latticeDensity >1$ by emphasizing the fact that the orthogonalization procedure destroys the time-frequency concentration of the initial pulse. Instead of maximizing the number of symbols per second per hertz via the orthogonal pulses, orthogonality is abandoned with Gabor frames, and sequence detector is applied at the receiver. Also, matching equations for the pulse and lattice adaptations, similar to \eqref{eq:matching}, are provided for the hexagonal geometry.

Note that it is not an easy task to develop a single lattice structure and a prototype filter for a scheme serving multiple users with different channel characteristics. To deal with the lattice and pulse design in a scheme, while one might consider the worst case scenario to optimize pulse and lattice, other might include multiple lattices and filter structures within the frame. In \cite{sahin_2012a}, 
unlike the conventional \ac{OFDMA} transmission frame structures that consider the worst case communication channel, multiple \ac{CP} and subcarrier spacings are employed by taking the statistics of the mobility and the range of the users into account. As a result, better frequency spread immunity and higher bandwidth efficiency are obtained. Indeed, the proposed method applies the pulse adaptation and the lattice adaptation for multiple users, based on rectangular filter. As a different strategy, based on the requirements of different users, a multi-mode scheme which include different access methods such as \ac{SC-FDMA}, \ac{FBMC}, or \ac{FB-S-FBMC} is proposed  in \cite{40_ihalainen2009filter}. Another approach based on rectangular filter for a different purpose, suppressing the sidelobes in frequency, is proposed in \cite{sahin_2011edge}. By allowing less \ac{CP} duration for edge subcarriers, more room for windowing duration is provided at the edge subcarriers. This approach inherently corresponds to use different filters for the subcarriers. In a later work, this approach is combined with scheduling to decrease the interference due to the lack of \ac{CP}, which relies on the fact that \ac{CP} duration required to communicate with nearby users is smaller  \cite{Sahin_2011_edgeSchedule}.

\subsection{Equalization}
\label{subsec:equalization}
In a wireless communication medium, the transmitted signal arrives at the receiver after passing though a time-varying multipath channel. The multipath environment and the mobility disperse the transmitted signal both in time and frequency, which causes self-interference between the symbol in the lattice. As shown in \cite{sahin_2013a} and analytically expressed in \eqref{Eq:QAM_interference}, the structure of the  self-interference is related with the transmit filter, time and frequency dispersion characteristics of the communication medium, and the receive filter. An equalizer deals with the problem of removing the  self-interference to accurately  extract the desired symbols.

In order to reduce the self-interference, one of the approaches is to limit the number of symbols that interfere with the desired symbol in time and frequency by exploiting well-localized filters. This rationale offers manageable \ac{ICI} and \ac{ISI} characteristics even in doubly dispersive channels.  In \cite{Ihalainen_2007} and \cite{PHYDYAS_D31_2008}, equalization methods for this approach are introduced under three categories: 
\begin{itemize}
	\item In the first approach, the effects of \ac{ICI} and \ac{ISI} on the neighboring symbols are assumed to be negligible, relying solely on the concentration of the prototype filters  in time and frequency. In this context, a basic single-tap equalization per subcarrier may be employed to recover the symbols \cite{ 105_Vahlin_TC_1996,77_kozek1998nonorthogonal, 88_jung2007wssus, 80_matz2007analysis, 81_trigui2007optimum, 75_schafhuber2002pulse, 89_amini2010isotropic}.  This approach reduces the self-interference without using a complex equalization technique at the receiver. 
	\item The second approach utilizes equalization filters at the receiver for each subcarrier, which operates at the symbol rate \cite{Ihalainen_2007,16_cherubini2002filtered}.  This approach deals with the
aliasing between the subcarriers if lattice staggering is considered \cite{Ihalainen_2007}.
	 \item The third approach exploits fractional sampling to equalize the symbols. The equalizer works at the fractional symbol rate considering the lattice staggering approach.
Compared to the second approach, it reduces the unwanted aliasing at the expense of a higher number of samples to be processed.
 In this category, it is possible to develop different types of equalizers. For example,  \ac{MMSE} equalizers \cite{PHYDYAS_D31_2008, Hirosaki_1980,Waldhauser_2008a,42_ikhlef2009enhanced}, \ac{MLSE} equalizers \cite{Baltar_2010}, low-complexity equalizers \cite{Ihalainen_2007,PHYDYAS_D31_2008,48_ihalainen2011channel}, and channel  tracking equalizers
\cite{Waldhauser_2008,Waldhauser_2009}  all have different optimization goals.
\end{itemize}
In \cite{sahin_2013a}, instead of the equalizer itself, the relation between equalization complexity and pulse shape is investigated. Composite effect of the transmit filter, the response of the communication medium, and the receive filter is discussed including the \ac{SNR} observed at the receiver. Using \ac{AIC},  number of effective interfering symbols (model order) is obtained for different prototype filters.
It is shown that while rectangular and sinc functions are immune to the dispersions in only one domain (either time or frequency) Gaussian function yields a better balance in doubly dispersive channels. Note that if a receiver has effective equalization capabilities, Nyquist criterion requirements in the prototype filter design can be relaxed. 


Bi-orthogonal schemes are closely related with the concept of equalization. For example, transmitting the symbols over a non-orthogonal basis and forcing the correlations between the symbols to be zero at the receiver is similar to zero-forcing equalization. Furthermore, one may harness bi-orthogonal schemes to obtain single-tap equalization. For example, \ac{CP}-\ac{OFDM} benefits from the extension of the rectangular filter  to yield a single-tap equalization in time-invariant multipath channels. 
A similar approach can also be obtained via \ac{FMT}. Consider a Gabor system at the transmitter, which utilizes sinc function that has a bandwidth greater than $\frequencySpacing$, e.g., $\frequencySpacing+F_{\rm cp}$. Also, assume that the subcarrier spacing is extended not to allow overlapping between the subcarriers, i.e., $\frequencySpacingVariable=\frequencySpacing+F_{\rm cp}$ and $\timeSpacingVariable=\timeSpacing$. If the Gabor system at the receiver is equipped with another sinc function that has a bandwidth $\frequencySpacing$, which is smaller than $\frequencySpacingVariable$,  a cyclic behavior within band of $F_{\rm cp}$ is obtained, similar to the one in time domain for \ac{CP}-\ac{OFDM}. Therefore, a single-tap equalization would be sufficient for this scheme to combat with time-selective channels (which only introduces frequency dispersion). This issue is also shown in \figurename~\ref{fig:amb}\subref{fig:esinc} using the ambiguity surface for the extended null regions.

For the schemes where $\latticeDensity>1$, as in faster-than-Nyquist, partial-response signaling, or Weyl-Heisenberg frames, equalizers heavily deal with the intentional overlapping between the symbols rather than channel itself. Hence, the equalizers tend to be complex in such scenarios. For these schemes, the equalization process is combined with approaches that utilizes the discreteness of symbol, e.g., \ac{SIC} \cite{Fettweis_2009,82_datta2011fbmc,Rusek_2006}, \ac{MLSE}  \cite{Rusek_TWC_2009}, or subspace classification \cite{117_Han_TSP_2007}.

\subsection{Time-Frequency Synchronization}

\begin{figure}[!t]
\centering
\subfloat[Carrier Frequency Offset.]{\includegraphics[width=3.5in]{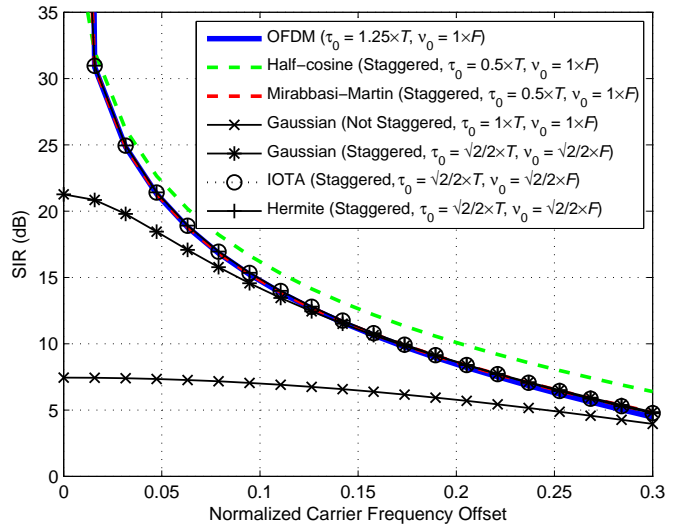}
\label{fig:CFO}}\\
\subfloat[Timing offset.]{\includegraphics[width=3.5in]{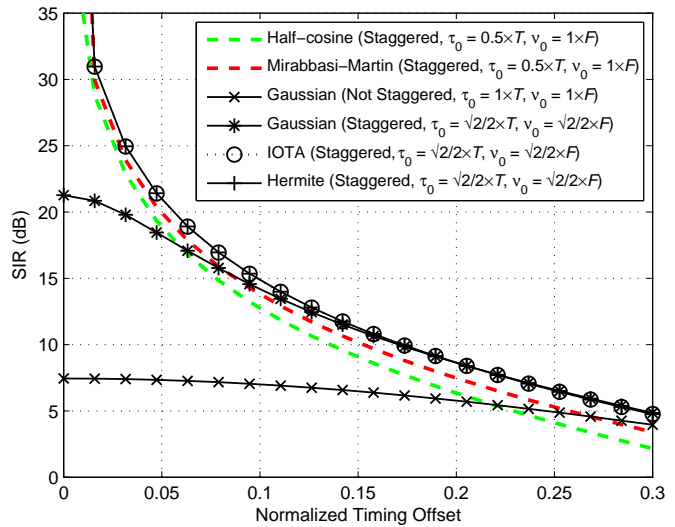}
\label{fig:timeoffset}}
\caption{SIR performances of different prototype filters under synchronization errors.}
\label{fig:SIRsynch}
\end{figure}

The robustness of the multicarrier schemes against time/frequency synchronization errors, i.e., \ac{CFO} and \ac{TO}, mainly depends on the null regions in the ambiguity surfaces. For example, \ac{CP} utilization of the conventional \ac{OFDM} scheme provides a wider null-region in time which increases the robustness of the scheme against timing errors. However, it does not provide any precaution against \ac{CFO}. In \figurename~\ref{fig:SIRsynch}, \ac{SIR} performances of various filters are given for different \ac{TO} and \ac{CFO} values, using the expression in \eqref{Eq:SIR}. Three main conclusions can be drawn from \figurename~\ref{fig:SIRsynch}:
\begin{itemize}
\item 
Filters are able to provide different immunity against \ac{CFO} and \ac{TO}. In \cite{36_schaich2010filterbank} and \cite{35_Ringset_FNMS_2010} through simulations, and in \cite{SaeediSourck20111604,37_fusco2008sensitivity} through theoretical analysis, similar results are obtained. Proper prototype filter utilization with lattice staggering may increase the robustness against synchronization errors. 
\item
The immunity against \ac{TO} is identical to the one against \ac{CFO} for Hermite, IOTA, and Gaussian functions. It is due to the fact that these filters provide identical responses in time and frequency when $\gaussianRollOff = 1$.
\item
Lattice staggering is helpful to improve the performance of prototype filters that do not satisfy Nyquist criterion in the presence of  \ac{CFO} and \ac{TO}. Although the Gaussian function does not fulfill the Nyquist criterion, inherent orthogonality due to the lattice staggering provides a significant performance improvement.
\end{itemize}

It is worth noting that  although some of the filters provide similar response in terms of \ac{SIR}, as given in \figurename~\ref{fig:SIRsynch}, the positions of interfering symbols in the lattice might be different, depending on the filter type. Exploiting this issue along with band-limited filters are particularly effective to combat against the degradation caused by the misalignments among the transmissions of different users for the uplink \cite{37_fusco2008sensitivity,33_saeedi2011complexity,Du2008}. For the discussions on synchronization methods, we refer the reader to
\cite{60_stitz2009cfo,61_stitz2010pilot,68_fusco2010joint} for preamble-based methods, to \cite{63_stitz2008practical,58_fusco2009data} for  training-based methods, and to \cite{Fusco_SPAWC_2007,Fusco_TSP_2007} for blind approaches. 

\subsection{Spatial Domain Approaches}
Lattice staggering introduces challenges on the implementation of multiple antenna systems. This arises due to the fact that lattice staggering exploits the orthogonality in the real domain, and imaginary part may appear as an interference  for spatial domain approaches (especially, for the spatial diversity). In other words,  multiple antenna techniques, which are easy to implement for the schemes that achieve orthogonality in complex domain, e.g. \ac{FMT} and \ac{CP-OFDM}, are not trivial for the multicarrier schemes that exploit lattice staggering. In \cite{44_bellanger2010fbmc},  an intuitive discussion on lattice staggering is provided considering multiple antennas as follow:
\begin{itemize}
    \item {\em Spatial Multiplexing:} In the case of spatial multiplexing, multiple data streams reach the receiver antennas over different channels. For instance, considering 2 transmit and 2 receive antennas and assuming single tap channels between the antennas, received symbols are expressed as
	\begin{align}
	\begin{bmatrix} 
	  y_{1}\\ 
	  y_{2}
	\end{bmatrix}
=
	\begin{bmatrix} 
	  h_{11}    & h_{21}\\ 
	  h_{12} &  h_{22}
	\end{bmatrix}
	\begin{bmatrix} 
	  d_1+ju_1\\ 
	  d_2+ju_2
	\end{bmatrix}
	\nonumber
	\end{align}
where, $d_1$ and $d_2$ are the desired symbols, $u_1$ and $u_2$ are the interfering parts due to the lattice staggering, $h_{11}$, $h_{12}$, $h_{21}$, and $h_{22}$ are the channel coefficients, and $y_{1}$ and $y_{2}$ are the received symbols. Since lattice staggering offers orthogonality in real domain, one can obtain the symbols as
	\begin{align}
	\begin{bmatrix} 
	  d_{1}\\ 
	  d_{2}
	\end{bmatrix}
	=\Re\left\{
	\begin{bmatrix} 
	  h_{11}    & h_{21}\\ 
	  h_{12} &  h_{22}
	\end{bmatrix}^{-1}
	\begin{bmatrix} 
	  y_{1}\\ 
	  y_{2}
	\end{bmatrix}
	\right\}\nonumber
	\end{align}
which corresponds to zero-forcing. In order not to enhance the noise, one can apply \ac{MMSE} and \ac{MLSE}. It is worth noting that lattice staggering provides an advantage for \ac{MLSE}, since the symbols are in the real domain. Also, the assumption of single tap requires that the equalizers remove \ac{ISI} and \ac{ICI} due to the channel dispersion before the processing for the spatial multiplexing. 
    \item {\em Spatial Diversity:} While the receiver diversity does not introduce any complication for lattice staggering, the transmit diversity results in samples where the real and imaginary parts are mixed up due to the complex channel coefficients. For example, considering delay diversity with 2 transmit antennas and 1 receive antenna, received symbol at the $m$th instant is given by
	\begin{align}
	y_{m} = h_{11} (d_{m}+ju_{m}) + h_{21} (d_{m-1}+ju_{m-1})~,
	\nonumber
	\end{align}
where $d_{m}$ and $d_{m-1}$ are the desired symbols and $u_{m}$ and $u_{m-1}$ are the interfering parts. Since $h_{11} $ and $h_{21} $ are complex coefficients, additional processing has to be applied at the receiver to remove $u_{m}$ and $u_{m-1}$.
\end{itemize}

In \cite{Lele_JASP_2010} and \cite{lele_2007cdma}, spatial diversity is achieved by exploiting the spreading approaches along with lattice staggering. Alamouti \ac{STBC} is combined with a scheme which allows complex symbol utilization via a \ac{CDMA} based spreading operation \cite{lele_2007cdma}. Similarly, in order to remove the imaginary parts at the receiver, a scheme, referred as FFT-FBMC, is proposed in \cite{Zakaria_TWC_2012} and applied to the multiple antenna systems. While   imaginary parts are canceled for the case of single-delay \ac{STTC} with two transmit and one receive antennas in \cite{Bellanger_ISWPC_2008},  later, this study is extended to multiple transmit antennas in \cite{45_chrislin2010decoding} by proposing an iterative decoding approach. All in all, there still needs to be more studies on the application of multiple antennas to lattice staggering approaches under practical implementation scenarios.

\subsection{Channel Estimation}
Channel estimation methods for the multicarrier schemes which allow orthogonality in complex domain, e.g., \ac{CP-OFDM}, are extensively available in the literature \cite{Ozdemir_2007}. However, there are limited number of studies for the schemes with lattice staggering, and further work is needed to develop effective algorithms. Conventional approach for channel estimation is to transmit known symbols in the form of pilots or preambles:

\begin{itemize}
    \item {\em Pilot-based Channel Estimation:}
	Pilots are the symbols that do not carry any data, and are scattered into the known positions in the transmission frame with a certain pattern. Considering the schemes with lattice staggering, a pilot-based channel estimation method is proposed in \cite{PHYDYAS_D31_2008,Javaudin_VTC_2003}. It exploits the so-called  auxiliary symbols, which are calculated based on the data symbols and the prototype filter in order to cancel the imaginary interference that affects the pilots. By canceling the imaginary interference at the transmitter, lattice staggering is able to accommodate complex channel estimation. However, in \cite{PHYDYAS_D31_2008}, it is noted that using auxiliary symbols may increase the \ac{PAPR} of the scheme. 
	\item {\em Preamble-based Channel Estimation:} Preambles generally reside at the beginning of frame and may also be utilized for other purposes (e.g., synchronization). Considering the schemes with lattice staggering, two channel estimation methods based on  preambles are proposed in~\cite{62_lele2008channel}. First approach is referred as {\em pairs of pilots} which relies on the real symbols placed at the two consecutive time positions. It obtains the channel coefficients via simple matrix operations, but suffers from noise enhancement.
Second method is known as \ac{IAM} and targets to obtain a complex pilot symbol at the receiver by approximating the intrinsic imaginary
interference from real valued neighboring pilots. It does not require a~priori knowledge of the prototype function. In \cite{116_lele_ICC_2008}, the real-valued symbols are replaced with imaginary ones to achieve additional improvements. In a later study, a general theoretical framework for \ac{IAM} preamble design is given in \cite{du_2009_preamble}. Also, a comparative study on the preamble-based \ac{LS} channel estimation is given for the schemes with/without lattice staggering,  considering the sparse preambles in \cite{Katselis_2010}.

\end{itemize}

\subsection{Hardware Impairments}

\begin{figure}[!t]
\centering
\subfloat[Before power amplifier (Ideal responses).]{\includegraphics[width=3.3in]{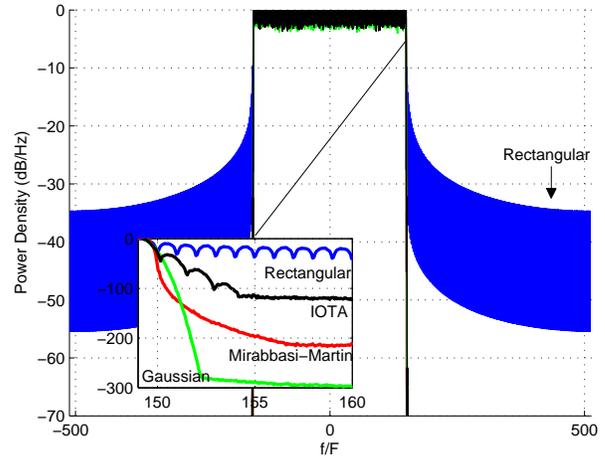}
\label{fig:before}}
\\\vspace{-3mm}
\subfloat[After power amplifier.]{\includegraphics[width=3.3in]{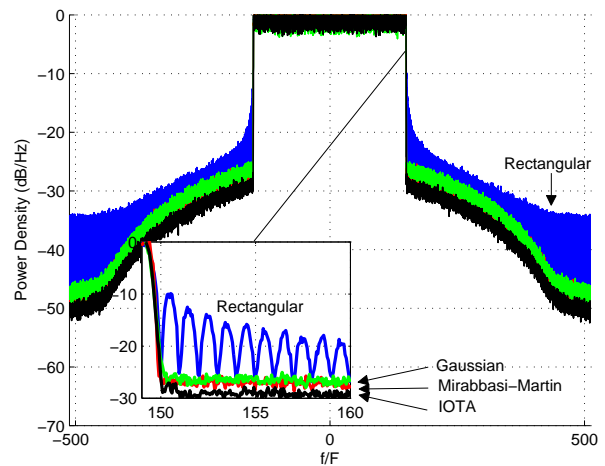}
\label{fig:after10}}\\\vspace{0mm}
\caption{Comparison of the power spectral densities of the schemes with different prototype functions.}
\label{fig:PSD}
\end{figure}

Desired properties of a waveform might not be maintained due to the non-linear characteristics of \ac{RF} front-ends. Such non-linear characteristics of the \ac{RF} front-ends might distort the signal and cause inter-modulation products.
One of the well-known metrics to anticipate the signal robustness against the distortions, especially due the power amplifier, is the \ac{PAPR}, which captures the ratio of the peak power to the average power. Generally, it is expressed via \acp{CCDF}. In \cite{9_waldhauser2006comparison,40_ihalainen2009filter, 47_Kollar_VTC,70_Viholainen_PHYDYAS},
\ac{PAPR} results are provided for the multicarrier schemes with/without lattice staggering, considering different prototype filters. Without any specific precautions  (e.g., DFT spreading), \ac{PAPR} of different multicarrier schemes are shown to be similar in~\cite{9_waldhauser2006comparison,47_Kollar_VTC}. Similarly,~\cite{40_ihalainen2009filter} shows that \ac{FBMC} and \ac{OFDM} have practically identical \ac{PAPR} curves. On the other hand, use of spreading approaches, e.g., \ac{DFT}-spread  \cite{40_ihalainen2009filter, 113_Gharba_VTC_2012} and filter-bank-spreading \cite{40_ihalainen2009filter,70_Viholainen_PHYDYAS}, lowers the \ac{PAPR} of the transmitted signals, yielding \acp{PAPR} that are close to those of \ac{SC-FDMA}.

Impact of power amplifiers on the \acp{PSD} of the multicarrier schemes with/without staggering are compared in~\cite{47_Kollar_VTC}. While the use of band-limited filters offers less sidelobes when the power amplifier operates in the linear region, the benefit of their utilizations diminishes when the power amplifier operates in the non-linear region. This is also shown in \figurename~\ref{fig:PSD}, considering a polynomial model for the distortion given by \cite{Lei_2004}
\begin{align}
\transmittedSignalD[\timeSymbol] = a_1\transmittedSignal[\timeSymbol]+a_3\transmittedSignal[\timeSymbol]\absOperator[{\transmittedSignal[\timeSymbol]}]^2 +a_5\transmittedSignal[\timeSymbol]\absOperator[{\transmittedSignal[\timeSymbol]}]^4~,
\nonumber
\end{align}
where $a_1 = 1.0108 + 0.0858i$, $a_3 = 0.0879 - 0.1583i$, $a_5 = -1.0992 - 0.8891i$. While the filter characteristics dominate the \acp{PSD} before power amplifier, the distortion due to the power amplifier heavily affects the \acp{PSD}, as can be seen in \figurename~\ref{fig:PSD}\subref{fig:after10}. Yet, filters determine the  sharpness of the decaying at the edge of the band.

Waveform design also might address the phase noise which is specifically an issue at high operating frequencies, e.g., 60 GHz. While \ac{CP-OFDM} copes well with the high frequency selectivity, it is known that it is susceptible to frequency dispersion, e.g., phase noise. 
In \cite{Moret_2008}, it is shown that \ac{FMT} equipped with an \ac{RRC} prototype filter provides robustness against the phase noise, compared to \ac{OFDM}. 
In addition, the approach introduced in Section \ref{subsec:equalization} via extending the sinc function in frequency, captured in \figurename~\ref{fig:amb}\subref{fig:esinc}, might be promising to handle phase noise.

\subsection{Cognitive Radio and Resource Sharing}

Due to its favorable properties, multicarrier schemes, especially \ac{CP-OFDM}, have been commonly considered as promising approaches for dynamic spectrum access and cognitive radio systems~\cite{Bansal_TWC_2008,Haykin_JSAC_2005,Akyildiz_CN_2006}.
However, \ac{CP-OFDM} has its own drawbacks when used in cognitive radio applications. For example, sidelobes of the \ac{CP-OFDM} subcarriers may cause large  adjacent channel interference. This prevents efficient utilization of unused portion of the spectrum by secondary users and limits the aggregate bandwidth efficiency of a cognitive radio system. Moreover,  
\ac{CP-OFDM} is sensitive to the asynchronous nature of the secondary users, which puts stringent constraints for dynamic spectrum access in cognitive radio networks. Hence, low spectral leakage property of the schemes that utilizes band-limited filters, e.g. \ac{SMT} and \ac{FMT}, makes them an attractive candidate for cognitive radio systems~\cite{57_farhang2008multicarrier}. While a subcarrier in \ac{SMT} overlaps with its immediate adjacent subcarriers only, \ac{FMT} utilizes non-overlapping subcarriers. Due to the minimization of spectral leakage to neighboring subcarriers via band-limited filters, secondary users in a cognitive radio network may efficiently utilize the spectrum opportunities. Also, this approach does not require tight synchronization between primary and secondary networks, which relaxes implementation complexity.

\begin{figure*}[!t]
\centering
\subfloat[Synthesizing single symbol without IDFT.]{\includegraphics[width=3.8in]{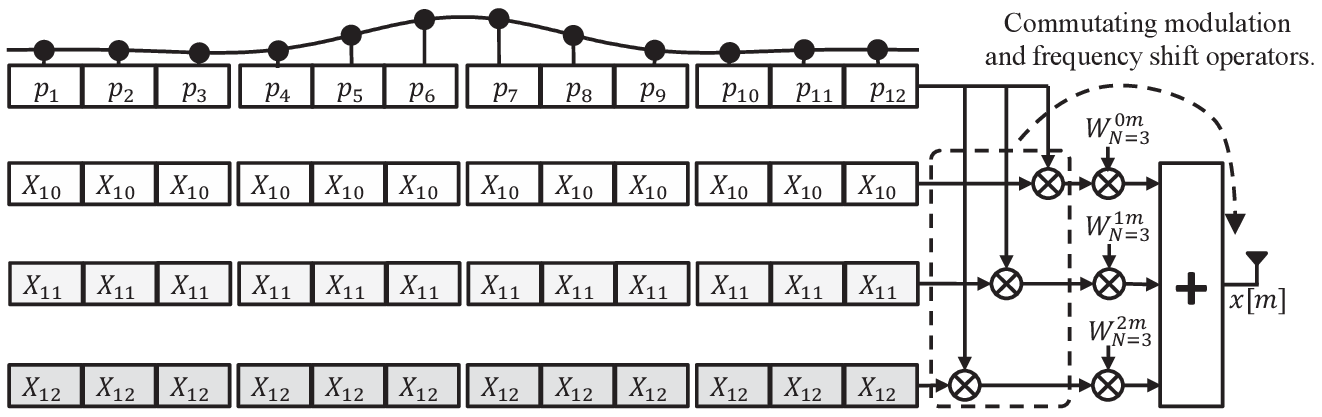}
\label{fig:s1}}
\subfloat[Synthesizing single symbol with IDFT, which generates the same output as in Fig. 13(a).
]{\includegraphics[width=3.2in]{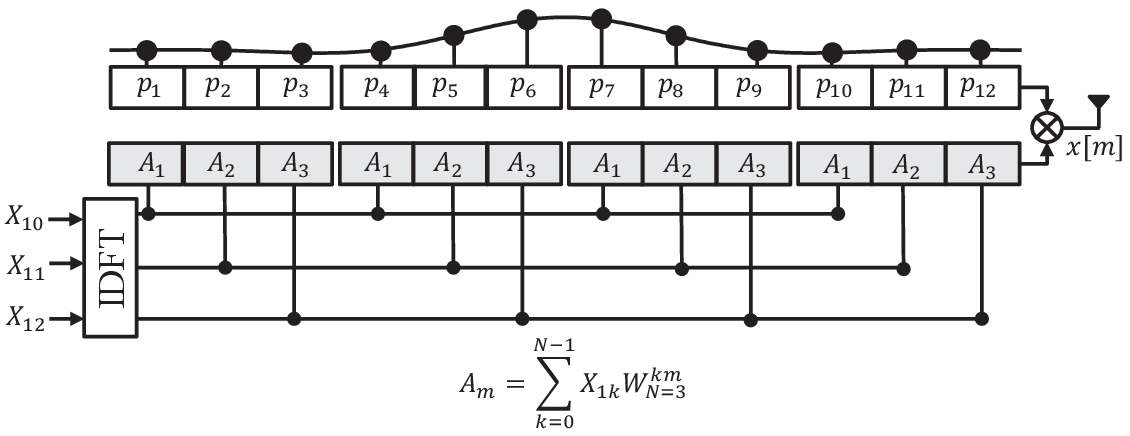}
\label{fig:s2}}\\
\subfloat[Combining multiple symbols.]{\includegraphics[width=7in]{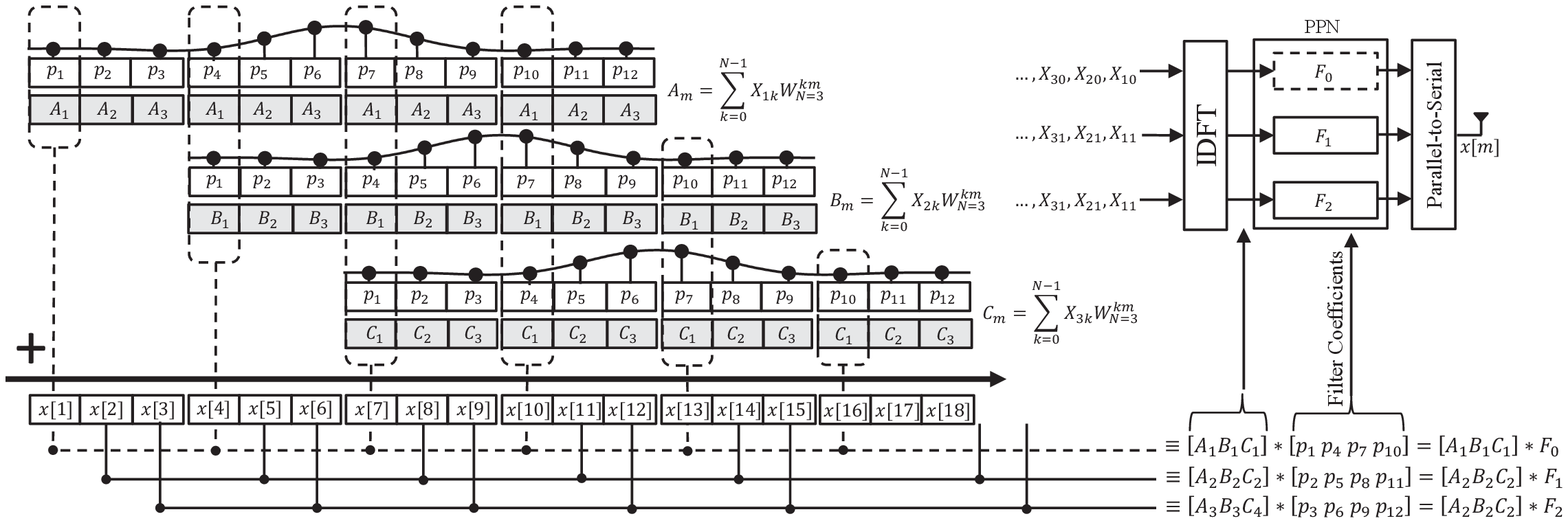}
\label{fig:s3}}
\caption{Synthesizing a multicarrier symbols with IDFT operation. Note that
$\symbols[\timeIndexTX][\frequencyIndexTX]$ is the symbol  on $\timeIndexTX$th time slot and $\frequencyIndexTX$th subcarrier, $[p_1, p_2,..., p_{\truncation\numberOfSubcarrier}]$ are the filter coefficients, and $F_i$ are the polyphase components of the prototype filter.
}
\label{Fig:ppn}
\end{figure*}

Downlink resource allocation in \ac{FBMC}-based cognitive radios is investigated in~\cite{32_shaat2010computationally}. Maximizing total capacity of the cognitive radio network is targeted by exploiting the low spectral leakage of band-limited filters, under certain constraints such as the total available power and low adjacent channel interference to primary users. 
Results show that an FBMC-based cognitive radio network yields higher aggregate bandwidth efficiency with significant gains under certain scenarios and yields lower interference to the primary users.

Uplink resource allocation in \ac{FBMC} is discussed in~\cite{46_payaroresource}. It is shown that \ac{FBMC} yields aggregate spectral efficiencies that are around $45\%$ higher than those delivered by \ac{CP-OFDM}. A similar conclusion regarding uplink spectral efficiency has been reached in~\cite{66_zhang2010spectral}, while~\cite{33_saeedi2011complexity} shows through simulations that \ac{FBMC} provides superior \acp{BER} when compared to \ac{CP-OFDM} in a multiuser uplink channel, at a much lower computational complexity. In \cite{40_ihalainen2009filter}, a multi-mode uplink access method is proposed. In this approach, depending on the specific requirements of different users, different access methods such as \ac{SC-FDMA}, \ac{FBMC}, or \ac{FB-S-FBMC} can be utilized by different users. Such a hybrid approach provides full flexibility in cognitive radio applications. 

From the view of spectrum sensing approaches, \ac{FFT} part of the \ac{CP-OFDM} demodulator can also be conveniently utilized to identify the presence of primary users in the vicinity~\cite{57_farhang2008multicarrier,Yucek_CST_2009}. On the other hand, instead of a rectangular filter, using a different prototype filter via \ac{FBMC} provides better sensing performance. For example,
\cite{57_farhang2008multicarrier} shows that the use of filter banks yields similar spectrum sensing accuracies when compared to those of optimum multi-taper based spectrum sensing method with less computational complexity.

\subsection{Poly-Phase Network}
\begin{figure*}[!t]
\centering
\subfloat[Analyzing $\frequencyIndexTX$th subcarrier along with PPN implementation.]{\includegraphics[width=7in]{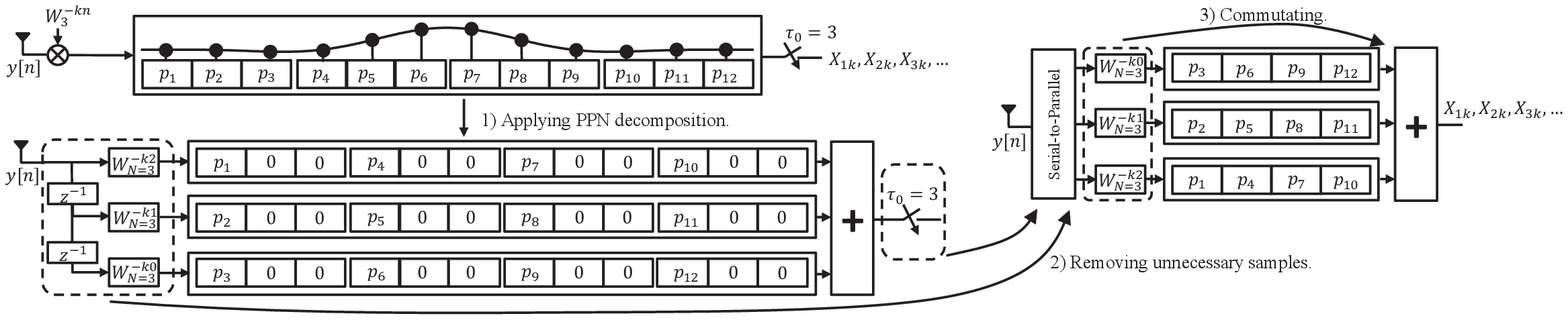}
\label{fig:a1}}\\
\subfloat[Considering all subcarriers with DFT operation.]{\includegraphics[width=7in]{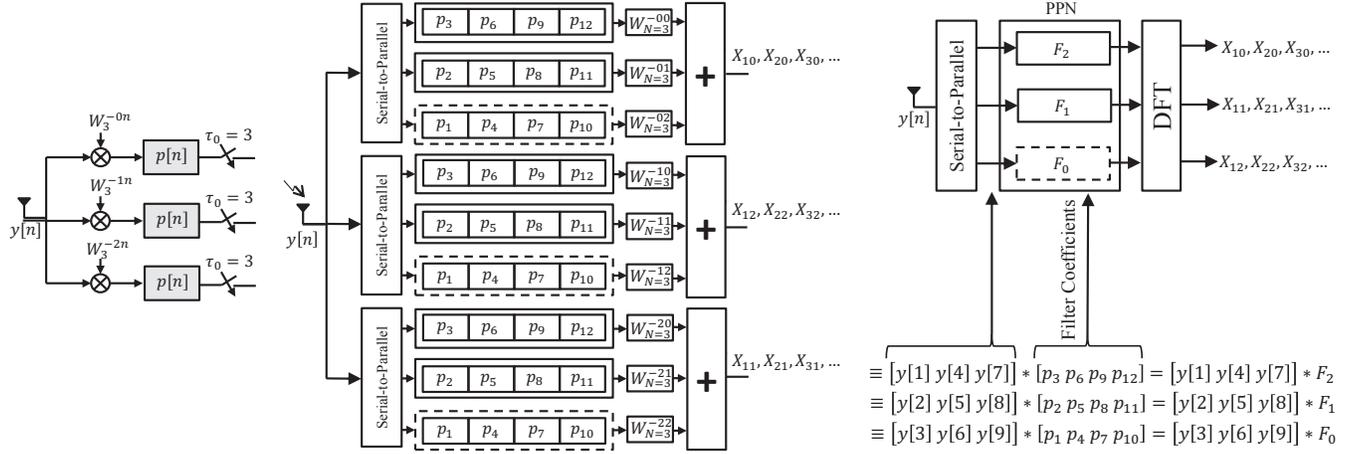}
\label{fig:a2}}
\caption{Analyzing multicarrier symbols with DFT operation.}
\label{Fig:ppna}
\end{figure*}

Employing generic prototype filters on multicarrier schemes may introduce high-complexity due to the extra filtering operations at the transmitter and the receiver. This complexity can be reduced by exploiting polyphase representations of filters and \ac{FFT} operations. 
An illustrative example for analyzing and synthesizing multicarrier symbols with 3 subcarriers are shown in \figurename~\ref{Fig:ppn} and \figurename~\ref{Fig:ppna}, respectively.
Filter length is considered as $4\times\numberOfSubcarrier$ where $\numberOfSubcarrier=3$. \ac{PPN} at the transmitter is given considering the following cases:
\begin{itemize}
    \item Trivial construction of a single multicarrier symbol is given in  \figurename~\ref{Fig:ppn}\subref{fig:s1}. After the prototype filter is multiplied with a modulation symbol in each branch, the multicarrier symbol is generated by combining the branches modulated via twiddle factors where $\exponentialTerm[\numberOfSubcarrier][\timeIndexTX][\frequencyIndexTX]=e^{j2\pi\timeIndexTX\frequencyIndexTX/\numberOfSubcarrier}$. 
	\item In \figurename~\ref{Fig:ppn}\subref{fig:s2}, an equivalent diagram of \figurename~\ref{Fig:ppn}\subref{fig:s1} is realized with \ac{IDFT} by applying the multipliers indicated in \figurename~\ref{Fig:ppn}\subref{fig:s1} after the summation operation. To make these substitutions, the repetition of exponential term $\exponentialTerm[\numberOfSubcarrier][\timeIndex][\frequencyIndex]$ for every other three samples is exploited. 
	\item In \figurename~\ref{Fig:ppn}\subref{fig:s3},
three multicarrier symbols are taken into consideration. Each symbol is shifted by symbol spacing, which is 3, and combined to generate the transmission frame. Indeed, this operation can be performed by convolution operations, as shown in \figurename~\ref{Fig:ppn}\subref{fig:s3}. Essentially, the output of \ac{IDFT} is filtered with some filter coefficients.
\end{itemize}

The basic idea behind the receiver consists of three basic stage:
1) shifting the desired subcarrier to baseband, 2) filtering the signal with receiving filter to  eliminate the impacts of other subcarriers, and 3)
sampling at correct instants. Considering these steps, analysis of a single subcarrier is given in \figurename~\ref{Fig:ppna}\subref{fig:a1}.
\ac{PPN} at the receiver is given considering the following steps:
\begin{itemize}
    \item First, polyphase decomposition is applied to filter using shift operation ($z^{-1}$). As the modulation symbols are constructed in every other 3 samples and the exponential term repeats itself around the unit circle, the exponential terms can be distributed to each phase as factors. Since the intermediate samples are not necessary, the implementation is simplified using a commutator and removing the zero coefficients at each branch.
	\item Then, by moving the positions of the exponential coefficients before summation operations and analyzing each subcarrier, \ac{DFT}  is obtained as in \figurename~\ref{Fig:ppna}\subref{fig:a2}. Similar to the transmitter, additional filtering operation is necessary before \ac{DFT} operations.
\end{itemize}

As indicated in this illustrative example, prototype filters may bring some additional filtering operations after/before \ac{IDFT} and \ac{DFT} operations. Note that these filters are simply equal to 1 for \ac{OFDM} without \ac{CP}. In addition, for a complete scheme with lattice staggering, the introduced structure has to be replicated individually and combined with proper time shift to construct the transmission frames.  Extended summaries on \ac{PPN} are also provided in  \cite{79_vaidyanathan1990multirate, 25_vaidyanathanmultirate, 28_farhang2010signal,80_matz2007analysis, A4_Bellanger_TC_1974, 6_rhee1998performance, 8_lee2004polyphase,78_bellanger1976digital, 75_schafhuber2002pulse}. 

\subsection{Complexity Analysis}

One of the important criteria for the adoption of FBMC in the future wireless standards is whether it can yield sufficiently better gains than existing techniques such as OFDM, at the cost of a reasonable complexity  increase. The complexity of FBMC has been investigated in the literature from different perspectives in~
\cite{Mehmood_2012,5_bellanger2008filter, 11_baltar2007out, 22_benvenuto2002equalization,29_hirosaki1981orthogonally,33_saeedi2011complexity,36_schaich2010filterbank, 48_ihalainen2011channel,50_zhang2008oversampled,112_Jiang_GLOBECOM_2012}. 

As discussed in~\cite{36_schaich2010filterbank}, a major factor for complexity increase in FBMC is due to the replacement of IFFT/FFT in OFDM with the filter banks. When the Split-Radix algorithm is used, number of real multiplications required to implement FFT/IFFT operations over $N$ subcarriers is given by~\cite{36_schaich2010filterbank,Murphy_TC_2002}
\begin{align}
C_{\rm FFT/IFFT} = N(\log_2(N)-3)+4~.\label{Eq:OFDM_Complx} 
\end{align}
On the other hand, for FBMC systems, complexity of synthesis and analysis filterbanks need to be separately investigated. The IFFT at the synthesis filterbank processes only real or imaginary samples (which are never complex), while the FFT in the analysis filterbank processes complex samples. Considering this difference, the number of real multiplications per complex symbol are approximately given by~\cite{36_schaich2010filterbank}
\begin{align}
C_{\rm SFB} &= \log_2(N/2)-3+4K~,\label{Eq:SFB_Complx}\\
C_{\rm AFB} &= 2\big(\log_2(N)-3\big)+4K~,\label{Eq:AFB_Complx} 
\end{align}
where SFB and AFB refer to the synthesis filterbank and the analysis filterbank, respectively, and $K$ denotes the filter length. For example, for $N=512$ and $K=3$, using \eqref{Eq:OFDM_Complx}-\eqref{Eq:AFB_Complx}, number of real multiplications {\em per complex symbol} for different approaches are given by $C_{\rm FFT/IFFT}=6$, $C_{\rm SFB}=17$, and $C_{\rm AFB}=24$, which implies a slightly higher complexity of FBMC than OFDM~\cite{36_schaich2010filterbank}.

Implementation of complex equalization and MIMO techniques in FBMC may also increase its complexity further, see e.g.~\cite{22_benvenuto2002equalization}, \cite{48_ihalainen2011channel}. For example, per subchannel equalization complexity of OFDM is compared with FMT using different equalization approaches in~\cite{22_benvenuto2002equalization}. While OFDM uses a single complex multiplication per subchannel due to one-tap equalization, number of complex multiplications per subchannel varies between $46$ and $108$ for different FMT equalization techniques with $N=128$. 

On the other hand, an additional complexity has to be introduced into OFDM transmitter in order to suppress out-of-band power leakage, which is handled naturally by FBMC. Number of complex multiplications required to handle inter-carrier interference with OFDM and FBMC are provided in~\cite{33_saeedi2011complexity} for an uplink scenario. The provided analysis shows at least an order of magnitude complexity reduction through using FBMC in typical scenarios, when compared with the implementation of parallel interference cancellation in OFDM.

\subsection{Testbeds and Extensions to Standards}

While there are many publications theoretically comparing \ac{FBMC} with \ac{CP-OFDM} and other waveforms, there are  limited number of works that discuss testbed implementation of schemes with lattice staggering. In~\cite{35_Ringset_FNMS_2010,Baranda_2011}, an \ac{FBMC} testbed, which is developed as a part of the \ac{PHYDYAS} project~\cite{27_farhang2010cosine},~\cite{PHYDYAS_D9_2009}, is described. The testbed is capable of real-time transmission and reception of \ac{FBMC} signals. Transmitter side involves a \ac{FPGA} that includes the physical layer only, and can operate both in \ac{OFDM} and in \ac{FBMC} modes. The receiver side is composed of an \ac{RF} front-end, a USRP motherboard (involving \acp{ADC} and \ac{FPGA}), and a software subsystem. 
In order to better assess the various implementation trade-offs between schemes with/without lattice staggering, further studies are needed on the testbed experimentation.

\section{Concluding Remarks}
\label{sec:conc}
In this survey, multicarrier schemes are examined based on a generalized framework which relies on Gabor systems. The framework categorizes the multicarrier schemes based on their lattices, filters, and symbols. It explains the conventional \ac{CP-OFDM} systems, while at the same time providing insights into multicarrier schemes different than \ac{CP-OFDM}. Hence, the survey provides a useful framework  to develop and analyze new types of waveforms, which may pave the way for further enhancements for the next generation radio access techniques.

The combined effect of filters, lattices, and symbols introduce different characteristics into the multicarrier schemes. Yet, for a given scenario, it is possible to addresses the system requirements via proper considerations within the provided framework. For example, \ac{CP-OFDM} addresses low-complexity receiver implementation and multiple antenna functionalities, effectively. However, utilization of \ac{CP-OFDM} may be an issue considering co-channel interference scenarios and operation under the frequency dispersive channels. One may provide solutions for these challenging issues via proper selections of lattices, filters, and  symbols.

\section*{Acknowledgment}
We would like thank the reviewers for their valuable comments and suggestions which significantly helped improving the survey. We also would like to thank Anas Tom and Ertugrul G\"{u}venkaya of University of South Florida, USA, for their helpful inputs. Part of this work has been done while Ismail Guvenc was at DOCOMO Innovations, Inc. This study has been supported by DOCOMO Innovations, Inc.

\def\classificationTableColumnSize{0.47in}
\def\classificationParboxSize{0.52in}
\def\classificationParboxTitleSize{0.56in}
\begin{table*}[!htbp]
\caption{Classification of papers related to prototype filters.} \label{Table:FBMC_Taxonomy}
\begin{center}
\scriptsize
\begin{tabular}{p{\classificationParboxTitleSize}||p{\classificationTableColumnSize}|p{\classificationTableColumnSize}|p{\classificationTableColumnSize}|p{\classificationTableColumnSize}|p{\classificationTableColumnSize}|p{\classificationTableColumnSize}|p{\classificationTableColumnSize}|p{\classificationTableColumnSize}|p{\classificationTableColumnSize}|p{\classificationTableColumnSize}}\parbox{\classificationParboxSize}{ \textbf{Topic}} &\parbox{\classificationParboxSize}{\centering \textbf{Rectangular}} &\parbox{\classificationParboxSize}{\centering \textbf{No Specific Filter}} &\parbox{\classificationParboxSize}{\centering \textbf{RRC}} &\parbox{\classificationParboxSize}{\centering \textbf{Mirabbasi-Martin}} &\parbox{\classificationParboxSize}{\centering \textbf{IOTA}} &\parbox{\classificationParboxSize}{\centering \textbf{Gaussian}} &\parbox{\classificationParboxSize}{\centering \textbf{Channel-based Pulses}} &\parbox{\classificationParboxSize}{\centering \textbf{PSWF}} &\parbox{\classificationParboxSize}{\centering \textbf{EGF}} &\parbox{\classificationParboxSize}{\centering \textbf{Hermite-Gaussian}} \\ \hline\hline \parbox{\classificationParboxTitleSize}{Filter Basics} &\parbox{\classificationParboxSize}{\cite{Hwang_2009,2_le1995coded,26_Du_KTH_2007,Du2008,28_farhang2010signal,24_farhang2011ofdm,Harris_1978,Geckinli_1978,54_Du_ICICS_07,54_Du_ICICS_07}} &\parbox{\classificationParboxSize}{} &\parbox{\classificationParboxSize}{\cite{2_le1995coded,26_Du_KTH_2007,Du2008,27_farhang2010cosine,28_farhang2010signal,24_farhang2011ofdm}} &\parbox{\classificationParboxSize}{\cite{106_Martin_TCS_1998,3_bellanger2001specification,Mirabbasi_2002}} &\parbox{\classificationParboxSize}{\cite{2_le1995coded,26_Du_KTH_2007,Du2008,27_farhang2010cosine,28_farhang2010signal,24_farhang2011ofdm,107_Siohan_TSP_2000,54_Du_ICICS_07}} &\parbox{\classificationParboxSize}{\cite{26_Du_KTH_2007,97_Sondergaard_Thesis_2007,Du2008,28_farhang2010signal,24_farhang2011ofdm,Harris_1978}} &\parbox{\classificationParboxSize}{} &\parbox{\classificationParboxSize}{\cite{28_farhang2010signal,24_farhang2011ofdm,Landau_1961,Slepian_1961,Landau_1962,Slepian_1964,Slepian_1978,Slepian_1983,Kaiser_1980,Halpern_1979,105_Vahlin_TC_1996,Walter2005432,Moore2004208,Harris_1978,Geckinli_1978,Nigam_2010}} &\parbox{\classificationParboxSize}{\cite{Roche_1997,26_Du_KTH_2007,Du2008,107_Siohan_TSP_2000,54_Du_ICICS_07}} &\parbox{\classificationParboxSize}{\cite{97_Sondergaard_Thesis_2007,27_farhang2010cosine,28_farhang2010signal,24_farhang2011ofdm,84_haas1997time}} \\ \hline\parbox{\classificationParboxTitleSize}{Lattice \\ Staggering} &\parbox{\classificationParboxSize}{\cite{54_Du_ICICS_07}} &\parbox{\classificationParboxSize}{\cite{feichtinger1998gabor,Daubechies_1991,Kutyniok_2005,88_jung2007wssus,26_Du_KTH_2007,Du2008,56_bolcskei1999design,gabor_book_chap,27_farhang2010cosine,28_farhang2010signal,24_farhang2011ofdm,Siclet_2000}} &\parbox{\classificationParboxSize}{\cite{2_le1995coded,A1_Chang_BST_1966,14_saltzberg1967performance,14_saltzberg1967performance,4_Siohan_TSP_2002,Ihalainen_2007,37_fusco2008sensitivity,Fusco_TSP_2007}} &\parbox{\classificationParboxSize}{\cite{106_Martin_TCS_1998,3_bellanger2001specification,Mirabbasi_2003,36_schaich2010filterbank,44_bellanger2010fbmc,61_stitz2010pilot,Zakaria_TWC_2012,Bellanger_ISWPC_2008,32_shaat2010computationally,46_payaroresource,66_zhang2010spectral}} &\parbox{\classificationParboxSize}{\cite{2_le1995coded,54_Du_ICICS_07,62_lele2008channel,59_lele2007preamble,Javaudin_VTC_2003,Tabach_ICC_2007,Lele_JASP_2010,Zakaria_TWC_2012,45_chrislin2010decoding,Mehmood_2012}} &\parbox{\classificationParboxSize}{} &\parbox{\classificationParboxSize}{\cite{89_amini2010isotropic}} &\parbox{\classificationParboxSize}{\cite{4_Siohan_TSP_2002,105_Vahlin_TC_1996}} &\parbox{\classificationParboxSize}{\cite{4_Siohan_TSP_2002,54_Du_ICICS_07}} &\parbox{\classificationParboxSize}{\cite{84_haas1997time}} \\ \hline\parbox{\classificationParboxTitleSize}{Equalization} &\parbox{\classificationParboxSize}{\cite{A10_Sari_GLOBECOM_1994,A11_Sari_ComMag_1995,A12_Falconer_ComMag_2002,Peled_1980,sahin_2013a,Ihalainen_2007,PHYDYAS_D31_2008,Waldhauser_2008a,Baltar_2010,48_ihalainen2011channel,Waldhauser_2008,Waldhauser_2009,61_stitz2010pilot}} &\parbox{\classificationParboxSize}{\cite{behrouz_2003}} &\parbox{\classificationParboxSize}{\cite{16_cherubini2002filtered,22_benvenuto2002equalization,Fettweis_2009,Rusek_2006,Rusek_TWC_2009,sahin_2013a,Ihalainen_2007,Hirosaki_1980,Waldhauser_2008a,48_ihalainen2011channel,Waldhauser_2008,Waldhauser_2009,63_stitz2008practical,82_datta2011fbmc}} &\parbox{\classificationParboxSize}{\cite{PHYDYAS_D31_2008,42_ikhlef2009enhanced,Baltar_2010,61_stitz2010pilot,5_bellanger2008filter,67_zakaria2010maximum}} &\parbox{\classificationParboxSize}{\cite{62_lele2008channel,59_lele2007preamble}} &\parbox{\classificationParboxSize}{\cite{Rusek_TWC_2009,sahin_2013a}} &\parbox{\classificationParboxSize}{\cite{77_kozek1998nonorthogonal,88_jung2007wssus,80_matz2007analysis,81_trigui2007optimum,75_schafhuber2002pulse}} &\parbox{\classificationParboxSize}{\cite{105_Vahlin_TC_1996}} &\parbox{\classificationParboxSize}{} &\parbox{\classificationParboxSize}{\cite{93_aldirmaz2010spectrally}} \\ \hline\parbox{\classificationParboxTitleSize}{Time \& Frequency Synchronization} &\parbox{\classificationParboxSize}{\cite{Hwang_2009,SaeediSourck20111604,Du2008,64_du2008pulse,36_schaich2010filterbank,35_Ringset_FNMS_2010,37_fusco2008sensitivity}} &\parbox{\classificationParboxSize}{\cite{Jung_Thesis_2007}} &\parbox{\classificationParboxSize}{\cite{SaeediSourck20111604,Du2008,64_du2008pulse,Ihalainen_2007,37_fusco2008sensitivity,58_fusco2009data,63_stitz2008practical,Fusco_SPAWC_2007,Fusco_TSP_2007}} &\parbox{\classificationParboxSize}{\cite{36_schaich2010filterbank,PHYDYAS_D31_2008,35_Ringset_FNMS_2010,61_stitz2010pilot,68_fusco2010joint,60_stitz2009cfo}} &\parbox{\classificationParboxSize}{\cite{SaeediSourck20111604}} &\parbox{\classificationParboxSize}{} &\parbox{\classificationParboxSize}{\cite{SaeediSourck20111604,33_saeedi2011complexity}} &\parbox{\classificationParboxSize}{} &\parbox{\classificationParboxSize}{\cite{Du2008,64_du2008pulse}} &\parbox{\classificationParboxSize}{\cite{SaeediSourck20111604}} \\ \hline\parbox{\classificationParboxTitleSize}{Filter \\ Adaptation} &\parbox{\classificationParboxSize}{\cite{Muquet_2002,64_du2008pulse,sahin_2012a,Sahin_2011_edgeSchedule,sahin_2011edge}} &\parbox{\classificationParboxSize}{\cite{feichtinger1998gabor,SPMAG2013,2_le1995coded}} &\parbox{\classificationParboxSize}{\cite{64_du2008pulse}} &\parbox{\classificationParboxSize}{} &\parbox{\classificationParboxSize}{\cite{39_strohmer2003optimal}} &\parbox{\classificationParboxSize}{\cite{117_Han_TSP_2007,86_han2009wireless,99_Kozek_SPAWC_1997}} &\parbox{\classificationParboxSize}{\cite{77_kozek1998nonorthogonal,88_jung2007wssus,80_matz2007analysis,Jung_Thesis_2007,81_trigui2007optimum,75_schafhuber2002pulse}} &\parbox{\classificationParboxSize}{} &\parbox{\classificationParboxSize}{\cite{Du2008,64_du2008pulse}} &\parbox{\classificationParboxSize}{\cite{77_kozek1998nonorthogonal,81_trigui2007optimum}} \\ \hline\parbox{\classificationParboxTitleSize}{Spatial \\ Domain \\ Approaches} &\parbox{\classificationParboxSize}{\cite{Hwang_2009,48_ihalainen2011channel,Tabach_ICC_2007,Zakaria_TWC_2012,45_chrislin2010decoding,46_payaroresource,52_xiang2011ici,31_estella2010ofdm}} &\parbox{\classificationParboxSize}{\cite{44_bellanger2010fbmc}} &\parbox{\classificationParboxSize}{\cite{48_ihalainen2011channel,52_xiang2011ici}} &\parbox{\classificationParboxSize}{\cite{Zakaria_TWC_2012,Bellanger_ISWPC_2008,46_payaroresource,10_payaro2010performance,31_estella2010ofdm}} &\parbox{\classificationParboxSize}{\cite{Tabach_ICC_2007,Lele_JASP_2010,Zakaria_TWC_2012,45_chrislin2010decoding}} &\parbox{\classificationParboxSize}{} &\parbox{\classificationParboxSize}{\cite{89_amini2010isotropic}} &\parbox{\classificationParboxSize}{} &\parbox{\classificationParboxSize}{} &\parbox{\classificationParboxSize}{} \\ \hline\parbox{\classificationParboxTitleSize}{Gabor \\ Analysis} &\parbox{\classificationParboxSize}{} &\parbox{\classificationParboxSize}{\cite{Daubechies_1992,feichtinger1998gabor,christensen2003,Benedetto_1994,Janssen_2001,heil_2007,SPMAG2013,Daubechies_1990,Daubechies_1991,janssen_1994_duality,95_Strohmer_ACHA_2001,Werther_2005,Kutyniok_2005,97_Sondergaard_Thesis_2007,Jung_Thesis_2007}} &\parbox{\classificationParboxSize}{} &\parbox{\classificationParboxSize}{} &\parbox{\classificationParboxSize}{} &\parbox{\classificationParboxSize}{\cite{Gabor_1946,117_Han_TSP_2007}} &\parbox{\classificationParboxSize}{\cite{77_kozek1998nonorthogonal,88_jung2007wssus,80_matz2007analysis}} &\parbox{\classificationParboxSize}{} &\parbox{\classificationParboxSize}{} &\parbox{\classificationParboxSize}{} \\ \hline\parbox{\classificationParboxTitleSize}{PPN \& DFT} &\parbox{\classificationParboxSize}{\cite{A2_Weinstein_TCT_1971}} &\parbox{\classificationParboxSize}{\cite{Siclet_2000,40_ihalainen2009filter,29_hirosaki1981orthogonally,78_bellanger1976digital,79_vaidyanathan1990multirate,4_Siohan_TSP_2002,36_schaich2010filterbank,44_bellanger2010fbmc,9_waldhauser2006comparison,behrouz_2003,48_ihalainen2011channel,33_saeedi2011complexity,61_stitz2010pilot,60_stitz2009cfo,32_shaat2010computationally,50_zhang2008oversampled,A4_Bellanger_TC_1974,6_rhee1998performance,8_lee2004polyphase}} &\parbox{\classificationParboxSize}{} &\parbox{\classificationParboxSize}{} &\parbox{\classificationParboxSize}{} &\parbox{\classificationParboxSize}{} &\parbox{\classificationParboxSize}{} &\parbox{\classificationParboxSize}{} &\parbox{\classificationParboxSize}{} &\parbox{\classificationParboxSize}{} \\ \hline\parbox{\classificationParboxTitleSize}{Spreading Approaches} &\parbox{\classificationParboxSize}{\cite{Hwang_2009,A10_Sari_GLOBECOM_1994,A11_Sari_ComMag_1995,A12_Falconer_ComMag_2002,A8_Hyung_VTMag_2006,A9_Berardinelli_WirelessComm_2008,40_ihalainen2009filter,102_Gharba_ISWCS_2010,113_Gharba_VTC_2012,47_Kollar_VTC}} &\parbox{\classificationParboxSize}{\cite{47_Kollar_VTC}} &\parbox{\classificationParboxSize}{\cite{40_ihalainen2009filter}} &\parbox{\classificationParboxSize}{\cite{Zakaria_TWC_2012,70_Viholainen_PHYDYAS}} &\parbox{\classificationParboxSize}{\cite{lele_2007cdma,Lele_JASP_2010,Zakaria_TWC_2012}} &\parbox{\classificationParboxSize}{} &\parbox{\classificationParboxSize}{} &\parbox{\classificationParboxSize}{} &\parbox{\classificationParboxSize}{\cite{102_Gharba_ISWCS_2010,113_Gharba_VTC_2012}} &\parbox{\classificationParboxSize}{} \\ \hline\parbox{\classificationParboxTitleSize}{WSSUS \\ Assumption} &\parbox{\classificationParboxSize}{\cite{sahin_2013a}} &\parbox{\classificationParboxSize}{\cite{SPMAG2013,Jung_Thesis_2007,64_du2008pulse,85_bello1963characterization,91_molnar1996wssus,90_matz2005statistical,99_Kozek_SPAWC_1997}} &\parbox{\classificationParboxSize}{\cite{sahin_2013a}} &\parbox{\classificationParboxSize}{} &\parbox{\classificationParboxSize}{\cite{39_strohmer2003optimal}} &\parbox{\classificationParboxSize}{\cite{117_Han_TSP_2007,86_han2009wireless,sahin_2013a}} &\parbox{\classificationParboxSize}{\cite{77_kozek1998nonorthogonal,88_jung2007wssus,80_matz2007analysis,81_trigui2007optimum,75_schafhuber2002pulse}} &\parbox{\classificationParboxSize}{} &\parbox{\classificationParboxSize}{\cite{Du2008}} &\parbox{\classificationParboxSize}{} \\ \hline\parbox{\classificationParboxTitleSize}{Complexity Analysis} &\parbox{\classificationParboxSize}{\cite{Murphy_TC_2002,Waldhauser_2008a,Waldhauser_2008,Waldhauser_2009,11_baltar2007out,50_zhang2008oversampled}} &\parbox{\classificationParboxSize}{\cite{22_benvenuto2002equalization,36_schaich2010filterbank,48_ihalainen2011channel,33_saeedi2011complexity,11_baltar2007out,50_zhang2008oversampled,70_Viholainen_PHYDYAS}} &\parbox{\classificationParboxSize}{\cite{Waldhauser_2008a,Waldhauser_2008,Waldhauser_2009}} &\parbox{\classificationParboxSize}{\cite{5_bellanger2008filter}} &\parbox{\classificationParboxSize}{\cite{Mehmood_2012}} &\parbox{\classificationParboxSize}{} &\parbox{\classificationParboxSize}{} &\parbox{\classificationParboxSize}{} &\parbox{\classificationParboxSize}{} &\parbox{\classificationParboxSize}{} \\ \hline\parbox{\classificationParboxTitleSize}{Channel \\ Estimation} &\parbox{\classificationParboxSize}{\cite{Hwang_2009,Du2008,116_lele_ICC_2008,du_2009_preamble}} &\parbox{\classificationParboxSize}{\cite{Katselis_2010}} &\parbox{\classificationParboxSize}{\cite{Du2008,63_stitz2008practical,du_2009_preamble}} &\parbox{\classificationParboxSize}{} &\parbox{\classificationParboxSize}{\cite{62_lele2008channel,59_lele2007preamble,Javaudin_VTC_2003,116_lele_ICC_2008}} &\parbox{\classificationParboxSize}{} &\parbox{\classificationParboxSize}{\cite{linhao_2009}} &\parbox{\classificationParboxSize}{} &\parbox{\classificationParboxSize}{\cite{Du2008,116_lele_ICC_2008,du_2009_preamble}} &\parbox{\classificationParboxSize}{} \\ \hline\parbox{\classificationParboxTitleSize}{Bi-orthogonality} &\parbox{\classificationParboxSize}{\cite{86_han2009wireless}} &\parbox{\classificationParboxSize}{\cite{christensen2003,janssen_1994_duality,39_strohmer2003optimal,Werther_2005,Siclet_2000}} &\parbox{\classificationParboxSize}{} &\parbox{\classificationParboxSize}{} &\parbox{\classificationParboxSize}{} &\parbox{\classificationParboxSize}{\cite{feichtinger1998gabor,Daubechies_1990,95_Strohmer_ACHA_2001,86_han2009wireless,gabor_book_chap}} &\parbox{\classificationParboxSize}{\cite{77_kozek1998nonorthogonal,80_matz2007analysis,81_trigui2007optimum,75_schafhuber2002pulse}} &\parbox{\classificationParboxSize}{} &\parbox{\classificationParboxSize}{} &\parbox{\classificationParboxSize}{} \\ \hline\parbox{\classificationParboxTitleSize}{Non-orthogonality} &\parbox{\classificationParboxSize}{} &\parbox{\classificationParboxSize}{} &\parbox{\classificationParboxSize}{\cite{Mazo_1975,Fettweis_2009,Liveris_2003,Rusek_2005,Rusek_2006,Rusek_TWC_2009,82_datta2011fbmc}} &\parbox{\classificationParboxSize}{} &\parbox{\classificationParboxSize}{\cite{Mehmood_2012}} &\parbox{\classificationParboxSize}{\cite{117_Han_TSP_2007,49_tonello2005performance,Rusek_TWC_2009}} &\parbox{\classificationParboxSize}{} &\parbox{\classificationParboxSize}{} &\parbox{\classificationParboxSize}{} &\parbox{\classificationParboxSize}{\cite{93_aldirmaz2010spectrally}} \\ \hline\parbox{\classificationParboxTitleSize}{Uplink \& Downlink Performance} &\parbox{\classificationParboxSize}{\cite{A8_Hyung_VTMag_2006,102_Gharba_ISWCS_2010,113_Gharba_VTC_2012,36_schaich2010filterbank,37_fusco2008sensitivity,66_zhang2010spectral}} &\parbox{\classificationParboxSize}{} &\parbox{\classificationParboxSize}{\cite{37_fusco2008sensitivity}} &\parbox{\classificationParboxSize}{\cite{36_schaich2010filterbank,66_zhang2010spectral,34_Medjahdi_TWC_2011}} &\parbox{\classificationParboxSize}{} &\parbox{\classificationParboxSize}{} &\parbox{\classificationParboxSize}{} &\parbox{\classificationParboxSize}{} &\parbox{\classificationParboxSize}{\cite{102_Gharba_ISWCS_2010,113_Gharba_VTC_2012}} &\parbox{\classificationParboxSize}{} \\ \hline\parbox{\classificationParboxTitleSize}{Orthogonaliz. Procedure} &\parbox{\classificationParboxSize}{} &\parbox{\classificationParboxSize}{\cite{56_bolcskei1999design,gabor_book_chap}} &\parbox{\classificationParboxSize}{} &\parbox{\classificationParboxSize}{} &\parbox{\classificationParboxSize}{\cite{2_le1995coded,39_strohmer2003optimal}} &\parbox{\classificationParboxSize}{\cite{feichtinger1998gabor,95_Strohmer_ACHA_2001,39_strohmer2003optimal}} &\parbox{\classificationParboxSize}{} &\parbox{\classificationParboxSize}{\cite{Halpern_1979,105_Vahlin_TC_1996,Nigam_2010}} &\parbox{\classificationParboxSize}{} &\parbox{\classificationParboxSize}{} \\ \hline\parbox{\classificationParboxTitleSize}{FMT} &\parbox{\classificationParboxSize}{\cite{49_tonello2005performance}} &\parbox{\classificationParboxSize}{\cite{28_farhang2010signal,24_farhang2011ofdm}} &\parbox{\classificationParboxSize}{\cite{15_cherubini1999filtered,16_cherubini2002filtered,22_benvenuto2002equalization,49_tonello2005performance,37_fusco2008sensitivity,Fusco_SPAWC_2007}} &\parbox{\classificationParboxSize}{} &\parbox{\classificationParboxSize}{} &\parbox{\classificationParboxSize}{\cite{49_tonello2005performance}} &\parbox{\classificationParboxSize}{} &\parbox{\classificationParboxSize}{} &\parbox{\classificationParboxSize}{} &\parbox{\classificationParboxSize}{} \\ \hline\parbox{\classificationParboxTitleSize}{Spectral Leakage} &\parbox{\classificationParboxSize}{\cite{47_Kollar_VTC,Ihalainen_2007,11_baltar2007out,50_zhang2008oversampled,112_Jiang_GLOBECOM_2012}} &\parbox{\classificationParboxSize}{\cite{47_Kollar_VTC}} &\parbox{\classificationParboxSize}{\cite{Ihalainen_2007,11_baltar2007out,50_zhang2008oversampled}} &\parbox{\classificationParboxSize}{\cite{112_Jiang_GLOBECOM_2012}} &\parbox{\classificationParboxSize}{} &\parbox{\classificationParboxSize}{} &\parbox{\classificationParboxSize}{} &\parbox{\classificationParboxSize}{} &\parbox{\classificationParboxSize}{} &\parbox{\classificationParboxSize}{} \\ \hline\parbox{\classificationParboxTitleSize}{Resource \\ Allocation} &\parbox{\classificationParboxSize}{\cite{A9_Berardinelli_WirelessComm_2008,40_ihalainen2009filter,32_shaat2010computationally,46_payaroresource,66_zhang2010spectral}} &\parbox{\classificationParboxSize}{} &\parbox{\classificationParboxSize}{\cite{40_ihalainen2009filter}} &\parbox{\classificationParboxSize}{\cite{32_shaat2010computationally,46_payaroresource,66_zhang2010spectral}} &\parbox{\classificationParboxSize}{} &\parbox{\classificationParboxSize}{} &\parbox{\classificationParboxSize}{} &\parbox{\classificationParboxSize}{} &\parbox{\classificationParboxSize}{} &\parbox{\classificationParboxSize}{} \\ \hline\parbox{\classificationParboxTitleSize}{PAPR} &\parbox{\classificationParboxSize}{\cite{40_ihalainen2009filter,9_waldhauser2006comparison,47_Kollar_VTC,Ihalainen_2007}} &\parbox{\classificationParboxSize}{\cite{47_Kollar_VTC}} &\parbox{\classificationParboxSize}{\cite{40_ihalainen2009filter,9_waldhauser2006comparison}} &\parbox{\classificationParboxSize}{\cite{70_Viholainen_PHYDYAS}} &\parbox{\classificationParboxSize}{} &\parbox{\classificationParboxSize}{} &\parbox{\classificationParboxSize}{\cite{69_farhang2008square}} &\parbox{\classificationParboxSize}{} &\parbox{\classificationParboxSize}{} &\parbox{\classificationParboxSize}{} \\ \hline\parbox{\classificationParboxTitleSize}{Lattice \\ Adaptation} &\parbox{\classificationParboxSize}{\cite{sahin_2012a}} &\parbox{\classificationParboxSize}{\cite{feichtinger1998gabor}} &\parbox{\classificationParboxSize}{} &\parbox{\classificationParboxSize}{} &\parbox{\classificationParboxSize}{\cite{39_strohmer2003optimal}} &\parbox{\classificationParboxSize}{\cite{117_Han_TSP_2007,86_han2009wireless}} &\parbox{\classificationParboxSize}{} &\parbox{\classificationParboxSize}{} &\parbox{\classificationParboxSize}{\cite{Du2008}} &\parbox{\classificationParboxSize}{} \\ \hline\parbox{\classificationParboxTitleSize}{Hardware \\ Impairments} &\parbox{\classificationParboxSize}{\cite{Ihalainen_2007,Moret_2008}} &\parbox{\classificationParboxSize}{\cite{Jung_Thesis_2007,70_Viholainen_PHYDYAS}} &\parbox{\classificationParboxSize}{\cite{Ihalainen_2007,Moret_2008}} &\parbox{\classificationParboxSize}{} &\parbox{\classificationParboxSize}{} &\parbox{\classificationParboxSize}{} &\parbox{\classificationParboxSize}{} &\parbox{\classificationParboxSize}{} &\parbox{\classificationParboxSize}{} &\parbox{\classificationParboxSize}{} \\ \hline\parbox{\classificationParboxTitleSize}{Implementation} &\parbox{\classificationParboxSize}{} &\parbox{\classificationParboxSize}{} &\parbox{\classificationParboxSize}{} &\parbox{\classificationParboxSize}{\cite{Baranda_2011,35_Ringset_FNMS_2010,PHYDYAS_D9_2009}} &\parbox{\classificationParboxSize}{\cite{Mehmood_2012}} &\parbox{\classificationParboxSize}{} &\parbox{\classificationParboxSize}{} &\parbox{\classificationParboxSize}{} &\parbox{\classificationParboxSize}{} &\parbox{\classificationParboxSize}{} \\ \hline\parbox{\classificationParboxTitleSize}{Spectrum Sensing} &\parbox{\classificationParboxSize}{} &\parbox{\classificationParboxSize}{} &\parbox{\classificationParboxSize}{} &\parbox{\classificationParboxSize}{} &\parbox{\classificationParboxSize}{} &\parbox{\classificationParboxSize}{} &\parbox{\classificationParboxSize}{} &\parbox{\classificationParboxSize}{\cite{ 57_farhang2008multicarrier}} &\parbox{\classificationParboxSize}{} &\parbox{\classificationParboxSize}{} \\ \hline\end{tabular}

\normalsize
\end{center}
\end{table*}

\newpage
\begin{acronyms*}
\acrodef{OFDM}{Orthogonal frequency division multiplexing}
\acrodef{BFDM}{Bi-orthogonal frequency division multiplexing}
\acrodef{FBMC}{Filter bank multicarrier}
\acrodef{FB-S-FBMC}{Filter-bank-spread-filter-bank multicarrier}
\acrodef{SC-FDMA}{Single carrier frequency division multiple accessing}
\acrodef{GFDM}{Generalized frequency division multiplexing}
\acrodef{CDMA}{Code division multiple accessing}
\acrodef{CP}{Cyclic prefix}
\acrodef{ZP}{Zero-padded}
\acrodef{FMT}{Filtered multitone}
\acrodef{SMT}{Staggered multitone}
\acrodef{CMT}{Cosine-modulated multitone}
\acrodef{VSB}{Vestigial side-band modulation}
\acrodef{CPM}{Continuous phase modulation}

\acrodef{HCF}{Half-cosine function}
\acrodef{RRC}{Root-raised cosine}
\acrodef{PSWF}{Prolate spheroidal wave function}
\acrodef{OFDP}{Optimal finite duration pulse}
\acrodef{IOTA}{Isotropic orthogonal transform algorithm}
\acrodef{EGF}{Extended Gaussian function}

\acrodef{WSSUS}{Wide-sense stationary uncorrelated scattering}
\acrodef{BER}{Bit error rate}
\acrodef{ICI}{Inter-carrier interference}
\acrodef{ISI}{Inter-symbol interference}
\acrodef{AIC}{Akaike information criterion}
\acrodef{TO}{Timing offset}
\acrodef{CFO}{Carrier frequency offset}
\acrodef{PN}{Phase noise}
\acrodef{ADC}{Analog-to-digital converter}
\acrodef{PAPR}{Peak-to-average-power ratio}
\acrodef{OOB}{Out-of-band radiation}
\acrodef{MIMO}{Multiple-input multiple-output}
\acrodef{STTC}{Space-time trellis coding}
\acrodef{STBC}{Space-time block coding}

\acrodef{IAM}{Interference approximation method}
\acrodef{MMSE}{Minimum mean square error}
\acrodef{MLSE}{Maximum likelihood sequence estimator}
\acrodef{FDE}{Frequency domain equalization}
\acrodef{SIC}{Successive interference cancellation}

\acrodef{FPGA}{Field-programmable gate array}
\acrodef{DFT}{Discrete Fourier transformation}
\acrodef{IDFT}{Inverse discrete Fourier transformation}
\acrodef{PPN}{Polyphase network}

\end{acronyms*}

\bibliographystyle{IEEEtran}
\bibliography{FBMC_Survey}

\end{document}